\pdfoutput=1
\documentclass{jfm}
\usepackage{graphicx}
\usepackage{natbib}
\usepackage{amsmath,calc,multicol}
\usepackage{subfigure}
\usepackage{amssymb}
\usepackage{latexsym,pdflscape}
\usepackage{epsfig,curves,rotating,graphicx}
\usepackage{wasysym}

\let\realverbatim=\verbatim
\let\realendverbatim=\endverbatim
\renewcommand\verbatim{\par\addvspace{6pt plus 2pt minus 1pt}\realverbatim}
\renewcommand\endverbatim{\realendverbatim\addvspace{6pt plus 2pt minus 1pt}}
\makeatletter
\newcommand\verbsize{\@setfontsize\verbsize{10}\@xiipt}
\renewcommand\verbatim@font{\verbsize\normalfont\ttfamily}
\makeatother
\ifCUPmtlplainloaded \else
  \checkfont{eurm10}
  \iffontfound
    \IfFileExists{upmath.sty}
      {\typeout{^^JFound AMS Euler Roman fonts on the system,
                   using the 'upmath' package.^^J}%
       \usepackage{upmath}}
      {\typeout{^^JFound AMS Euler Roman fonts on the system, but you
                   don't seem to have the}%
       \typeout{'upmath' package installed. jfm.cls can take advantage
                 of these fonts,^^Jif you use 'upmath' package.^^J}%
      }
  \else
  \fi
\fi

\ifCUPmtlplainloaded \else
  \checkfont{msam10}
  \iffontfound
    \IfFileExists{amssymb.sty}
      {\typeout{^^JFound AMS Symbol fonts on the system, using the
                'amssymb' package.^^J}%
       \usepackage{amssymb}%
         \let\leq=\leqslant
         \let\geq=\geqslant
      }{}
  \fi
\fi

\ifCUPmtlplainloaded \else
  \IfFileExists{amsbsy.sty}
    {\typeout{^^JFound the 'amsbsy' package on the system, using it.^^J}%
     \usepackage{amsbsy}}
    {}
\fi

\newsavebox{\astrutbox}
\sbox{\astrutbox}{\rule[-5pt]{0pt}{20pt}}

\usepackage{amsmath}
\usepackage{amsfonts}
\usepackage{bm}
\usepackage{euscript}
\usepackage{wasysym}
\usepackage{graphicx}
\usepackage{subfigure}
\usepackage[usenames,dvipsnames]{color}
\usepackage{lscape}
\usepackage{rotating}
\usepackage{framed}
\usepackage{appendix}
\usepackage[english]{babel}

\graphicspath{{figs_pdf/}}

\newcommand{\mathd}{\mathrm{d}}

\newcommand{\Diff}{\mathrm{D}}

\newcommand{\veca}{\bm{v}}
\newcommand{\vmode}{\bm{v}_\alpha}
\newcommand{\philaplace}{\phi_{\alpha\,\omega}}

\newcommand{\ai}{\alpha_{\mathrm{i}}}
\newcommand{\ar}{\alpha_{\mathrm{r}}}
\newcommand{\armax}{\alpha_{\mathrm{r,max}}}
\newcommand{\omi}{\omega_{\mathrm{i}}}
\newcommand{\omitemp}{\omega_{\mathrm{i}}^{\mathrm{temp}}}
\newcommand{\omrtemp}{\omega_{\mathrm{i}}^{\mathrm{temp}}}
\newcommand{\omizero}{\omega_{\mathrm{i}0}}
\newcommand{\omr}{\omega_{\mathrm{r}}}
\newcommand{\cg}{c_{\mathrm{g}}}

\newcommand{\Reint}{Re_\mathrm{int}}

\newcommand{\thickness}{\epsilon}

\newcommand{\dgas}{d_G}
\newcommand{\dliq}{d_L}

\newcommand{\grav}{F}
\newcommand{\surft}{S}
\newcommand{\gravpre}{F_0}
\newcommand{\surftpre}{S_0}

\newcommand{\imag}{\mathrm{i}}
\newcommand{\mathe}{\mathrm{e}}
\newcommand{\biberg}{\mathcal{F}}
\newcommand{\geom}{\theta}

\title[Journal of Fluid Mechanics]{Absolute linear instability in laminar and turbulent gas/liquid two-layer channel flow}

\author[L. \'O N\'araigh, P. D. M. Spelt and S. J. Shaw]{L\ls E\ls N\ls N\ls O\ls N\ns \'{O}\ns N\ls \'{A}\ls R\ls A\ls I\ls G\ls H$^1$,
\ns P\ls E\ls T\ls E\ls R\ns D.\ns M.\ns S\ls P\ls E\ls L\ls T$^2$\ \and S\ls T\ls E\ls P\ls H\ls E\ls N\ns J.\ns S\ls H\ls A\ls W$^3$ \ns}

\affiliation{
$^{1}$School of Mathematical Sciences, University College Dublin, Belfield, Dublin 4, Ireland\\
$^{2}$D\'{e}partement M\'{e}canique, Universit\'{e} de Lyon 1 and Laboratoire de la M\'{e}canique des Fluides \& d'Acoustique (LMFA), CNRS, Ecole Centrale Lyon, Ecully, France\\
$^{3}$Department of Mathematical Science, Xi'an Jiaotong-Liverpool University, 111 Ren Ai Road, Dushu Lake Higher
Education Town, Suzhou, Jiangsu, 215123, China.
}

\volume{000}
\pagerange{\pageref{firstpage}--\pageref{lastpage}}
\doi{}

\begin{document}

\label{firstpage}
\maketitle

\date{\today}

\begin{abstract}
We study two-phase stratified flow where the bottom layer is a thin laminar liquid and the upper layer is a fully-developed gas flow.  The gas flow can be laminar or turbulent.  To determine the boundary between convective and absolute instability, we use Orr--Sommerfeld stability theory, and a combination of linear modal analysis and ray analysis.   For turbulent gas flow, and for the density ratio $r=1000$, we find large regions of parameter space that produce absolute instability.  These parameter regimes involve viscosity ratios of direct relevance to oil/gas flows.   If, instead, the gas layer is laminar, absolute instability persists for the density ratio $r=1000$, although the convective/absolute stability boundary occurs at a viscosity ratio that is an order of magnitude smaller than in the turbulent case.
Two further unstable  temporal modes exist in both the laminar and the turbulent cases, one of which can exclude absolute instability. We compare our results with an experimentally-determined flow-regime map, and discuss the potential application of the present method to non-linear analyses.
\end{abstract}
\maketitle
%
%
%
%\newpage
\section{Introduction}
\label{sec:intro}
%
%\commentpdms{The linear temporal instability of two-layer fluid flows has attracted much attention, either for oceanographic or industrial (oil/gas transport) applications. We are here concerned with confined systems where the main, long-standing problem is to develop understanding and modelling of the onset of and route towards droplet entrainment in pressure-driven gas/liquid flows~\citep{Hewitt1970}. The understanding and modelling of temporal linear instability have well advanced for these flows, ~\citep{Miesen1995,Boomkamp1996,Boomkamp1997,Ozgen1998,Onaraigh2011a,Onaraigh2011b}, but the spatio-temporal instability characteristics are not well understood at all. Yet we anticipate this to be an important ingredient in any subsequent nonlinear analysis. Although it may seem a reasonable assumption that disturbances mainly be introduced at the inlet of a pipe or channel, it would appear important to know whether such disturbance is merely advected downstream whilst it is amplified, or destabilizes the entire flow from there on. Also, a perturbation may be introduced through droplet impact onto a liquid layer further downstream. Our recent work \cite{Valluri2010} has investigated this problem, but only in laminar liquid/liquid systems. The current work aims to fill in these two gaps in the literature (gas/liquid systems, and turbulence) and introduces modifications and extensions of existing methodologies (developed previously for single-phase flows or for temporal stability analysis only) that are potentially of interest in other areas.}

We investigate linear absolute and convective instability for a liquid film sheared by
laminar and turbulent gas streams in a channel. In the oil/gas industries, this approach
serves as a model that can be used to predict the onset of droplet entrainment~\citep{Hewitt1970}. 
The motivation for this work is twofold: previous work on the turbulent case focussed uniquely on temporal stability analysis~\citep{Miesen1995,Boomkamp1996,Boomkamp1997,Ozgen1998},
while previous work on the laminar case~\citep{Valluri2010} omitted large regions of parameter space that
are relevant to the oil-and-gas industries, and which are found herein to be absolutely
unstable. The current work aims to fill in these two gaps in the literature and introduces
modifications and extensions of existing methodologies (developed previously for single-phase flows or for temporal stability analysis only) that are potentially of interest in other areas.

The route to droplet entrainment from a liquid layer into a gas stream in pipe and channel flows is still unclear. The idealized system of fully-developed flow with a flat gas/liquid interface is  linearly unstable to infinitesimally small perturbations. For a laminar base state, a stratification in dynamic viscosity produces instability; for a turbulent base state, the mechanism of~\citet{Miles1957} may also dominate for deep liquid layers at large Froude numbers~\citep{Onaraigh2011b}.  Other mechanisms, such as a Tollmien--Schlichting mode in the liquid~\citep{Boomkamp1996} may also be important in certain parameter regimes. Although linear (temporal) instability is arguably a necessary condition for droplet entrainment in the gas, it does not provide much insight regarding whether a localized disturbance grows whilst it is merely convected downstream or whether instead the disturbance destabilizes the entire system. To answer these questions a spatio-temporal analysis is required. A recent spatio-temporal analysis by~\citet{Valluri2010} has revealed a region in parameter space wherein the laminar base state is absolutely unstable, indicating that the system does not merely act as an amplifier but also as a generator of disturbances. This was found to include a large range of practically useful viscosity ratios, but was limited to a density ratio of $O(1)$. 

The laminar density-matched problem studied by~\citet{Valluri2010} has only limited applicability in oil/gas transport, where the density ratio is large, and where the operating conditions produce turbulence in the gas layer, or in both layers (see, e.g., the visualisation of droplet entrainment events by~\citet{Lecoeur2010}). The present study therefore investigates the corresponding problem for a turbulent base state, but the laminar case is also revisited.
Although replacement of the base state by a turbulent one~\citep{Onaraigh2010,Onaraigh2011a} in a linear modal spatio-temporal analysis may seem trivial, the results turn out to be difficult to interpret, due to the presence of multiple unstable modes. Therefore, in this study, we have developed a twin-track approach, in which modal analysis and ray analysis are combined to locate and characterize absolute instability.  The ray analysis used herein extends the work of~\citet{Chomaz1998b} for single-phase flows.
This approach yields surprising results.
In particular, it has revealed significant regions in parameter space where the turbulent base state is absolutely unstable for large density ratios. This also holds for the laminar base state, thereby contradicting \cite{Valluri2010}, who only found absolute instability for density ratios of $O(1)$. We have therefore revisited the spatio-temporal work of~\citet{Valluri2010}, and have established using both ray and modal analyses that the laminar system  is indeed absolutely unstable at large density ratios for a substantial range of viscosity ratios and  liquid-film depths. Reasons for the oversight in the previous work are  given.

The paper is organized as follows. The turbulence model and the linear stability analysis are
formulated in Section~\ref{sec:model}. We discuss some theoretical and numerical aspects of the linear
stability analysis in Section~\ref{sec:postprocess}, paying close attention to the development of a ray analysis
for two-phase flows. We apply this technique to the turbulent base state in Section~\ref{sec:turb},
while the laminar case is revisited in Section~\ref{sec:laminar}. In Section~\ref{sec:conc} we argue for the importance of
using the ray analysis and the modal analysis simultaneously, for complete and accurate
results.   We also compare the flow-regime boundaries identified herein with those found in experiments, and discuss the generalisation of our work to non-linear and non-parallel flows.

\section{Linear stability analysis}
\label{sec:model}
In this section we review a model of turbulent channel flow used elsewhere by the
authors~\citep{Onaraigh2010,Onaraigh2011a}. This is a Reynolds-averaged
model describing co-current flow in a stably-stratified system where the upper layer is a
turbulent gas and the lower layer a laminar liquid film.  We also recall the Orr--Sommerfeld
\begin{table}
  \begin{center}
\vspace{\baselineskip}
  \begin{tabular}{p{5cm}p{3cm}p{3cm}}
  \hline
  Physical Quantity            &           Symbol                  & Numerical value \\
  \hline
  Gas-layer dynamic viscosity  & $\mu_G$                           &$1.8\times 10^{-5}\,\mathrm{Pa}\cdot\mathrm{s}$\\
  Liquid-layer dynamic viscosity & $\mu_L$                         &\\
  Viscosity ratio              & $m=\mu_L/\mu_G$                   &$50$--$10000$\\
  Gas-layer density            & $\rho_G$                          &$1\,\mathrm{kg}\,\mathrm{m}^{-3}$\\
  Liquid-layer density         & $\rho_L$                          &\\
  Density ratio                & $r=\rho_L/\rho_G$                 &$10^3$\\
  Liquid film thickness        & $\dliq$                           &$10^{-3}$--$10^{-2}$ m\\
  Gas-layer depth              & $\dgas$                           &\\
  Ratio of layer depths        & $\thickness=\dliq/\dgas$          &$0.01$--$0.2$\\
  Acceleration due to gravity  & $g$                               &$9.8\,\mathrm{m}\,\mathrm{s}^{-2}$\\
  Surface tension              & $\gamma$                          &$0.07\,\mathrm{N}\,\mathrm{m}^{-1}$\\
  \hline
  \end{tabular}
  \caption{Table of parameters in the two-phase problem and their typical values.   The range of values of the liquid-layer dynamic viscosity $\mu_L$, the liquid-layer density $\rho_L$, and the liquid-film thickness $d_L$ can be backed out from the gas-layer analogues and the ratios $m$, $r$, and $\thickness$, respectively.}
  \label{tab:values0}%
  \end{center}%
\end{table}%
technique used to determine the stability of the interface in this two-layer system.  For reference, typical values of the problem parameters are given in Table~\ref{tab:values0}, where the subscripts $G,L$ indicate the gas and liquid, respectively.

\subsection{The base state}
\label{sec:model:basestate}
\begin{figure}
\begin{center}
\includegraphics[width=0.7\textwidth]{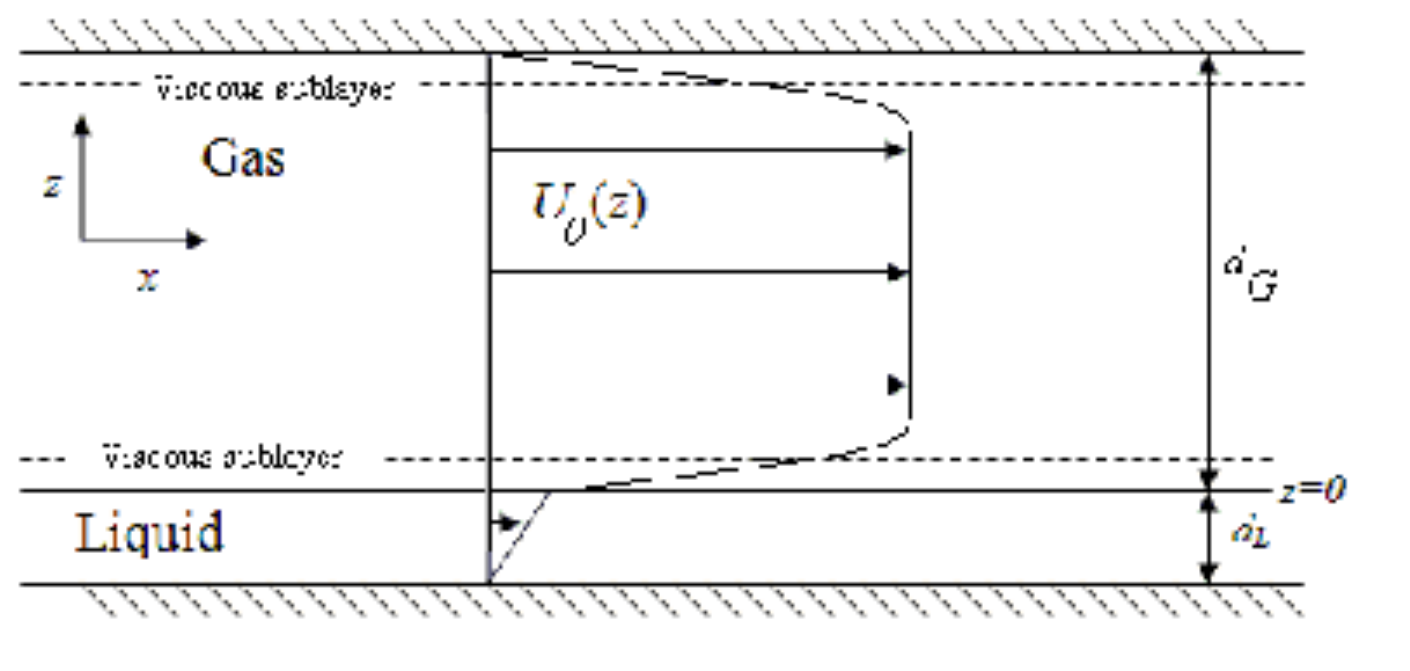}
\end{center}
\caption{A schematic diagram of the base flow.  The liquid layer is laminar,
while the gas layer exhibits fully-developed turbulence, described here by
a Reynolds-averaged velocity profile.  A pressure gradient in the $x$-direction
drives the flow.}
\label{fig:schematic0}
\end{figure}

We consider a flat-interface base state in two-layer stratified flow (Figure~\ref{fig:schematic0}). The bottom layer is a thin, laminar, liquid layer, and the top
layer is gaseous, turbulent and fully-developed. A pressure gradient is applied along the
channel. The base-state profile of the system is a uni-directional flow in the horizontal, $x$-
direction as a function of the cross-flow coordinate $z$. In the bottom layer, the profile is determined by balancing the viscous and the
pressure forces; in the top layer, the viscosity in the balance law must be supplemented
by the turbulent eddy viscosity:
\begin{equation} \mu_G\frac{\partial U_0}{\partial
z}+\tau_{0}=\tau_{\mathrm{i}}+\frac{dP}{dL}z,
\label{eq:rans0}
\end{equation}
where $U_0(z)$ is the base-state velocity in the gas, $\tau_\mathrm{i}$ is the interfacial shear stress, and $dP/dL$ is the applied pressure gradient.  Moreover, $\tau_{0}=-\rho_G\langle u' w'\rangle$ is the
turbulent shear stress due to the averaged effect of the turbulent
fluctuating velocities, $u^{'}$ and $w^{'}$. In channel flows, it is appropriate to model
this term using an eddy-viscosity model~\citep{TurbulenceMonin}.
 In mixing-length theory, the eddy viscosity depends on the local
 rate of strain~\citep{Bradshaw1974}, which means that the turbulent shear
 stress depends on the square of the rate of strain.  Instead, as in the work of~\citet{Onaraigh2011a} and~\citet{Biberg2007}, we use an interpolation function for the eddy viscosity.  This mimics the ordinary mixing-length theory near the interface
 and near the wall, and transitions smoothly  from the near-wall and near-interfacial regions to the  zone surrounding the gas centreline. Thus, the
 turbulent shear stress is linear in the rate of strain, and
\begin{equation}
\tau_{0}=\mu_T\frac{\partial U_0}{\partial z},
\qquad\mu_T=\kappa\rho_G \dgas U_{*\mathrm{w}}\biberg\left(s\right)\psi_{\mathrm{i}}\left(s\right)\psi_{\mathrm{w}}\left(1-s\right),\qquad s=z/\dgas,
\label{eq:tss0}
\end{equation}
where
$\mu_T$ is the eddy viscosity and  $\kappa$ is the von K\'arm\'an constant, taken as 0.4.   Additionally, $U_{*\mathrm{w}}$ is the friction velocity at the upper wall.  The corresponding stress is $\tau_\mathrm{w}$, with $U_{*\mathrm{w}}=\sqrt{|\tau_\mathrm{w}|/\rho_G}$.  Similarly, the interfacial friction velocity is defined as $U_*=\sqrt{\tau_\mathrm{i}/\rho_G}$.
The function $\biberg$ is the interpolation function described in~\citet{Onaraigh2010} and~\citet{Onaraigh2011a}:
\begin{equation}
\biberg\left(s\right)=s\left(1-s\right)\left[\frac{s^3+|R|^{5/2}\left(1-s\right)^3}{R^2\left(1-s\right)^2+Rs\left(1-s\right)+s^2}\right],\qquad R=\tau_{\mathrm{i}}/\tau_{\mathrm{w}}.\\
\end{equation}
Finally, $\psi_{\mathrm{i}}$ and $\psi_{\mathrm{w}}$ are
interface and wall functions respectively, which damp the effects of
turbulence to zero rapidly near the interface and the wall.  These are given below.

We non-dimensionalize the problem on the gas-layer depth $\dgas$, the gas-layer density $\rho_G$, the gas-layer viscosity $\mu_G$, and the velocity scale $U_p$, where
\[
\rho_G U_p^2=\dgas|dP/dL|.
\]
Then, integration of Equation~\eqref{eq:rans0} yields the non-dimensional base state:
\begin{equation}
U_0\left(z\right)=\begin{cases}\frac{\mu_G}{\mu_L}\left[-\tfrac{1}{2}Re\left(z^2-\thickness^2\right)+\frac{Re_*^2}{Re}\left(z+\thickness\right)\right],&-\thickness\leq z\leq0,\\
\frac{\mu_G}{\mu_L}\left(\tfrac{1}{2}\thickness^2 Re+\thickness\frac{Re_*^2}{Re}\right)
+\frac{Re_*^2}{Re}\int_0^{z}\frac{\left(1-\frac{Re^2}{Re_*^2}s\right)ds}{1+\frac{\kappa Re_{*}}{\sqrt{|R|}}g\left(s\right)\psi_\mathrm{i}\left(s\right)\psi_\mathrm{w}\left(1-s\right)},&0\leq z\leq1,\end{cases}
\label{eq:U_base}
\end{equation}
where $\thickness=\dliq/\dgas$, $Re=\rho_G U_p \dgas /\mu_G$, and where $Re_*=(U_*/U_p)Re$.  Knowledge of $Re_*$ amounts to knowledge of the interfacial shear stress.  This is not known \textit{a priori} as a function of the externally-imposed parameters.  However, it is available within the model, and  the root-finding procedure
\begin{equation}
U_0\left(1;Re_*\right)=0
\label{eq:U_base_restar}
\end{equation}
yields $Re_*$ as a function of the parameters   $\left(Re,\thickness,\mu_L/\mu_G\right)$.
For completeness, we also list the interfacial and wall functions:
\begin{subequations}
\begin{eqnarray}
\psi_\mathrm{i}\left(s\right)&=&1-e^{-C_A Re_*^2s^2},\\
\psi_\mathrm{w}\left(s\right)&=&1-e^{-C_A Re_*^2s^2/R^2},
\end{eqnarray}%
%\label{eq:psi}%
\end{subequations}
where $C_A$ is a constant fixed such that the interfacial and wall viscous sublayers are five wall-units in extent~\citep{TurbulencePope}.
The functional forms for $\biberg$ and the wall functions
are confirmed by the excellent agreement between the model predictions of the
base state and experiments and numerical simulation~\citep{Onaraigh2011a}.
Having constituted the base state, we now introduce the theory necessary to determine its stability.

\subsection{The perturbation equations}
\label{sec:model:perturbations}
We base the dynamical equations for the interfacial motion on the Reynolds-averaged
Navier--Stokes (RANS) equations. The turbulent velocity is decomposed into averaged
and fluctuating parts. We make the quasi-laminar approximation, which means that the
fluctuations are only considered in the base state, where they are modelled using the eddy viscosity (Section~\ref{sec:model:basestate}).   Before deriving equations for perturbations induced by small waves at the interface, we discuss the dynamics of these perturbations with respect to their interactions with the turbulent eddies in the flow.

In a realisation of the three-dimensional turbulent two-layer flow with small-amplitude waves, a Fourier mode decomposition can be made of the interface elevation and field variables. Here, ensemble-averaged Fourier modes are assumed to be predominantly two-dimensional. The correspondingly averaged equations of motion are linearized in terms of wave amplitude. The linearized problem contains wave-induced Reynolds stress terms (WIRSs), but these have been found recently not to be significant in two-layer flows such as those studied here~\citep{Onaraigh2011a,Onaraigh2011b}. This can also be seen from an order-of-magnitude estimate of the WIRSs terms compared to inertial terms in the perturbation momentum equation.  Further theoretical justification exists for the case of viscosity-contrast instability, where the instability is dominated by conditions close to the interface, a zone where the perturbation turbulent stresses
are damped rapidly to zero by the existence of viscous sublayers. In practical terms, the quasi-laminar approximation, wherein the WIRSs are ignored and the effect of turbulence is assumed to be entirely through the base-state velocity profile, while brutal in its simplicity, yields similar results to other turbulence models that explicitly include the WIRSs. It also predicts critical Reynolds numbers for the onset of wavy flow that agree with the laboratory experiments of~\citet{CohenHanratty} and~\citet{Craik}. The reader is referred to the papers by~\citet{Onaraigh2011a,Onaraigh2011b} for further details.

Here, we study linear spatio-temporal instability. Although the above-mentioned prior findings are limited to temporal Fourier modes, there is an analytical connection between spatio-temporal and temporal modes~\citep{Onaraigh2012a} (see also Appendix~\ref{app:quadratic}), such that the properties of the temporal study are inherited by the spatio-temporal one. Therefore, we make here also the quasi-laminar approximation. Thus, a small disturbance $z=\eta\left(x,t\right)$ centred around the flat interface $z=0$ gives rise to disturbances in the velocity and pressure fields, which satisfy the following linearized equations of motion in the $j^{\text{th}}$ phase ($j=L,G$):

%Thus, a small disturbance $z=\eta\left(x,t\right)$ centred around the flat interface $z=0$ gives rise to disturbances in the velocity and pressure fields \commentpdms{(denoted here as ($\delta u,\delta w$ and $\delta p$, respectively}, which satisfy the following linearized equations of motion in the $j^{\text{th}}$ phase ($j=L,G$):
%
%
%
\begin{subequations}
\begin{eqnarray}
r_j\left[\frac{\partial}{\partial t}\delta u+U_0\frac{\partial}{\partial x} \delta u+\frac{\mathd
U_0}{\mathd{z}}\delta w\right]&=&-\frac{\partial}{\partial x}\delta p+\frac{m_j}{Re}\left(\frac{\partial^2}{\partial
x^2}+\frac{\partial^2}{\partial z^2}\right)\delta u,\\
r_j\left[\frac{\partial}{\partial t}\delta w+U_0\frac{\partial}{\partial x}\delta w\right]&=&-\frac{\partial}{\partial
z}\delta p+\frac{m_j}{Re}\left(\frac{\partial^2}{\partial x^2}+\frac{\partial^2}{\partial
z^2}\right)\delta w,\\
\frac{\partial}{\partial x}\delta u+\frac{\partial}{\partial z}\delta w&=&0,
\end{eqnarray}%
\label{eq:ns_linear}%
\end{subequations}%
where $\left(r_L,r_G\right)=\left(r,1\right)$ and $\left(m_L,m_G\right)=\left(m,1\right)$.  Using the incompressibility condition, this system of equations reduces to a single equation in the streamfunction $\phi$.  Further simplification occurs when the streamfunction is written as a sum of normal modes:
\begin{equation}
\phi\left(x,z,t\right)=\frac{1}{2\pi}\int_{C_\alpha=\mathbb{R}}\mathd\alpha\, \mathe^{\imag\alpha x}\phi_\alpha\left(z,t\right),
\label{eq:fourier}
\end{equation}
which in turn can be written in Laplace-transform notation:
\begin{equation}
\phi\left(x,z,t\right)=\frac{1}{4\pi^2}\int_{C_\alpha}\mathd\alpha\int_{C_\omega}\mathd\omega\, \mathe^{\imag\left(\alpha x-\omega t\right)}\phi_{\alpha\,\omega}\left(z\right),
\label{eq:laplace}
\end{equation}
where $C_\omega$ is the Bromwich contour~\citep{ArfkenWeber}.  If the $\omega$-singularities in the function $\phi_{\alpha\,\omega}\left(z\right)$ lie below the real axis in the complex-$\omega$ plane, then the integral~\eqref{eq:laplace} is an ordinary double Fourier integral.
Using the Fourier and Laplace decompositions, Equations~\eqref{eq:ns_linear} reduce to the Orr--Sommerfeld equation:
%
%`
%
\begin{equation}
\imag\alpha r_j\left[\left(U_0-\frac{\omega}{\alpha}\right)\left(\Diff^2-\alpha^2\right)\philaplace-\frac{\mathd^2U_0}{\mathd z^2}\philaplace\right]=\frac{m_j}{Re}\left(\Diff^2-\alpha^2\right)^2\philaplace,
\label{eq:os}
\end{equation}
where $\Diff=\mathd/\mathd z$.  Equation~\eqref{eq:os} only holds in the interior parts of the domain, $z\in\left(-\thickness,0^-\right)\cup\left(0^+,1\right)$.  To close the Equation~\eqref{eq:os}, 
no-slip and no-penetration conditions are applied at $z=-\thickness$ and $z=1$:
\begin{equation}
\philaplace\left(-\thickness\right)=\Diff\philaplace\left(-\thickness\right)=\philaplace\left(1\right)=\Diff\philaplace\left(1\right)=0,
\label{eq:bcs}
\end{equation}
and the streamfunction is matched across the interface $z=0$, where the following conditions hold (we use the notation $c=\omega/\alpha$):
\begin{subequations}
\begin{align}
\phi_{L}&=\phi_{G},\\
\Diff\phi_{L}&=\Diff\phi_{G}+\frac{\phi_G}{c-U_0}\left(\frac{\mathd U_{0}}{\mathd{z}}\bigg|_{0^+}-\frac{\mathd
U_{0}}{\mathd{z}}\bigg|_{0^-}\right),\\
m\left(\Diff^2+\alpha^2\right)\phi_L&=\left(\Diff^2+\alpha^2\right)\phi_{G},
\end{align}
\vskip -0.2in
\begin{multline}
m\left(\Diff^3\phi_L-3\alpha^2\Diff\phi_L\right)+\imag\alpha rRe\left(c-U_0\right)\Diff\phi_L+\imag\alpha
rRe  \frac{\mathd U_0}{\mathd z}\bigg|_{0^-}\phi_L
%
%-\frac{\imag\alpha r Re}{c-U_0}\frac{\grav(r-1)+\alpha^2\surft}{Re^2}\phi_L
%
-\frac{\imag\alpha r Re}{c-U_0}\left(\grav+\alpha^2 \surft\right)\phi_L
\\
=\left(\Diff^3\phi_{G}-3\alpha^2\Diff\phi_{G}\right)+\imag\alpha  Re\left(c-U_0\right)\Diff\phi_{G}+\imag\alpha Re\frac{\mathd U_{0}}{\mathd z}\bigg|_{0^+}\phi_{G}.
\label{eq:ic_normal}%
\end{multline}%
\label{eq:ic_interface}%
\end{subequations}%
Here $\grav$ and $\surft$ denote parameters that encode the effects of gravity and surface tension, respectively; they are defined here for the first time as
\begin{eqnarray}
\grav =\frac{g \dgas}{\left(\mu_G/\rho_G \dgas\right)^2}\frac{r-1}{Re^2}&:=&\gravpre(r-1)/Re^2,\\
\surft =\frac{\gamma}{\mu_G^2/\rho_G \dgas}\frac{1}{Re^2}&:=&\surftpre/Re^2,
\label{eq:fr_values0}%
\end{eqnarray}
where $g$ is acceleration due to gravity and $\gamma$ is surface tension. The appropriate range of values for $\gravpre$ and $\surftpre$  is discussed in Sec.~\ref{sec:turb}.
We abbreviate the Orr-Sommerfeld (or OS) equation~\eqref{eq:os} and the matching conditions~\eqref{eq:bcs}--\eqref{eq:ic_interface} using operator notation,
\begin{equation}
\mathcal{L}_{\alpha\,\omega}\philaplace=\imag\omega \mathcal{M}_{\alpha\,\omega}\philaplace.
\label{eq:os_abbrv}
\end{equation}
This equation amounts to an eigenvalue equation, which we solve numerically by introducing a trial solution:
\begin{subequations}
\begin{eqnarray}
\philaplace\left(z\right)&\approx&\sum_{n=0}^{N_1}a_n T_n\left(\frac{2z}{\thickness}+1\right),\qquad -\thickness \leq z\leq0,\\
\philaplace\left(z\right)&\approx&\sum_{n=0}^{N_2}b_n T_n\left(2z-1\right),\qquad 0 \leq z\leq 1,
\end{eqnarray}%
\label{eq:ansatz}%
\end{subequations}%
where $T_n(\cdot)$ is the $n^{\mathrm{th}}$ Chebyshev polynomial.
We substitute Equation~\eqref{eq:ansatz} into Equation~\eqref{eq:os_abbrv} and evaluate the result at $N_1+N_2-6$ interior points.  The ansatz~\eqref{eq:ansatz} is also substituted into the eight boundary and interfacial conditions.  This yields $N_1+N_2+2$ linear equations in as many unknowns.  In matrix terms, we have to solve
\begin{equation}
L_{\alpha\,\omega}\veca=\imag\omega M_{\alpha\,\omega}\veca,
\label{eq:os_discretized}
\end{equation}
where $\veca=\left(a_0,..,a_{N_1},b_0,..,b_{N_2}\right)^T$.  Such an equation is readily solved using linear-algebra packages.  This method is described in more detail and its implementation is tested against benchmarks in another paper by the present authors~\citep{Valluri2010}.  The number of collocation points $(N_1+1,N_2+1)$ is adjusted until convergence is achieved.  The application of this numerical method will be the subject of the following sections.

\section{Further numerical methods and postprocessing}
\label{sec:postprocess}
In this section we revisit the basic definition of absolute instability, namely that the
streamfunction response to a localized disturbance should grow exponentially in time at
the origin of the disturbance. Solving the associated Cauchy problem  gives
a quick and clear method to characterize the instability. This approach also enables us to pinpoint the source of the
instability through an energy-budget analysis.  We also review herein an equivalent method to determine absolute instability, namely modal analysis.

\subsection{Modal analysis}
\label{sec:postprocess:modal}

A purely temporal analysis involves the solution of the eigenvalue problem~\eqref{eq:os_abbrv}, where we write $\alpha=\ar+\imag\ai$, $\omega=\omr+\imag\omi$, for $\alpha=\ar$ only.  This gives a dispersion relation 
\[
\left(\omrtemp(\ar),\omitemp(\ar)\right)=\left(\omr(\ar,\ai=0),\omi(\ar,\ai=0)\right),
\]
with associated group velocity $\cg=d\omrtemp/d\ar$.  The pair $(\ar,\omitemp(\ar))$ that maximizes $\omi$ is called the {\textit{most dangerous mode}}.
The flow is  linearly unstable if $\omitemp>0$ for the most dangerous mode.
Unstable parallel flows are further classified as convectively unstable if
initially localized pulses are amplified in at least one moving frame of reference but
are damped in the laboratory frame, and absolutely unstable if such pulses lead to
growing disturbances in the entire domain in the laboratory frame.  To describe  such an instability, we use the description of~\citet{huerre90a}.   An unstable parallel flow is absolutely unstable if the following
criteria have all been met: (i)  $\omizero:=\omi(\alpha_0)>0$, where $\alpha_0$ is the wave number at
which the complex derivative $d\omega/d\alpha$ is zero, (ii)  the corresponding saddle point
$\alpha_0$ in the complex $\alpha$-plane is the result of the coalescence of spatial branches that
originate from opposite half-planes at a larger and positive value of $\omi$ and (iii) the saddle
point pinches at $\omizero$; this is verified by locating a cusp at $\omizero$ in the complex $\omega$
plane~\citep{Lingwood1997,Valluri2010} and ensuring that the complex wave number corresponding to the pinching point coincides with $\alpha_0$.

\subsection{Ray analysis}

The linear stability equations~\eqref{eq:ns_linear} in streamfunction form (Equation~\eqref{eq:os}), together with the boundary and initial conditions~\eqref{eq:bcs}--\eqref{eq:ic_interface} can be neatly encoded in linear-operator form:
\begin{equation}
\mathcal{L}\overline{\phi}=\mathcal{M}\partial_t\overline{\phi},
\label{eq:os_linop}
\end{equation}
where we study $\overline{\phi}(x,z,t)$, the filtered streamfunction containing only positive real wave numbers~\citep{huerre90a}; here $\mathcal{L}$ and $\mathcal{M}$ are linear operators (cf. Equation~\eqref{eq:os_abbrv}).  When an impulsive, localized force is applied to the streamfunction, Equation~\eqref{eq:os_linop} is modified:
\begin{equation}
\mathcal{L}\overline{\phi}+\delta(x)\delta(z)\delta(t)=\mathcal{M}\partial_t\overline{\phi},
\label{eq:os_impulse}
\end{equation}
in terms of the Dirac delta function $\delta(.)$.
The solution of Equation~\eqref{eq:os_impulse} (determined here in a domain with periodic boundaries at $x=\pm L_x/2$) can be used to characterize the instability that develops from the impulse.
We use the following algorithm developed by~\citet{Chomaz1998a} and~\cite{Chomaz1998b}, here applied to two-phase flows:
\begin{enumerate}
\item Compute the complex-valued filtered streamfunction $\overline{\phi}\left(x,z,t\right)$;
\item Form the $L^2$-norm
\begin{equation}
A\left(x,t\right)=\sqrt{\int_{-\thickness}^1\mathd{z}\,|\overline{\phi}\left(x,z,t\right)|^2};
\label{eq:pseudo_energy}
\end{equation}
\item Examine the norm along rays, $A(v,t)=A(x=vt,t)$.  If $A(0,t)$ is a decreasing function of time, the instability is convective.
\end{enumerate}
It suffices to consider positive and zero ray velocities only, since this enables a classification of the instability as either absolute or convective and, furthermore, in the convective case, gives information about the leading- and trailing-edge velocities of the downstream-propagating disturbance.

Additional information can be extracted from the evolution of the norm, provided the contributions to the growth of the streamfunction are dominated by a single eigenmode.  This caveat does not appear in the single-phase work of~\citet{Chomaz1998a} and~\citet{Chomaz1998b}: those problems contain a simple means of projecting the streamfunction on to the eigenmode of interest; the spatial symmetries that produce this projection do not exist in the current problem, and this approach is therefore not applicable.  We therefore assume that the evolution is dominated by a single eigenmode, and justify this assumption \textit{a posteriori}.
Thus, along rays $x=vt$, we assume that $A\left(x=vt,t\right)\sim t^{-1/2}e^{\sigma\left(v\right)t}$, where $\sigma$ is the spatiotemporal growth rate of the dominant eigenmode.  Therefore, we extract the finite-time estimate of the spatiotemporal growth rate as follows:
\begin{subequations}
\begin{equation}
\sigma\left(v\right)=\frac{\ln A\left(vt_2,t_2\right)-\ln A\left(vt_1,t_1\right)}{t_2-t_1}+\tfrac{1}{2}\frac{\ln t_2-\ln t_1}{t_2-t_1},
\label{eq:sigma_v}
\end{equation}
where $t_1$ and $t_2$ are large but finite times and $t_2>t_1$.
The complex wave number and frequency along the ray  $x=vt$ also follow from this analysis~\citep{Chomaz1998a,Chomaz1998b}:
\begin{eqnarray}
\alpha_i(v)&=&-\frac{\mathd\sigma}{\mathd v},
\label{eq:aiv}\\
\omi(v)&=& \sigma(v)+\ai v,\\
\ar(v)&=&\Re\left(\frac{-\imag}{\overline{\phi}}\frac{\partial\overline{\phi}}{\partial x}\right)_{z=0,x=vt},
\label{eq:arv}
\end{eqnarray}%
\label{eq:tdns_growth}%
\end{subequations}%
where Equation~\eqref{eq:arv} holds because the right-hand side is independent of time as $t\rightarrow\infty$.

\subsection{Transient direct numerical simulations}

Transient direct numerical simulation (DNS) of Equation~\eqref{eq:os_impulse} is complicated by the fact that the operator $\mathcal{M}$ is non-invertible. To solve this equation in an optimal way, we have developed our own numerical method, which we outline here.
As in Section~\ref{sec:model}, we write $\overline{\phi}\left(z,t\right)$ as a finite sum of Chebyshev polynomials:
\[
\overline{\phi}\left(x,z,t\right)=\sum_{\alpha>0} e^{\imag\alpha x}
\begin{cases}\sum_n a_{\alpha\,n}T_n\left(\frac{2z}{\thickness}+1\right),&-\thickness\leq z\leq 0,\\
\sum_n b_{\alpha\,n}T_n\left(2z-1\right),&0\leq z\leq 1,
\end{cases}
\]
or more compactly,
\begin{equation}
\overline{\phi}\left(x,z,t\right)=\sum_{\alpha>0} \sum_n e^{\imag\alpha x} v_{\alpha\,n} T_n\left(\eta_j\right),\qquad j=L,G.
\label{eq:phi_ansatz}
\end{equation}
We substitute Equation~\eqref{eq:phi_ansatz} into Equation~\eqref{eq:os_impulse}.  This yields the following equation for the normal mode $\alpha$:
\begin{equation}
M_\alpha\frac{\mathd\vmode}{\mathd t}=L_\alpha\vmode,\qquad t>0,
\label{eq:dae}
\end{equation}
where $M_\alpha$ and $L_\alpha$ are the Orr--Sommerfeld matrices described in Equation~\eqref{eq:os_discretized}.
The matrix $M_\alpha$ is not invertible: it has rows of zeros corresponding to the no-slip boundary conditions, the continuity of the streamfunction at the interface, and the continuity
of the tangential stress at the interface. Equation~\eqref{eq:dae} is therefore a differential algebraic
equation (DAE).

There are several standard methods for solving DAEs with computational software packages~\citep{Shampine1999}.   For a singular matrix $M$, the DAE $M(t,y)y'=f(t,y)$  has a solution only when the initial condition $y_0$  is consistent, that is, if there is an initial slope $y_{p0}$ such that  $M(t_0,y_0)y_{p0}=f(t_0,y_0)$.   In general, computational packages for solving DAEs demand not only that  the initial data be consistent, but also that the slope be prescribed as an input to the numerical solver~\citep{Shampine1999}.   We develop herein a numerical method for linear DAEs that removes the necessity to specify the slope.
 Moreover, long-time integrations of DAEs using computational packages can be costly, especially for the modal decomposition~\eqref{eq:phi_ansatz}, which contains a large number of wave numbers.   Thus, we resort to a semi-analytical solution method that holds for constant-coefficient DAEs such as Equation~\eqref{eq:dae}.  We re-write Equation~\eqref{eq:dae} as 
\[
\frac{\mathd}{\mathd t}M_\alpha\vmode=L_\alpha\vmode,
\]
and integrate it over a numerical time step $\Delta t$:
\[
M_\alpha \vmode\left(t+\Delta t\right)-M_\alpha\vmode\left(t\right)=\int_{t}^{t+\Delta t}L_\alpha\vmode\left(s\right)\mathd s.
\]
For a sufficiently small timestep, the integral on the right-hand side may be approximated by the trapezoidal rule.  Thus, the equation is approximated by
\[
M_\alpha \vmode\left(t+\Delta t\right)-M_\alpha\vmode\left(t\right)=\tfrac{1}{2}\Delta t\left[L_\alpha\vmode\left(t+\Delta t\right)+L_\alpha\vmode\left(t\right)\right].
\]
Re-arrangement gives
\[
\left[M_\alpha-\tfrac{1}{2}\Delta t L_\alpha\right]\vmode^{p+1}=\left[M_\alpha+\tfrac{1}{2}\Delta t L_\alpha\right]\vmode^{p},
\]
where $\vmode^{p+1}=\vmode\left(t+\Delta t\right)$ and $\vmode^p=\vmode\left(t\right)$.
The solution at an arbitrary time $p$ is therefore available from the initial condition using only three matrix operations:
\begin{equation}
\vmode^{p}=X_\alpha^p\bm{v}_{0\alpha},\qquad X_\alpha=\left[M_\alpha-\tfrac{1}{2}\Delta t L_\alpha\right]^{-1}\left[M_\alpha+\tfrac{1}{2}\Delta t L_\alpha\right].
\label{eq:exact_dae}
\end{equation}
Equation~\eqref{eq:exact_dae} becomes an exact solution to the DAE~\eqref{eq:dae} in the limit when $\Delta t\rightarrow 0$:
\[
\vmode\left(t\right)=\tilde{X}_{\alpha}\left(t\right)\bm{v}_{0\alpha},\qquad \tilde{X}_{\alpha}\left(t\right)=
\lim_{\Delta t\rightarrow 0\atop t=p\Delta t}
X_{\alpha}^p.
\]
Finally, to approximate a delta-function impulse, we fix the initial condition $\bm{v}_{0\alpha}$ as follows:
\[
\overline{\phi}\left(x,z,t=0\right)=\sum_{\alpha>0} \mathe^{\imag\alpha x}\mathe^{-\alpha^2 w_x^2}
\begin{cases}
\sum_n\frac{N_n}{\sqrt{\pi w_z^2}}T_n(\eta_L)\int_{-1}^1 \mathd s\,\mathe^{-\left[\thickness\left(s-1\right)/2\right]^2/w_z^2}\frac{T_n\left(s\right)}{\sqrt{1-s^2}},&z\leq 0,\\
\sum_n\frac{N_n}{\sqrt{\pi w_z^2}}T_n(\eta_G)\int_{-1}^1\mathd s\,\mathe^{-\left[\left(s+1\right)/2\right]^2/w_z^2}\frac{T_n\left(s\right)}{\sqrt{1-s^2}},&z\geq 0,\\
\end{cases}
\]
where $N_n=1/\pi$ if $n=0$ and $2/\pi$ otherwise.  The coefficients $w_x$ and $w_z$ are the widths of the approximate delta functions in the $x$- and $z$-directions respectively. 

This solution method is appropriate for the kind of long-time simulations performed herein because the time required on a computer to implement Equation~\eqref{eq:exact_dae} depends only very weakly on $p$: the computation time required to raise an arbitrary square matrix to the $p^\mathrm{th}$ power depends (in a worst-case-scenario) only logarithmically on $p$~\citep{Ar1994}.
Thus a DNS of 1000 time units takes (at most) only three times longer than  a DNS of 10 time units.  

We have validated the solution method by computing the growth rate for  delocalized monochromatic initial conditions, and checking that the dispersion curve agrees with standard (temporal) eigenvalue analysis: the results are identical in each case and are not reported further here.
We have further validated the spatio-temporal transient DNS by using Equations~\eqref{eq:tdns_growth} and computing the growth rate and wave number at the most dangerous spatio-temporal mode.  This is necessarily the most dangerous temporal mode~\citep{huerre90a}.  Thus, if our transient DNS is correct, then the maximum growth rate computed in this manner must agree with standard temporal eigenvalue analysis.  The results for this comparison are shown in Table~\ref{tab:max_sigma0}, and 
\begin{table}
\centering
\begin{tabular}{ccccc}
$m$ & $\sigma_{\mathrm{max}}$(ST) & $\sigma_{\mathrm{max}}$(Modal) & $\alpha$(ST) & $\alpha$(Modal)\\
\hline
400 & 1.52 & 1.52 & 20 & 21\\
1300& 1.39 & 1.39 & 17 & 17\\
\end{tabular}
\caption{Comparison between modal and spatio-temporal analyses at various values of $m$, with $r=1000$, $\thickness=0.1$, and $Re=2500$. 
The parameters $\surft$ and $\grav$ are given in Equation~\eqref{eq:fr_values}.  The DNS parameters are $N_1=21$, $N_2=60$, $L_x=150$, $\Delta x=0.01$, and $\Delta t=0.01$.
}
\label{tab:max_sigma0}
\end{table}
%It is not possible to compute this wave number for all of the $\thickness$-values, since the finite-difference discretization of Equation~\eqref{eq:arv} is not always accurate.  Nevertheless, 
%
 confirm the correctness of our transient DNS.  We now apply this method to classify completely the instability in turbulent two-phase stratified flow.

\section{Results for the turbulent base state}
\label{sec:turb}

\subsection{Modal analysis and flow-regime maps}
\label{sec:turb:scaling}

In this section we examine the spatio-temporal instability wherein the upper layer is turbulent.  We base the parameter values on  an upper layer of air of depth $\dgas=5\,\mathrm{cm}$, together with the values of surface tension and gravitational acceleration given in Table~\ref{tab:values0} (these parameter values were used in~\citet{Onaraigh2011a}).  Substituting these values into Equation~\eqref{eq:fr_values0}, we obtain 
\begin{equation}
\gravpre= 3.7809\times 10^6,\qquad \surftpre= 1.1420\times 10^7.
\label{eq:fr_values}
\end{equation}

\begin{figure}
\centering
\subfigure[$\,\,Re=2350,m=10000$]{\includegraphics[width=0.49\textwidth]{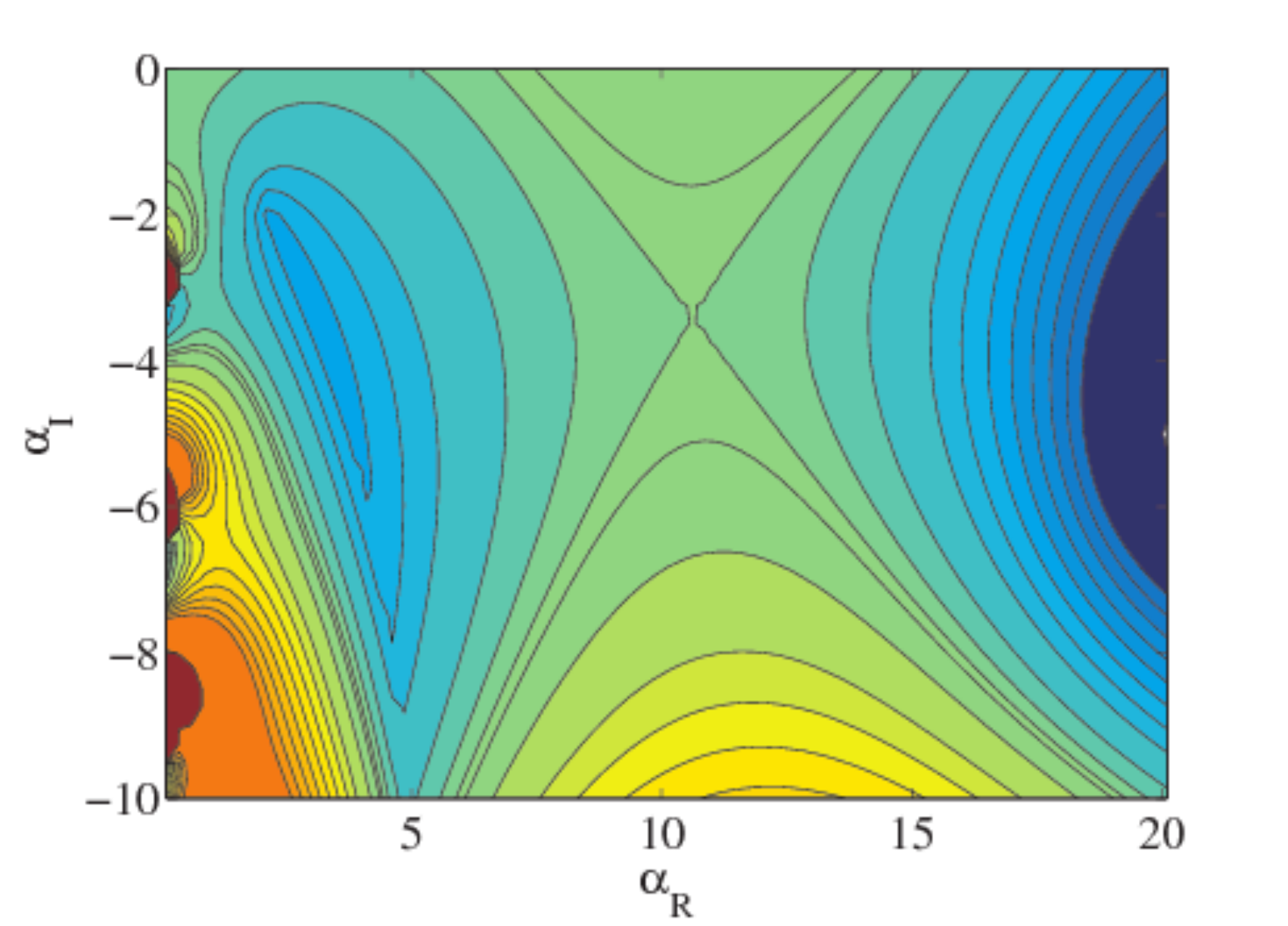}}
\subfigure[$\,\,Re=2500,m=810$]{\includegraphics[width=0.49\textwidth]{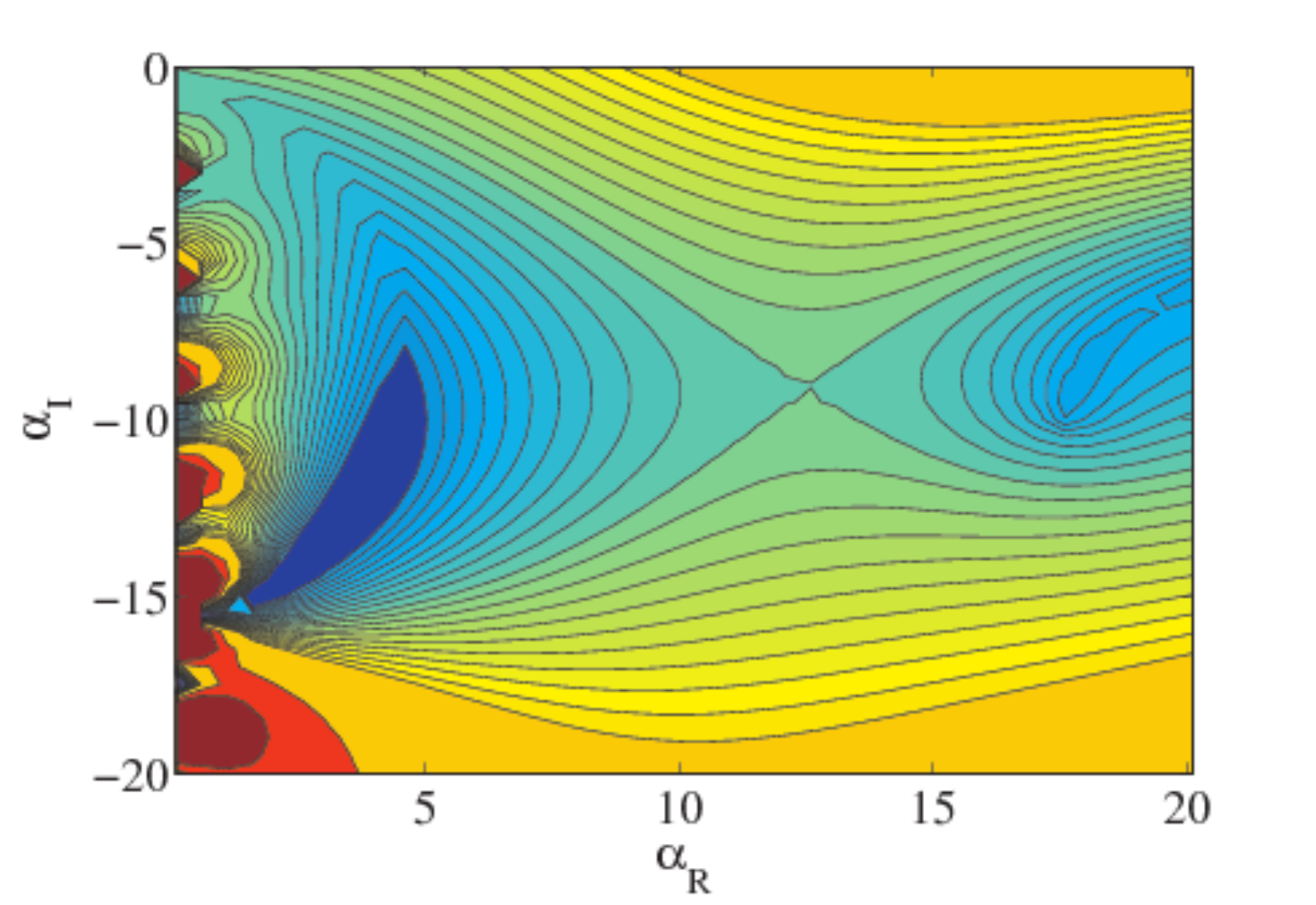}}
\subfigure[$\,\,Re=5000,m=1100$]{\includegraphics[width=0.49\textwidth]{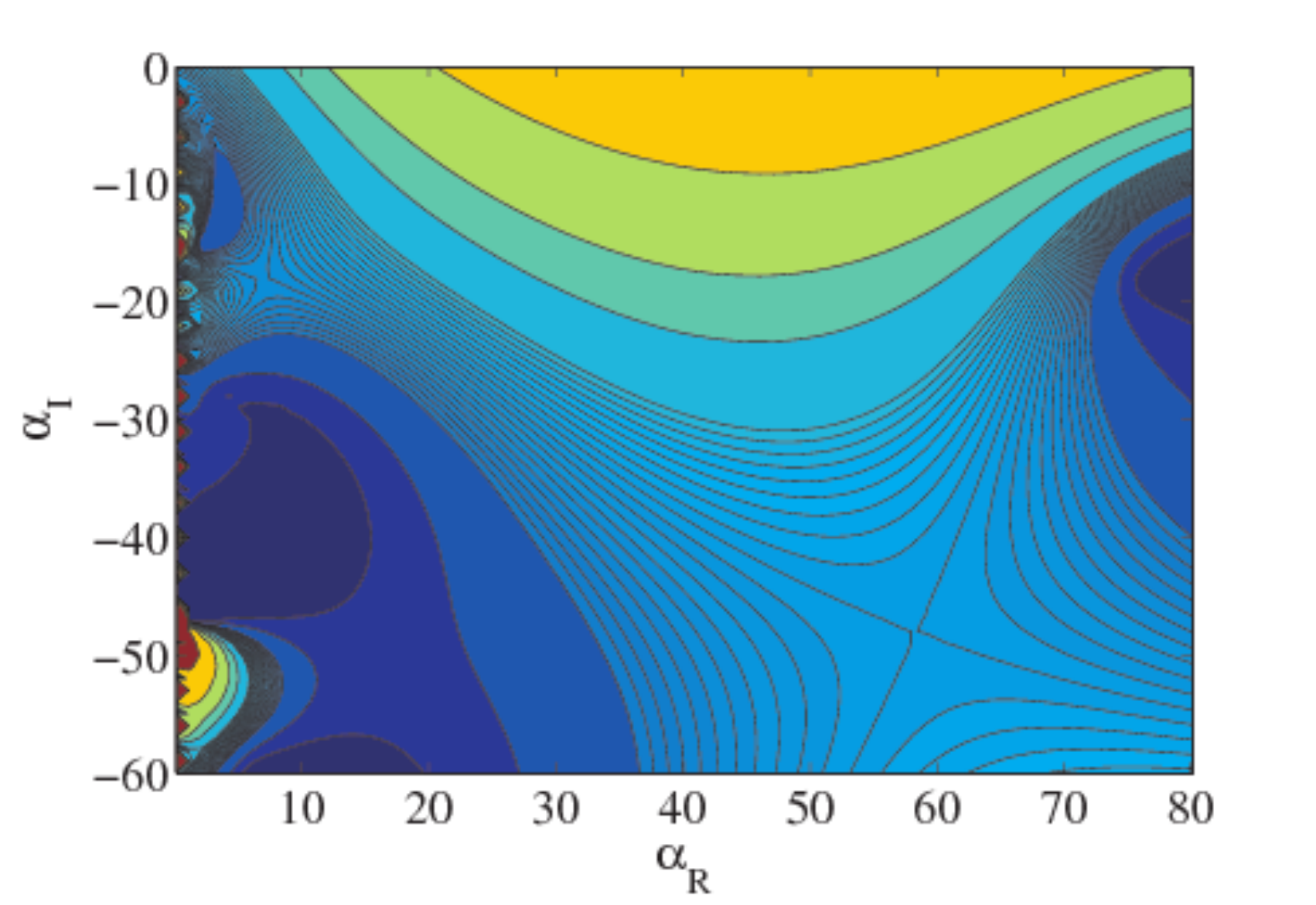}}
\caption{Contour plot in the complex wave number plane of the largest $\omega$ value.  In all cases, $r=1000$, $\thickness=0.1$, and $\gravpre$ and $\surftpre$ are given by Equation~\eqref{eq:fr_values}.}
\label{fig:CAturbplots}
\end{figure}

Starting with the base state described in Section~\ref{sec:model}, we carry out a \textit{modal} analysis for the turbulent case.  The results are shown in  Figure~\ref{fig:CAturbplots}.  Here, we have fixed $Re=2350$ and $\thickness=0.1$, and have
chosen the value of $m$  to give a convective/absolute (C/A) transition, such that the main saddle point in the contour plot for the largest value of $\omi$ has just become positive.   The other conditions described in Section~\ref{sec:postprocess:modal} that are required for absolute instability also apply.  However, from Figure~\ref{fig:CAturbplots}, it is not certain that the system is indeed absolutely unstable, given the odd features near the imaginary axis caused by different eigenmodes being dominant in different parts of the complex wave number plane.

\begin{figure}
\centering
\subfigure[]{\includegraphics[width=0.49\textwidth]{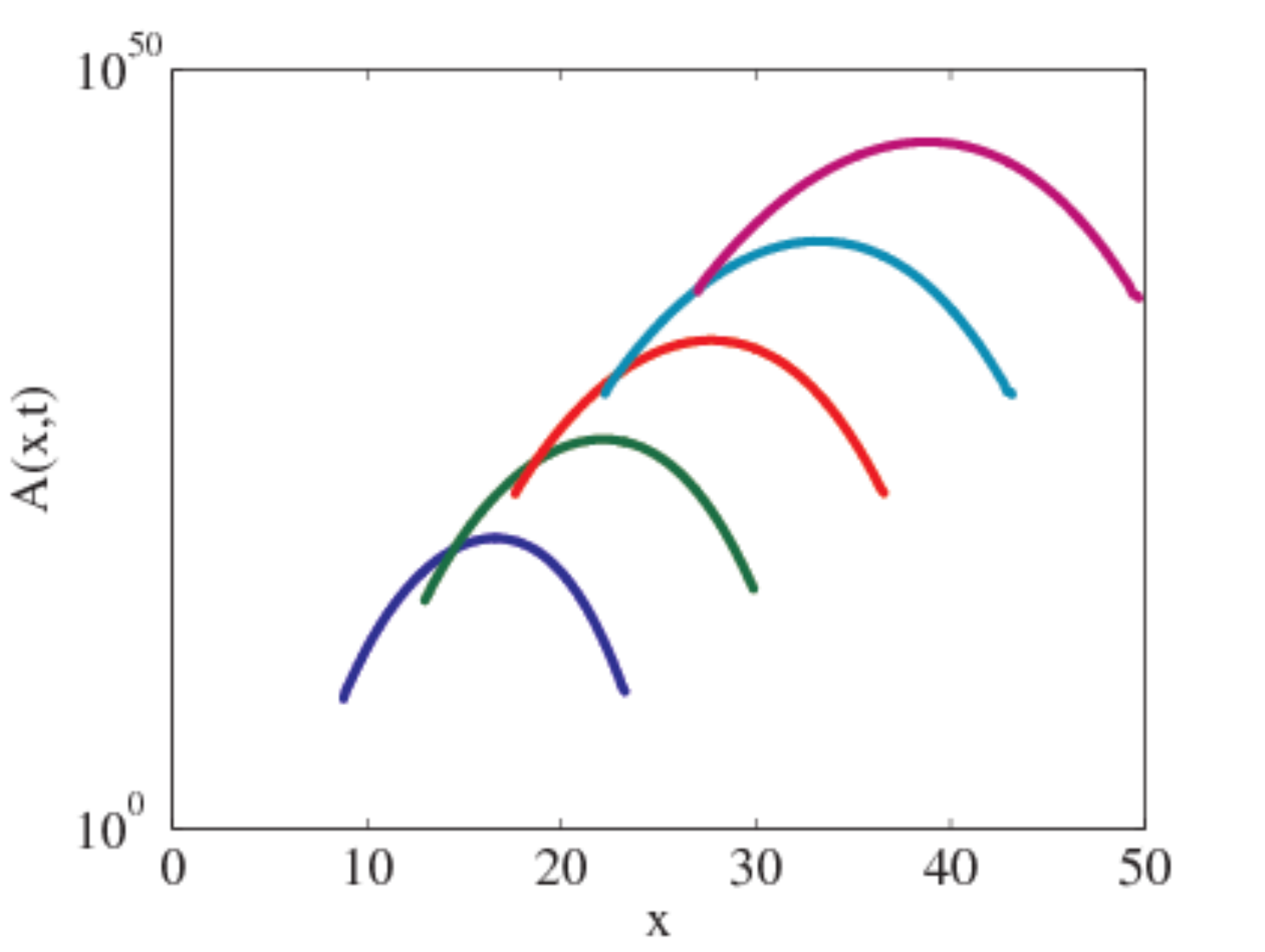}}
\subfigure[]{\includegraphics[width=0.49\textwidth]{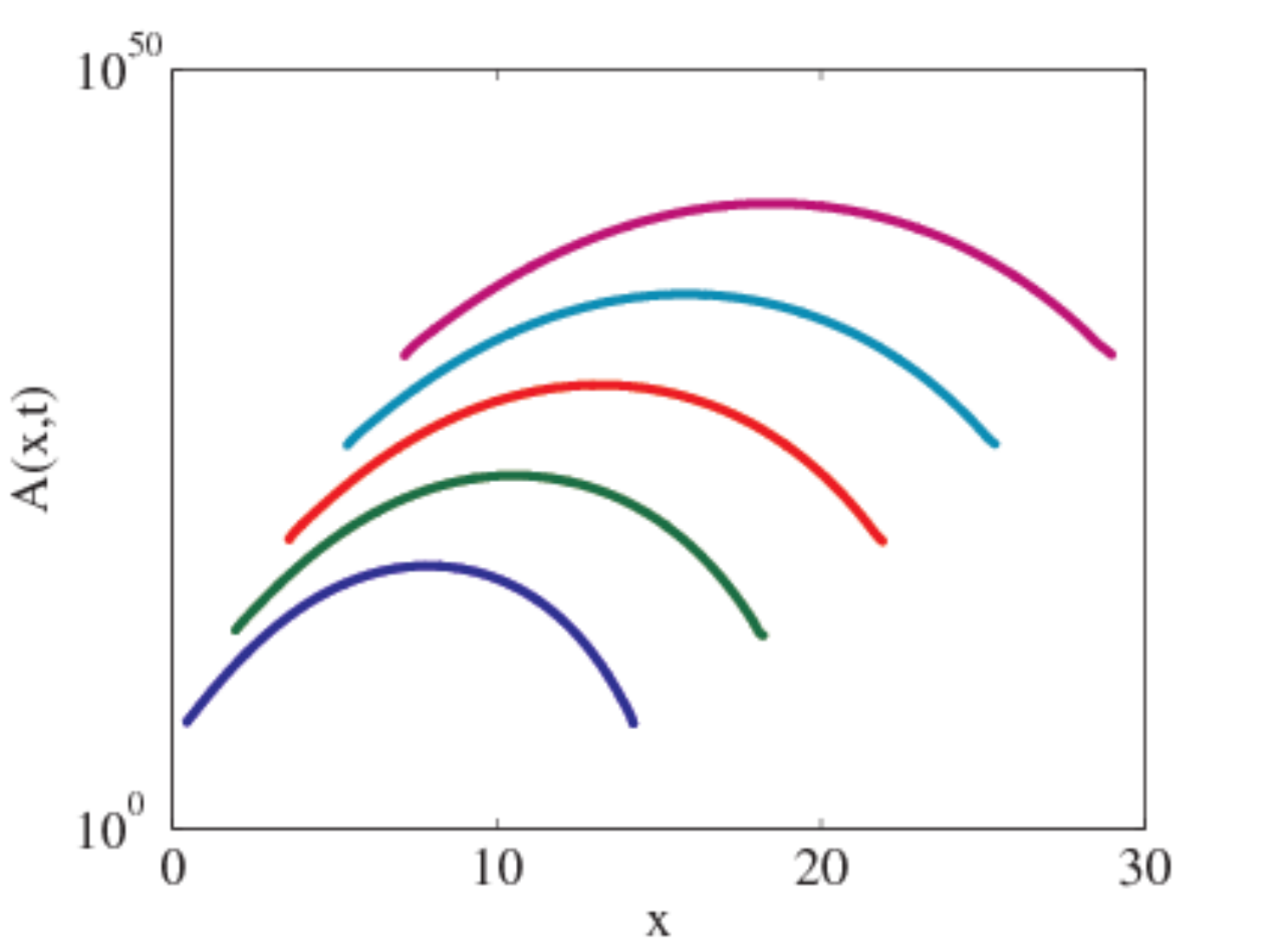}}
\caption{The norm $A(x,t)$ for (a) $m=400$ (convective); (b) $m=1300$ (absolute), at $t=30$, 40, 50, 60, 70 (from bottom to top).  The other parameters are $r=1000$, $Re=2500$, $\thickness=0.1$, and $\surft$ and $\grav$ are given in Equation~\eqref{eq:fr_values}.  The DNS parameters are $N_1=21$, $N_2=60$, $L_x=150$, $\Delta x=0.01$, and $\Delta t=0.01$.
Only a fraction of the $x$ domain is shown. }
\label{fig:normturb}
\end{figure}

Therefore, we turn to the ray analysis, representative results from which are shown in
Figure~\ref{fig:normturb}. The two cases are chosen for clarity, since they lie away from (and at opposite
sides of) the C/A transition such that the respective convective and absolute behaviour
is clearly visible. In this figure, the amplification of a pulse is very large. This poses a
practical problem when trying to infer a C/A transition: when starting from an absolutely
unstable case and repeating the analysis for a slightly different value of the parameter
$m$, a convective instability is found when the left tail of the signal decreases as time
goes by. But the tails of the pulse are close to the part of the $x$-domain that is
affected by numerical error. Moreover, for large $m$, the strong amplification of all parts of the pulse
makes a comparison of signal tails difficult.   The accuracy with which a C/A transition
can be determined with the ray analysis is therefore limited. That being as it may be,
close inspection of the ray analysis results near the C/A transitions inferred from a modal
analysis confirms the latter.

\begin{figure}
%\hspace*{0cm}\vspace{-1.5cm}
%
\centering
\subfigure[]{\includegraphics[width=0.49\textwidth]{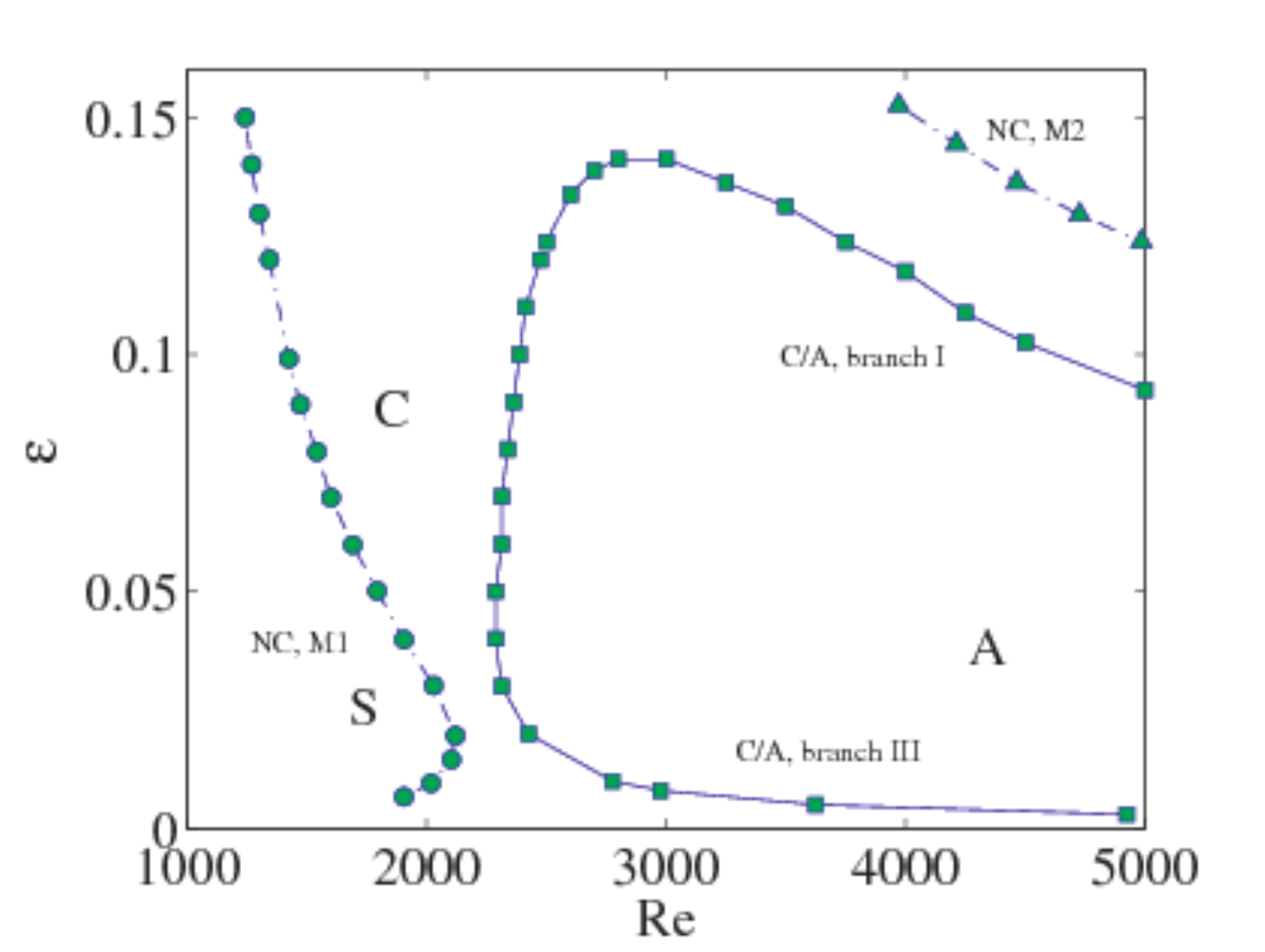}}
\subfigure[]{\includegraphics[width=0.49\textwidth]{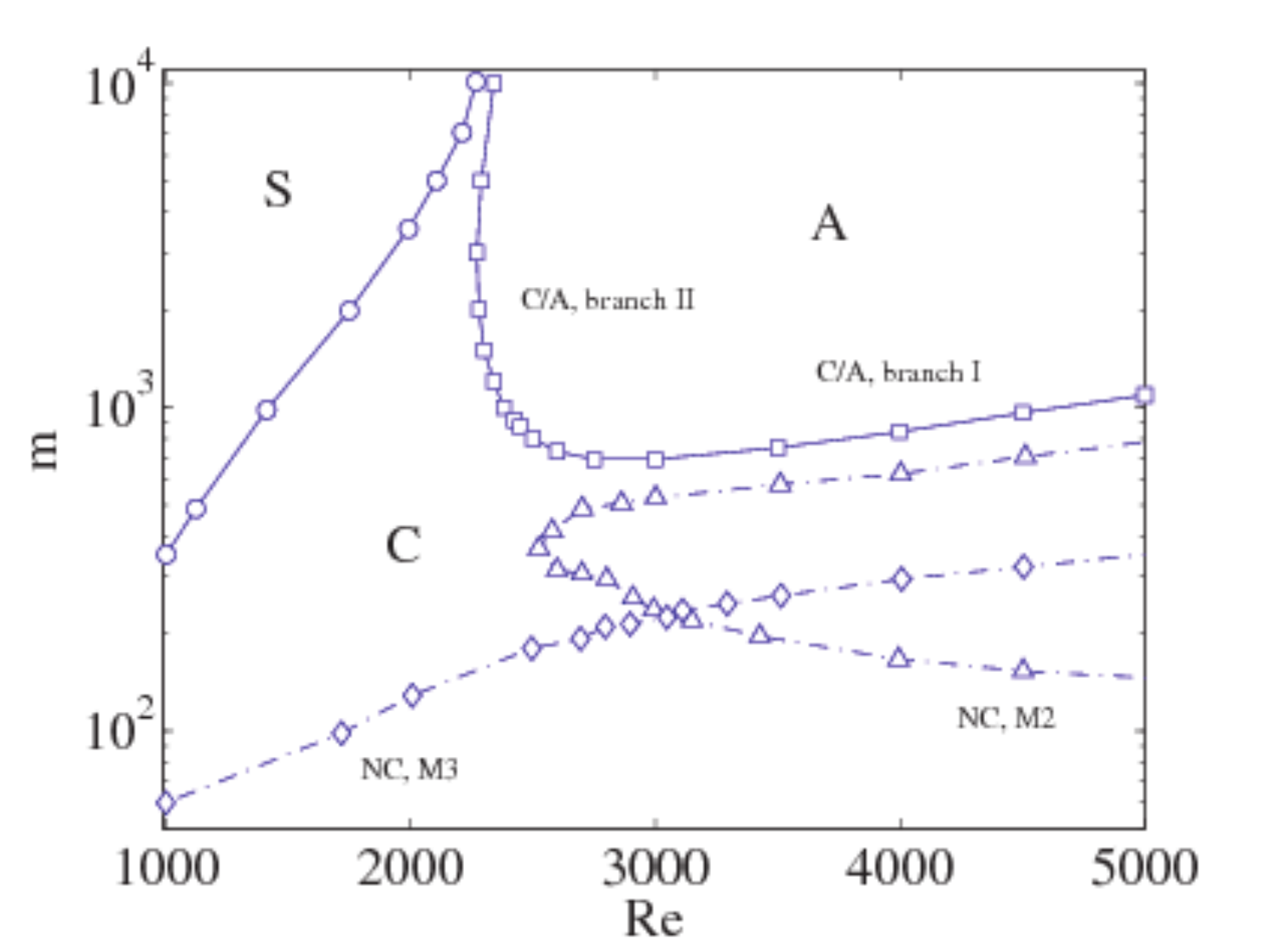}}\\
%\hspace*{0cm}\vspace{-2.5cm}
%
\caption{Flow-regime map for the turbulent base state.  Here, $r=1000$ and  $\surft$ and $\grav$, are given in Equation~\eqref{eq:fr_values}.   Mode M1 can be stable (S, neutral curve marked with circles), convectively unstable (C), or absolutely unstable (A, C/A transition curve marked with squares).  Further modes, called M2 and M3, can be convectively unstable, with neutral curves (`NC') given by dashed lines and labelled by triangles and diamonds respectively.
(a) Variations in $\thickness$ at fixed $m=1000$; (b) Variations in $m$ at fixed $\thickness=0.1$. 
}
\label{fig:CAturb}
\end{figure}

Carrying on from these studies, we have constructed a flow-regime map using the
modal analysis. We have carefully followed the dominant saddle points such as those
in Figure~\ref{fig:CAturbplots}, and have performed the necessary checks required for confirming absolute
instability (Section~\ref{sec:postprocess:modal}). Our modal studies have furthermore been confirmed by
the ray analysis discussed above. In Figure~\ref{fig:CAturb}(a), we see that if $Re$ is increased for a fixed
value of $m$, the system generally goes from a stable state, through a convectively unstable
to an absolutely unstable state.  Figure~\ref{fig:CAturb}(b) shows that the convectively unstable regime
can disappear entirely, when large values of $m$ are used. The same figure also shows that
the neutral curve `pushes' the C/A transition at large values of $m$ back to higher $Re$-values; this is discussed further below. We note in Figure~\ref{fig:CAturb}(b) the presence of two
additional modes of instability whose neutral curves are marked `NC:M2' and `NC:M3'; hence, three modes are convectively
unstable. A standard temporal energy-budget analysis conducted at $Re = 5000$, $m = 600$, confirms that M2 and M3 both derive most of the destabilizing energy from the interfacial region, and
a small contribution from the liquid layer. The profile of the wave-induced Reynolds stress has also confirmed that they are both conventional `internal modes' in the liquid~\citep{Boomkamp1996,Onaraigh2011a}.  However, these modes differ in one respect: M2 has a speed comparable to, but less than, the interfacial velocity whereas M3 is much slower, especially at large Reynolds numbers.
%
%\begin{figure}
%
%
%\centering
%
%\subfigure[]{\includegraphics[width=0.49\textwidth]{alphaM2M3turb}}
%\subfigure[]{\includegraphics[width=0.49\textwidth]{cM2M3turb}}
%\caption{The  wave number and the wave speed  along the neutral stability curves of the M2 and M3 modes.   Here $\Reint=rU_\mathrm{int}\thickness/m$, $U_\mathrm{int}$ is the base-state velocity at the interface. Open symbols: parameter variation as in Figure~\ref{fig:CAturb}(a); filled symbols: parameter variation as in Figure~\ref{fig:CAturb}(b).}
%\label{fig:M2M3_speeds}
%\end{figure}
%
%

\begin{figure}
\centering
\subfigure[]{\includegraphics[width=0.48\textwidth]{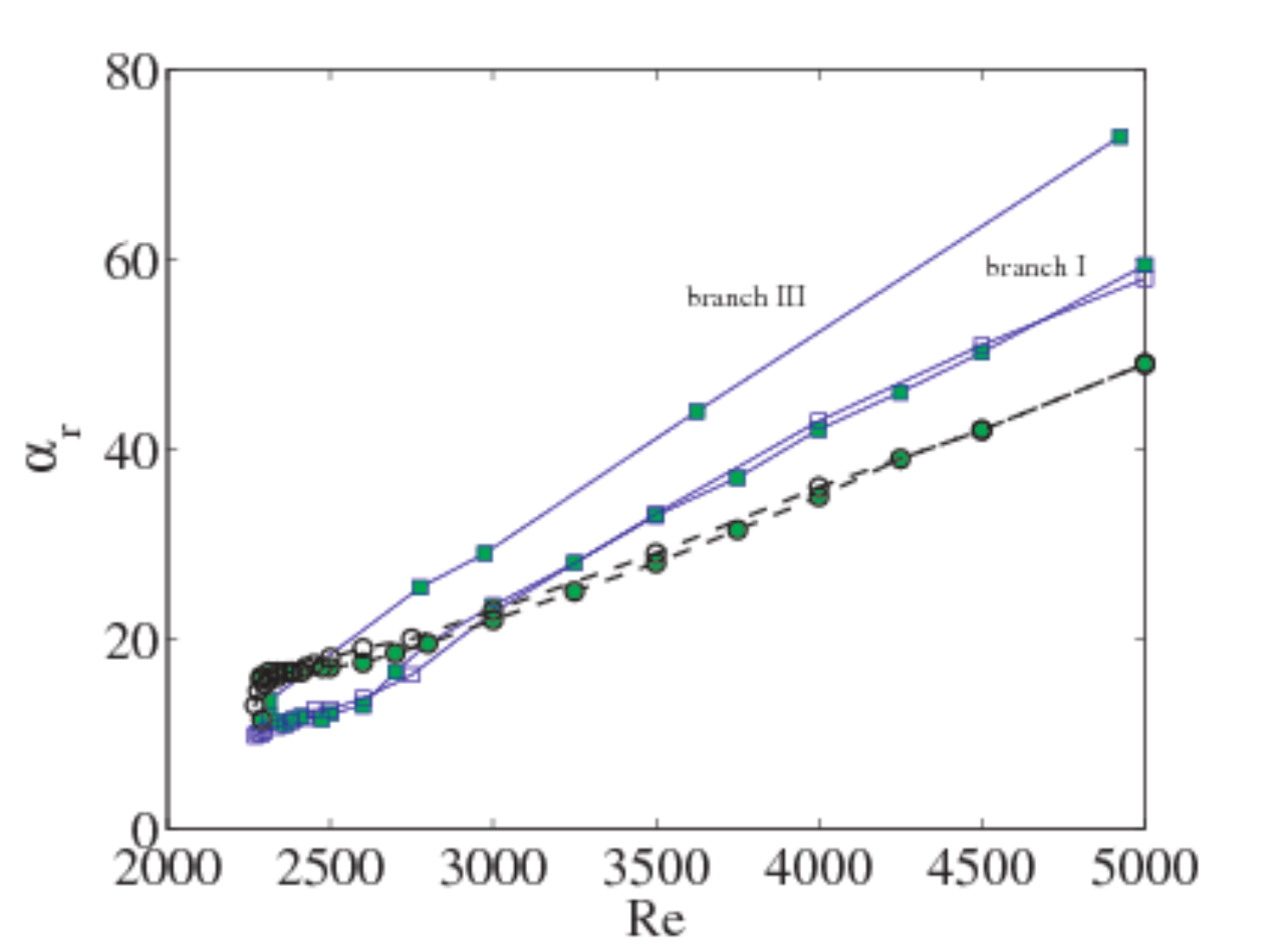}}
\subfigure[]{\includegraphics[width=0.48\textwidth]{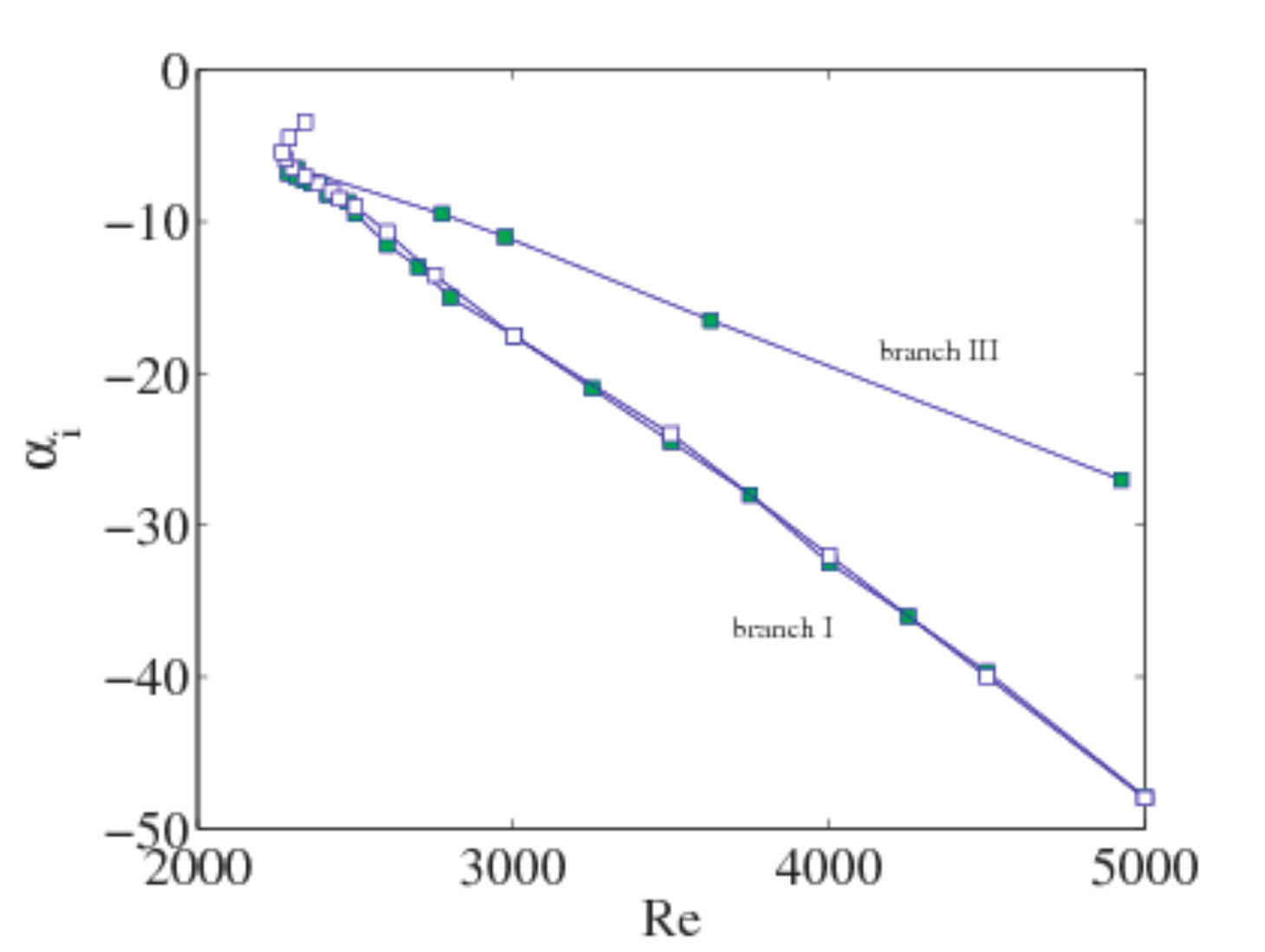}}
\subfigure[]{\includegraphics[width=0.48\textwidth]{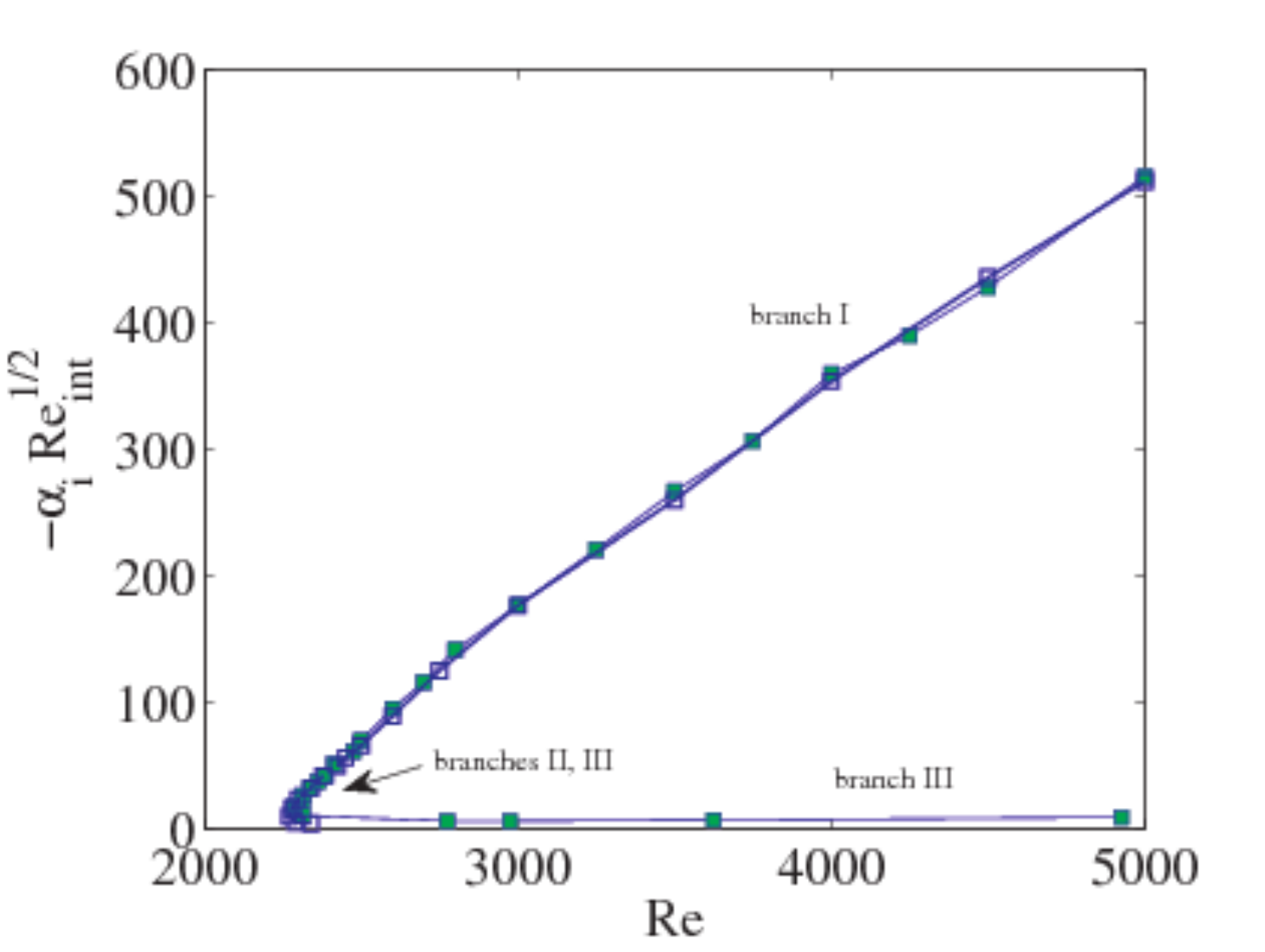}}
\caption{The complex wave number along the C/A transition in Figure~\ref{fig:CAturb} as a function of $Re$ (squares, solid lines).
In (a),  the wavenumber of the most dangerous temporal mode is also shown (circles, dashed line).
Filled squares: variation in $\thickness$ at fixed $m$; empty squares: variation in $m$ at fixed $\thickness$.  Panel (c) is a rescaled version of the imaginary wave number in (b).
}
\label{fig:CAalphaturb}
\end{figure}

In Figure~\ref{fig:CAalphaturb}, we examine the complex wave number $(\ar,\ai)$ at the C/A transition. 
We have plotted $\ar$ as a function of $Re$ for various $(\thickness,m)$ combinations; we have also plotted $\ai(\Reint)^{1/2}$ in the same manner.  Here $\Reint=rU_\mathrm{int}\thickness/m$ is the liquid Reynolds number based on the base-state velocity at the interface, $U_\mathrm{int}$ (the reason for this rescaling will become apparent in what follows).
Figure~\ref{fig:CAalphaturb}(a,b) demonstrates that $\ar$  at the transition is governed  primarily by $Re$.  The real and imaginary parts of the wave number both increase significantly with $Re$.  By rescaling the wavenumbers with respect to the lower-layer depth (i.e. letting $\ar\rightarrow \thickness \ar$), this variation is given some context: for $Re=1000$ the wavelength is comparable to the liquid-layer depth, while for $Re=5000$ it is approximately 10 times smaller than the liquid-layer depth.
 Figure~\ref{fig:CAalphaturb}(a) also indicates that $\ar$ at the saddle point along the neutral C/A curve
closely follows $\ar$ for the most-dangerous temporal mode.   This confirms the observation in Figure~\ref{fig:CAturbplots} that the saddle point lies almost directly below the most dangerous temporal mode.  This is explained as follows: for all the cases considered here, the group velocity $d\omr/d\ar$ calculated for temporal modes depends only weakly on $\ar$, and a straightforward application of the Cauchy--Riemann conditions to the analytic function $\omega=\omr(\ar,\ai)+\imag\omi(\ar,\ai)$ shows that for such group velocities, the $\ar$-values for the most-dangerous temporal mode and the saddle-point mode coincide.

\begin{figure}
\centering
\includegraphics[width=0.7\textwidth]{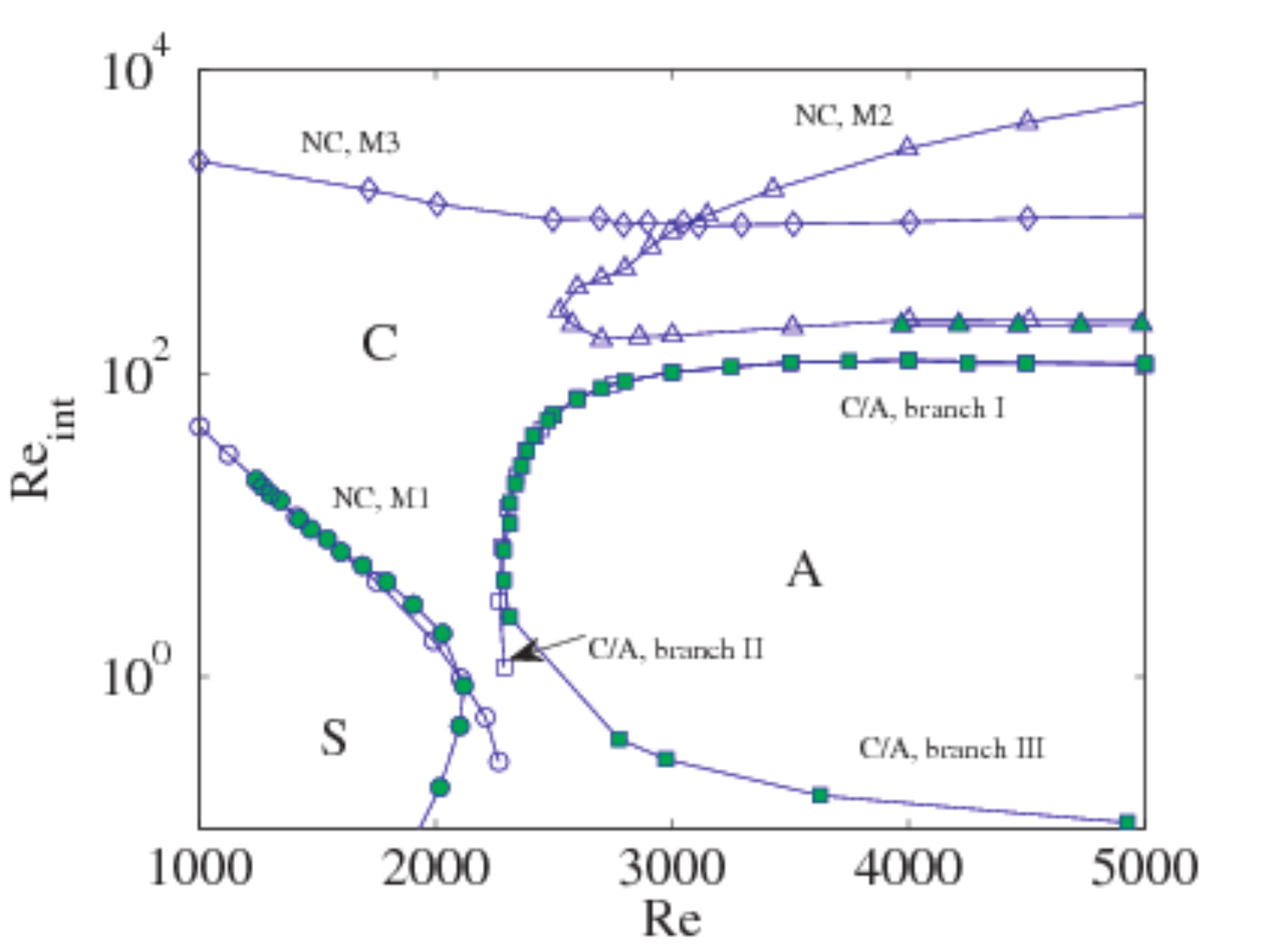}
%\subfigure[]{\includegraphics[width=0.49\textwidth]{turbtrans}}
%\subfigure[]{\includegraphics[width=0.49\textwidth]{Froude3p5}}
\caption{Reduced flow-regime map for the turbulent base state.
Filled symbols: variation in $\thickness$ at fixed $m$; empty symbols: variation in $m$ at fixed $\thickness$.
%
%\commentpdms{PDMS to improve clarity of symbols}\commentlon{I think it is okay!}
}
\label{fig:CAturb_collapse}
\end{figure}

Most of the results in Figure~\ref{fig:CAturb}(a,b) collapse into
one plot, namely Figure~\ref{fig:CAturb_collapse}(a). 
Almost all corresponding boundaries obtained for Figure~\ref{fig:CAturb}(a,b) collapse. Especially of interest is the large-$Re$/large-$\Reint$ behaviour of the C/A transition (branch I), which corresponds to a critical value of $\Reint$ that is virtually independent of $Re$. 
It is possible to explain this scaling behaviour using the theory developed by~\cite{Onaraigh2012a} (see also Appendix~\ref{app:quadratic}).  There, the spatio-temporal growth rate $\omi(\ar,\ai)$ is prescribed in terms of a Taylor series in $\ai$, where the coefficients in the Taylor series are derived from the purely temporal linear stability analysis, and depend on $\ar$.  We have verified that the C/A transition curves in Figure~\ref{fig:CAturb_collapse} are described accurately by a quadratic truncation of this Taylor series (Appendix~\ref{app:quadratic}).  In this `quadratic approximation', the saddle point occurs at $\ar=\ar^*$, where $\ar^*$ solves
\begin{equation}
\frac{d\omitemp}{d\ar}\frac{d^2\omitemp}{d\ar^2}\Big|_{\ar^*}=-\cg(\ar^*)\frac{d\cg}{d\ar}\Big|_{\ar^*}.
\label{eq:saddle}
\end{equation}
In the current application, the group velocity is approximately constant, hence  $\ar^*\approx\armax$, the location of the most-dangerous temporal mode.  The same quadratic approximation gives the following condition for the onset of absolute instability:
\begin{equation}
-\frac{d^2\omitemp}{d\ar^2}\bigg|_{\ar^*}\omitemp(\ar^*)=\tfrac{1}{2}\cg^2(\ar^*).
\label{eq:saddle_sign}
\end{equation}

We now approximate each term in Equation~\eqref{eq:saddle_sign}.  Consider first of all the right-hand side.  Previous work~\citep{Onaraigh2011a} demonstrates that the group velocity $\cg$ is only slightly in excess of the interfacial velocity $U_\mathrm{int}$ (this is also consistent with experiments, e.g.~\citet{CohenHanratty}).  Thus, we approximate the group velocity as $\cg\approx U_\mathrm{int}$.
Moreover, since the liquid layer is  thin and viscous, the base-state velocity in the liquid is close to linear shear flow.  In addition, for thin films, $Re_*=\geom Re$, where $\geom$ is a geometric factor independent of the flow parameters~\citep{Onaraigh2011a}.  Thus, we have the following string of equalities:
\begin{eqnarray*}
\frac{U_\mathrm{int}}{U_0}\approx \frac{\tau_{\mathrm{i}}d_L}{\mu_L U_0}=\frac{\thickness}{m}\frac{Re_*^2}{Re}
=\frac{\thickness}{m}\geom^2 Re.
\end{eqnarray*}
But
\begin{eqnarray*}
\Reint = \frac{\rho_L U_\mathrm{int}d_L}{\mu_L}=\frac{r}{\geom^2} \left(\frac{U_\mathrm{int}}{U_0}\right)^2,
\end{eqnarray*}
hence
\[
\frac{U_\mathrm{int}}{U_0}=\left(\geom/r^{1/2}\right)\Reint^{1/2},
\]
and
\begin{equation}
\cg^2\approx \left(\geom^2/r\right)\Reint.
\label{eq:cg_approx}
\end{equation}
Next, we consider the left-hand side.  We use the quadratic approximation
\[
\omitemp=A\alpha-\tfrac{1}{2}B\alpha^2,\qquad A,B>0.
\]
Note that this is not  a long-wave approximation, but is instead  a fit to the data around the most-dangerous temporal mode, where the fitting parameters $A$ and $B$ are selected with respect to the actual, computed values of the maximum growth rate, expressed as $\max(\omitemp)=A^2/(2B)$, and the cutoff wave number, written as $\alpha_0=2A/B$.  We now consider three parameter regimes and investigate the functional dependence of $A$ and $B$ on the Reynolds number $Re$.

\begin{figure}
\centering
\subfigure[]{\includegraphics[width=0.49\textwidth]{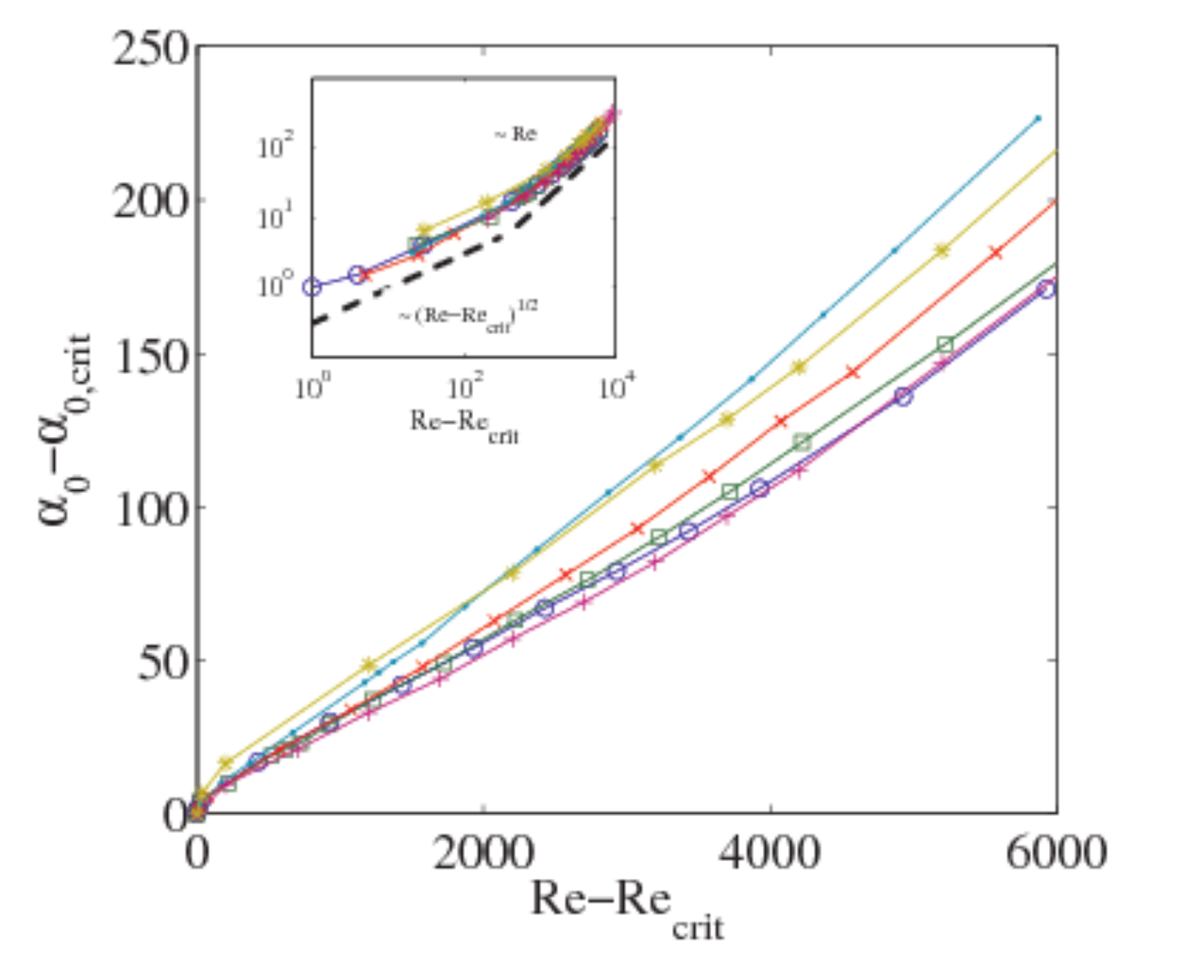}}
\subfigure[]{\includegraphics[width=0.49\textwidth]{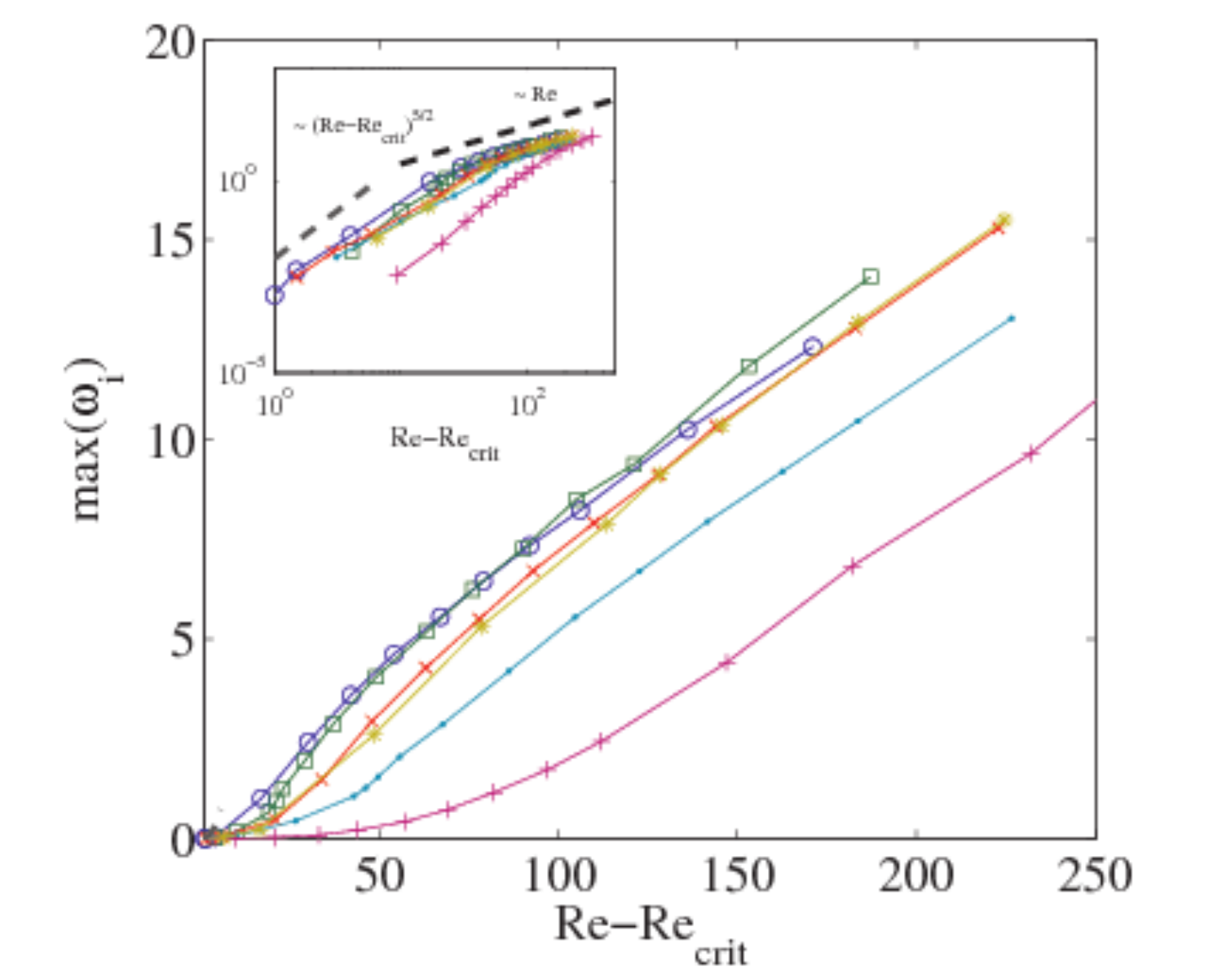}}
\caption{Dependence of the \textit{temporal} problem on the system parameters.  (a)  The cutoff wavenumber $\alpha_0$, for which $\omitemp(\alpha_0)=0$; (b) the maximum growth rate $\max(\omitemp)$.  The parameters to be varied are the pair $(m,\thickness)$.  Circles: (4000,0.1), Squares: (2000,0.1),  Crosses: (1000,0.1);  Diamonds: (500,0.1); Dashes: (1000,0.005); Stars: (1000,0.05).  Trendlines have been added to the insets.}
\label{fig:temp_props}
\end{figure}

\textit{Functional dependence for fixed $\thickness\approx 0.1$ and varying $m$:}
At large Reynolds numbers, both the growth rate and the cutoff wave number increase linearly with $Re$.   We write
\[
\max(\omitemp)=k_1 Re,\qquad \alpha_0=k_2 Re,
\]
where $k_1$ and $k_2$ are constants of proportionality; these are measured to be $m$-independent (Figure~\ref{fig:temp_props}).  In other words,
\[
\tfrac{1}{2}(A^2/B)=k_1Re, \qquad 2A/B=k_2 Re.
\]
Hence, $A^2=2k_1Re(2A/Re k_2)$, or $A=4k_1/k_2$.
At maximum growth,
\begin{equation}
-\omitemp\frac{d^2\omitemp}{d\ar^2}\bigg|_{\alpha_{\mathrm{r},\mathrm{max}}}=\tfrac{1}{2}A^2
%=\tfrac{1}{2}(16k_1^2/k_2^2)
=8k_1^2/k_2^2,
\label{eq:lhs_approx1}
\end{equation}
and the LHS of the C/A transition criterion is independent of $Re$ for large values of $Re$.  
Combining Equations~\eqref{eq:saddle_sign},~\eqref{eq:cg_approx}, and~\eqref{eq:lhs_approx1}, we have the following criterion for the onset of absolute instability:
\[
\Reint=C_1,\qquad  C_1=\frac{16 rk_1^2}{k_2^2\geom^2},
%8k_1^2/\ell^2=(\gamma^2/2r)\Reint,
\]
where $C_1$ is a parameter-independent constant.  Hence, at large $Re$, there is a critical Reynolds number $\Reint$ for the onset of absolute instability.  This is consistent with figure ~\ref{fig:CAturb_collapse} (branch I).

On the other hand, for smaller values of $Re$, Figure~\ref{fig:temp_props} gives
\[
\max(\omitemp)\propto |Re-Re_\mathrm{c}|^p,\qquad p>1,
\]
where $Re_\mathrm{c}$ is approximately independent of $m$.  We have also measured the cutoff wave number as
\[
\alpha_{0}=\alpha_{0\mathrm{c}}+k_3 |Re-Re_\mathrm{c}|^{1/q},\qquad q>1,
\]
where $\alpha_{0\mathrm{c}}$ and $k_3$ are approximately independent of $m$.   (We have found $p\approx 5/2$, $q\approx 2$; see Figure~\ref{fig:temp_props}.)   Putting these facts together, we get
\[
-\omitemp\frac{d^2\omitemp}{d\ar^2}\bigg|_{\mathrm{max}}\propto \frac{1}{\alpha_{0\mathrm{c}}^2}|Re-Re_\mathrm{c}|^{2p},\qquad Re\rightarrow Re_\mathrm{c},
\]
and the criterion for the onset of absolute instability is therefore
\[
\Reint=C_2 |Re-Re_\mathrm{c}|^{2p},
\]
where $C_2$ is independent of $m$.  This relation is consistent with figure ~\ref{fig:CAturb_collapse} (branch II).

\textit{Functional dependence for fixed $m$ and varying $\thickness$:}
At large Reynolds numbers and relatively large values $\Reint$ such that $\thickness\approx 0.1$, both the growth rate and the cutoff wave number increase linearly with $Re$ (Figure~\ref{fig:temp_props}).  Thus, the scaling arguments employed for fixed $\thickness$ and varying $m$ pertain here also, and
 the criterion for absolute instability is again $\Reint=C_1$, where $C_1$ is a parameter-free constant.   This relation is again consistent with Figure~\ref{fig:CAturb_collapse} (branch I).
For smaller values of $Re$, but for  still relatively large values of $\Reint$, Figure~\ref{fig:temp_props} gives
\[
\max(\omitemp)\propto |Re-Re_\mathrm{c}(\thickness)|^p,\qquad p>1,
\]
where $Re_\mathrm{c}(\thickness)$ is approximately independent of $m$ but depends on $\thickness$.  We have also measured the cutoff wave number as
\[
\alpha_{0}=\alpha_{0\mathrm{c}}(\epsilon)+k_3 |Re-Re_\mathrm{c}(\epsilon)|^{1/q},\qquad q>1,
\]
where $\alpha_{0\mathrm{c}}(\epsilon)$ is approximately independent of $m$ but dependent on $\thickness$, and where $k_3$ is independent of $m$ and $\thickness$.  Putting these facts together, we get
\[
-\omitemp\frac{d^2\omitemp}{d\ar^2}\bigg|_{\alpha_{\mathrm{r},\mathrm{max}}}\propto \frac{1}{\alpha_{0\mathrm{c}}^2(\epsilon)}|Re-Re_\mathrm{c}(\epsilon)|^{2p},\qquad Re\rightarrow Re_\mathrm{c},
\]
and the criterion for the onset of absolute instability is therefore
\[
\Reint=\frac{C_2}{\alpha_{0\mathrm{c}}(\thickness)} |Re-Re_\mathrm{c}(\thickness)|^{2p},
\]
where $C_1$ is independent of $m$.  This relation is consistent with Figure~\ref{fig:CAturb_collapse} (branch III, large $\Reint$), in the sense that the stability boundary depends on $\Reint$, $Re$, and $\thickness$.

As $\thickness$ is reduced, the Reynolds number $\Reint$ is also reduced.  In such a regime, and for large values of $Re$,
we have
\[
\max(\omitemp)=k_4 \thickness Re^2/m,\qquad \alpha_0=k_5 Re,
\]
where $k_4$ and $k_5$ are constants of proportionality. In other words,
\[
\tfrac{1}{2}(A^2/B)=k_4\thickness Re^2/m, \qquad 2A/B=k_5 Re,
\]
Hence, $A=4k_4\epsilon Re/(k_5 m)$, and $B=8k_4\epsilon /(k_5 m)$.
Thus, at maximum growth,
\[
-\omitemp\frac{d^2\omitemp}{d\ar^2}\bigg|_{\mathrm{max}}=\tfrac{1}{2}a^2
%=\tfrac{1}{2}\frac{16k^2\epsilon^2 Re^2}{\ell^2 m^2}
%=\frac{8k^2\epsilon^2 Re^2}{\ell^2m^2},
=\frac{8k_4^2\thickness^2 Re^2}{k_5^2 m^2}\propto \Reint^2.
\]
Therefore, in this case, the LHS of the  C/A transition criterion is proportional to $Re_{\rm int}^2$, while the RHS is proportional to $\Reint$.  Therefore, there is a \textit{minimum} value of $\Reint$ (rather than a maximum value) for absolute instability.  This corresponds precisely to the lower-branch C/A transition curve in Figure~\ref{fig:CAturb_collapse} (branch III, small $\Reint$).

These results have been presented for fixed values of $\gravpre$ and $\surftpre$.  We briefly sketch the effect of varying these parameters.  Increasing $\gravpre$ is stabilizing, and raises the critical Reynolds number $Re_c$ for the onset of temporal instability.  This suggests that  branches II and III should shift to the right under such an increase.  Similarly, increasing $\surftpre$ makes the temporal disturbances more stable, the main effect of which is (counter-intuitively) to \textit{increase} $-\omitemp d^2\omitemp/d\ar^2$ at $\ar=\alpha_{\mathrm{r},\mathrm{max}}$, which implies that the critical Reynolds number $\Reint$ for the onset of absolute instability at large $Re$ and $\Reint$ should be \textit{raised}.  
%
%\commentlon{Peter, I have omitted the gory details here, such as fitting a parabola `by eye' \&c.  Please check that whatever remains here is correct.} 
%
 We have performed some detailed calculations for the onset of absolute instability with the `full' dispersion relation, the results of which agree with this description provided by the quadratic approximation.

We also comment on the scaling behaviour for the imaginary part of the wave number at the C/A transition.  The value of $\ai$ at transition in the quadratic approximation is
\[
\ai=-2\max(\omitemp)/\cg.
\]
Approximating $\cg\propto \sqrt{\Reint}$, we have $\ai\sqrt{\Reint}\propto -\max(\omitemp)$.  For large values of the Reynolds numbers $Re$ and $\Reint$, we have $\max(\omitemp)\propto Re$, where the constant of proportionality is parameter-free.  This gives $-\ai\sqrt{\Reint}\propto Re$, in agreement with Figure~\ref{fig:CAalphaturb}.

\begin{figure}
\centering
\subfigure[$\,\,m=600$]{\includegraphics[width=0.49\textwidth]{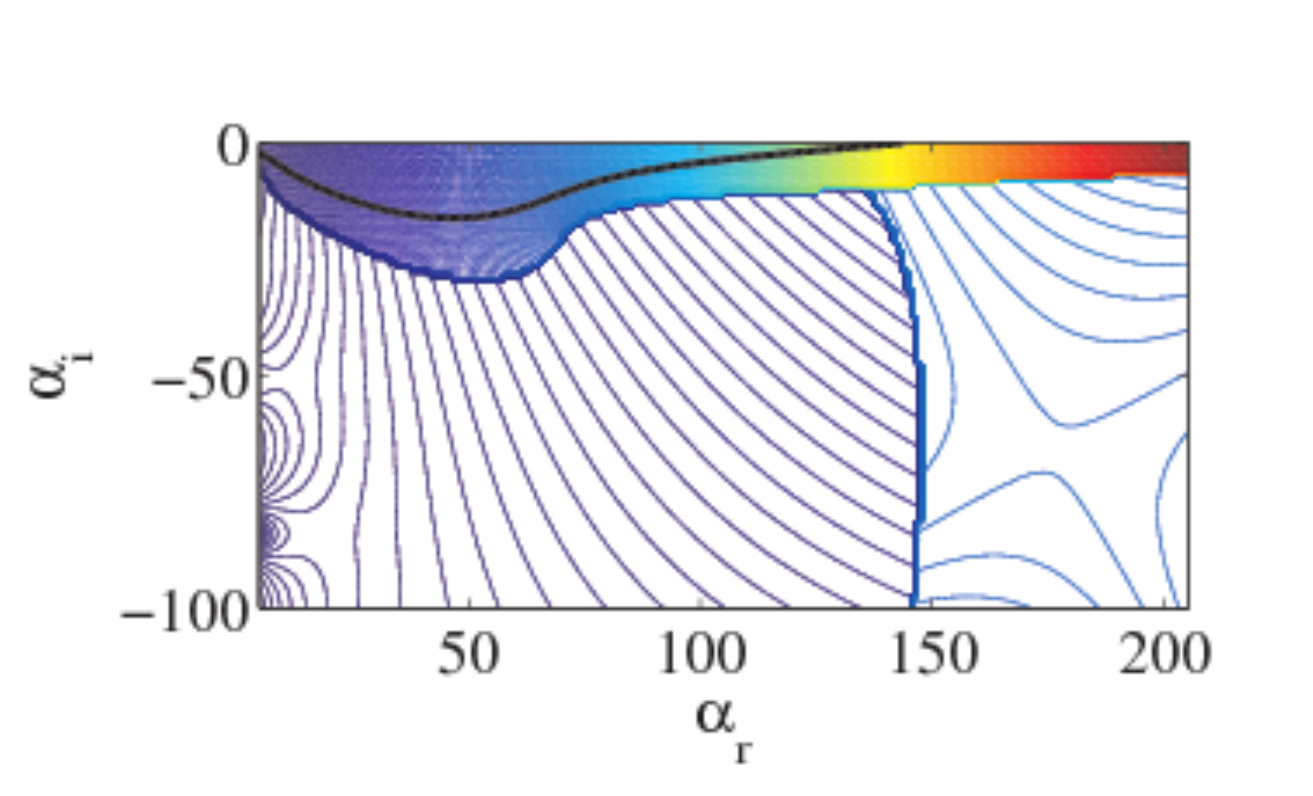}}
\subfigure[$\,\,m=700$]{\includegraphics[width=0.49\textwidth]{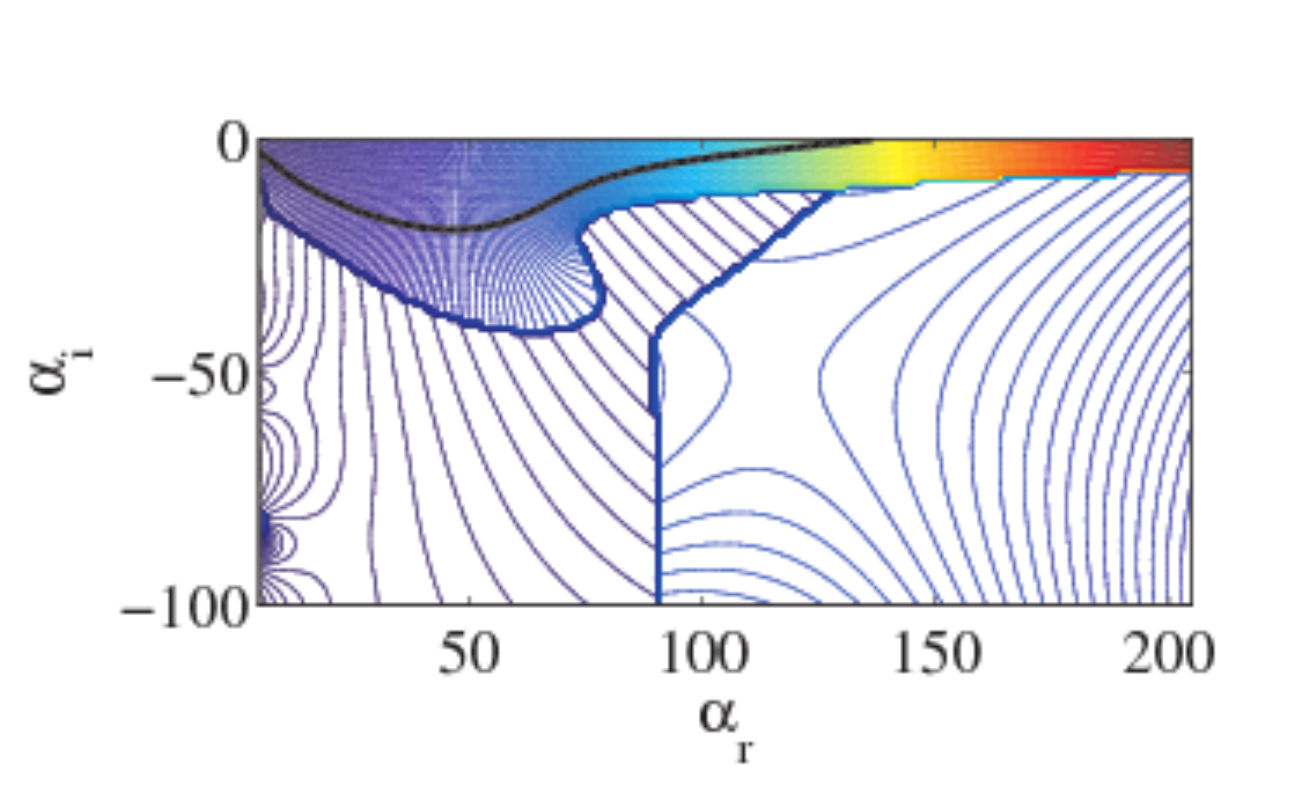}}
\subfigure[$\,\,m=710$]{\includegraphics[width=0.49\textwidth]{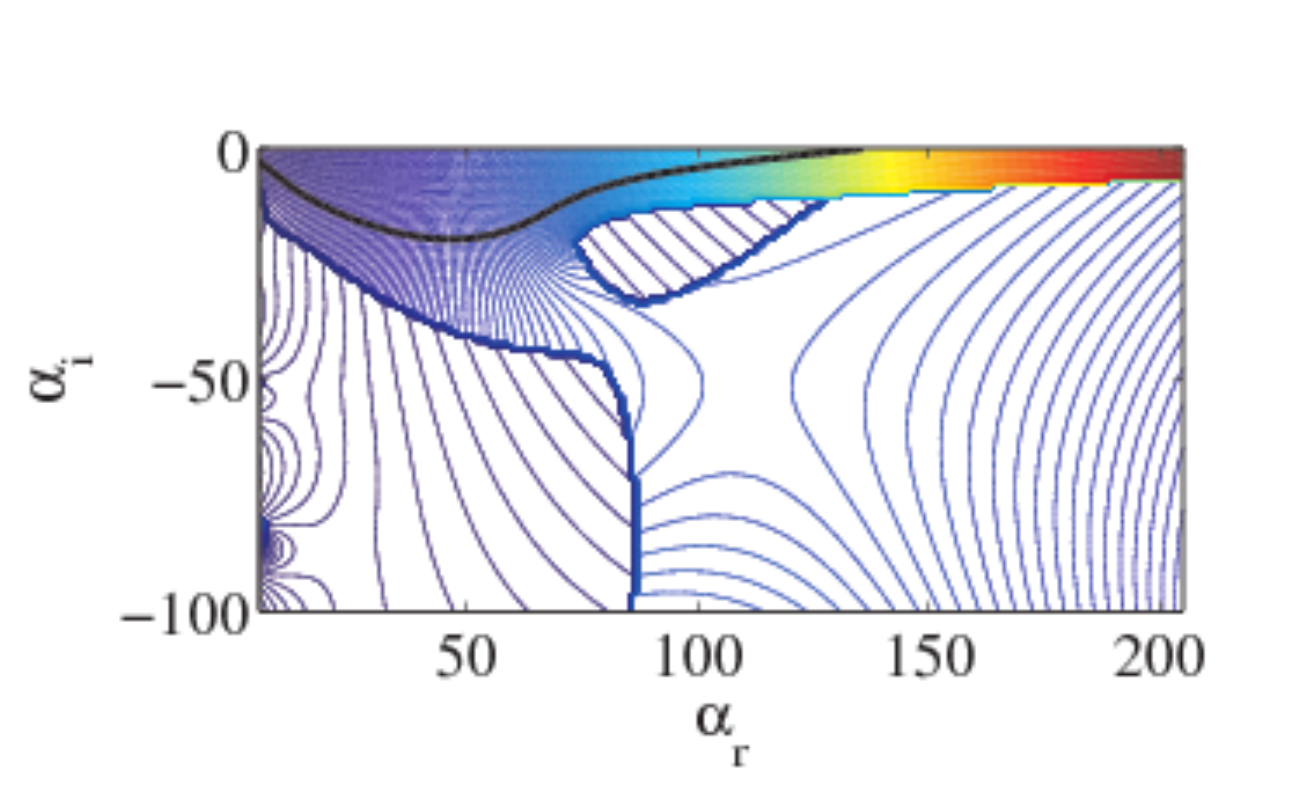}}
\subfigure[$\,\,m=900$]{\includegraphics[width=0.49\textwidth]{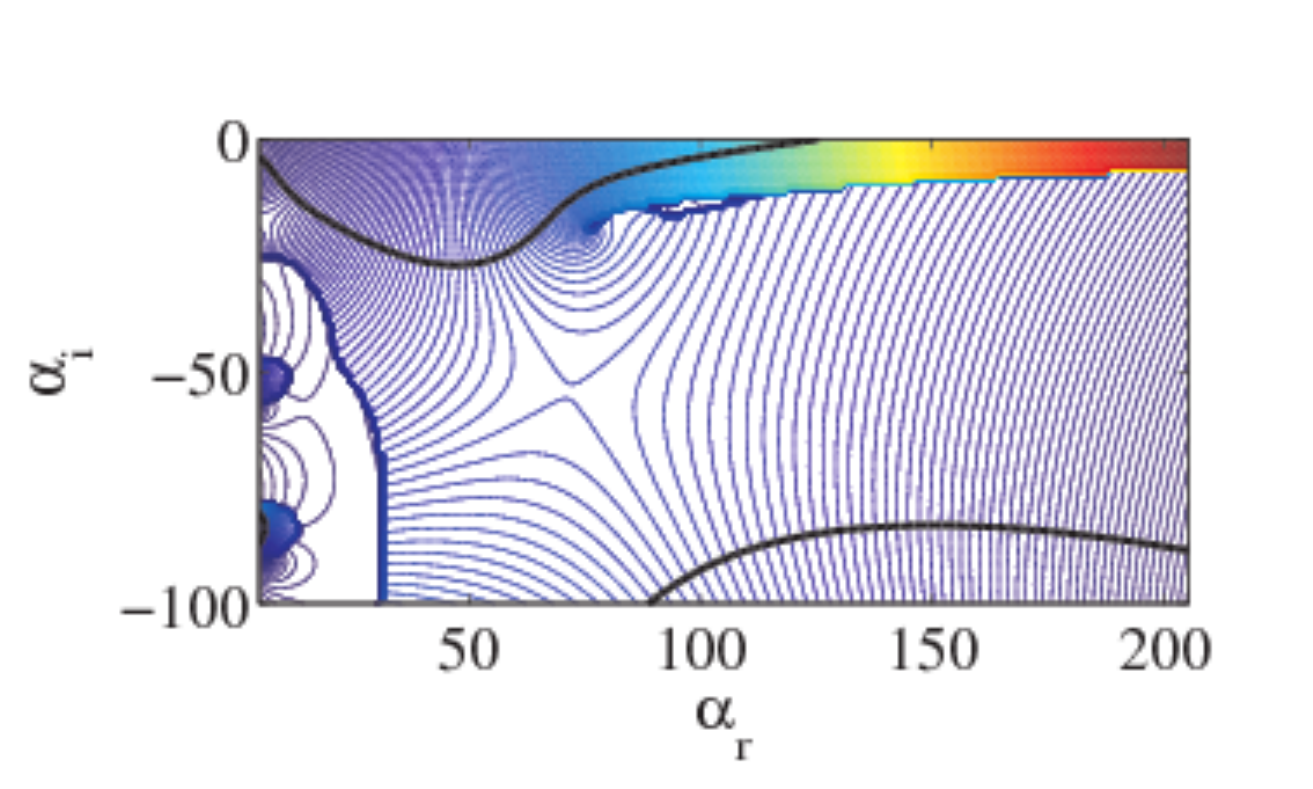}}
\caption{Spatio-temporal mode competition and the C/A transition: contour plots of $\omr$ for the most dangerous mode in the complex $\alpha$-plane.  The black lines represent the spatial curve $\omi=0$ of M1.   Here $Re=5000$, and $\thickness=0.1$.   The M2 temporal mode is unstable  for $m < 800$, and the C/A transition occurs near $m=1100$.   The sharp jumps in $\omr$ represent mode competition between M1 and M2, and the saddle point in (a),(b) does not correspond to M1.}
\label{fig:competition}
\end{figure}

One further result concerns the alignment of branch I (main C/A transition,  Figure~\ref{fig:CAturb_collapse}) with the neutral temporal stability curve of M2.  It is as if the absolute instability is quenched when M2 becomes unstable. This is due to mode competition, which
occurs in a convectively unstable regime close to the C/A boundary, in which both M1 and M2 are unstable, with a wave length comparable to the liquid-layer depth.
In Figure~\ref{fig:competition}, the
contours of $\omr$ are shown for the least stable mode at each complex $\alpha$. This case involves
the parameter values $Re = 5000$, $\thickness = 0.1$, and $m = 600$. The spatial curve $\omi = 0$ of the most
dangerous mode is identified with M1 by a Gaster-type analysis (see~\citet{Gaster1962} and Appendix~\ref{app:spatial}). The
contours $\omr = \mathrm{Const.}$ that connect orthogonally to the spatial curve are identified also
with M1. The orthogonality of contours of the real and imaginary parts of an analytic
function is a straightforward consequence of the Cauchy--Riemann conditions. In this
way, we have established in Figure~\ref{fig:competition}(a) that the saddle point (a necessary precursor for absolute instability) does not originate from the M1-eigenmode. However, upon increasing
$m$ (Figs.~\ref{fig:competition}(b--d)), the region of $\alpha$-space in which M1 dominates fills out, and the spatial
curve of M1 connects to a saddle: it is as if the dominant saddle point `switches' from non-M1 to M1. Going to
higher values of $m$, the sign of $\omi$ at the saddle point becomes positive, and absolute instability
ensues.

We have  investigated in more depth the identity of the  rival saddle points by using the quadratic approximation, which provides a direct connection between the spatio-temporal and temporal modes. 
By reconstructing Figure~\ref{fig:competition} from the quadratic approximation (not shown),
we have confirmed the spatio-temporal mode competition. Care has been taken to ensure correct identification of the dispersion curves by double-checking the continuity in $\ar$ of the corresponding $\omega_{\rm r}$ curves.  
In a reconstruction of the $\omi$-landscape of M1, a saddle point exists at $\ar\approx \alpha_{\mathrm{r},\mathrm{max}}$.  In a further reconstruction, two additional saddle points occur, and are connected with M2: one lies near $\alpha_{\mathrm{r},\mathrm{max}}$, while a further saddle point resides at large $\ar$.  Finally, we have verified that as $m$ increases, the values of $\omi$ associated with M1 increase (especially near the saddle point), eventually leading to absolute instability; the other saddles are simultaneously overwhelmed by this increase and no longer produce mode competition.

%The ray analysis provides yet further insight into the C/A transition. In Figure~\ref{fig:sigmaturb}, increasing the value of $m$ at a fixed value of $Re=2500$ (see Figure~\ref{fig:CAturb}(b)) causes the spatio-temporal growth rate $\sigma(v)$ to shift to lower values of the ray velocity $v$, leading eventually to a positive value of $\sigma (v=0)$ and a C/A transition. This is for 
%
%
%\begin{figure}
%\centering
%\includegraphics[width=0.49\textwidth]{sigmax2}
%\caption{The spatio-temporal growth rate for the turbulent case, with $r=1000$, $Re=2500$, $\thickness=0.1$ and the value of $m$ as indicated.  The values of $\surft$ and $\grav$ are given after Equation~\eqref{eq:fr_values}, $L_x=150$, $\Delta t=0.01$, and $\Delta x=0.01$.}
%\label{fig:sigmaturb}
%\end{figure}
%
%
%the branch to the right of the point on the C/A transition in Figure~\ref{fig:CAturb}(b) where a minimum value of $m$ is reached, which corresponds to the upper branch in Figure~\ref{fig:CAturb}(a).  The ray analysis also provides a basis for a classification of the instability according to sources and sinks in an energy budget.

\subsection{Energy budget}
\label{sec:turb:budget}

The purpose  of applying an energy-budget analysis is to infer the mechanism for instability generation.
In view of the connection between spatio-temporal and temporal modes exemplified in the quadratic approximation considered in Section~\ref{sec:turb:scaling}, a spatio-temporal energy-budget analysis necessarily inherits the properties of the corresponding temporal analysis.  Nevertheless, we consider the energy-budget analysis in this section, and investigate how such a study applies to the growth of a pulse (as opposed to a normal mode).  The ray analysis  provides a means of examining such pulses.

We examine the mechanism by which the pulse grows,  noting that this does not have to be the same  throughout the pulse.
We introduce the kinetic energy density
\begin{equation}
K\left(x,t\right)=\tfrac{1}{2}\int_{-\thickness}^0\mathd{z}\,r_L|\delta\bm{u}|^2+\tfrac{1}{2}\int_{0}^1\mathd{z}\,r_G|\delta\bm{u}|^2,
\label{eq:k_budget}
\end{equation}
where $\delta\bm{u}=(\delta u,\delta w)$ is the perturbation velocity.
We differentiate this expression with respect to time at a fixed location $x$, and apply the equations of motion~\eqref{eq:ns_linear} and Gauss's divergence theorem to obtain the following flux-conservation equation:
\begin{equation}
\frac{\partial K}{\partial t}+\frac{\partial F_K}{\partial x}=s_B\left(x,t\right)+s_I\left(x,t\right),
\label{eq:flux}
\end{equation}
where the source/sink terms $s_B$ and $s_I$, and the flux $F_K$ are described in what follows.  First, we introduce the following notation for the perturbative contribution to the viscous stress tensor $\delta T_{ij}$:
\[
\delta T_{ij}=-\delta_{ij}\delta p+\frac{m_j}{Re}\left(\frac{\partial }{\partial x_i}\delta u_j+\frac{\partial }{\partial x_j}\delta u_i\right);
\]
we also denote the separate fluid domains by $\left(a_L,b_L\right)=\left(-\thickness,0\right)$ and $\left(a_G,b_G\right)=\left(0,\dgas\right)$.
The flux $F_K$ can then be written as
\begin{multline*}
F_K\left(x,t\right)=\int_{a_L}^{b_L}\mathd{z}\left[\tfrac{1}{2}r_L U_0\left(z\right)|\delta\bm{u}|^2-\delta u \,\delta T_{xx}-\delta w\,\delta T_{xz}\right]\\
+
\int_{a_G}^{b_G}\mathd{z}\left[\tfrac{1}{2}r_G U_0\left(z\right)|\delta\bm{u}|^2-\delta u\,\delta T_{xx}-\delta w\,\delta T_{xz}\right].
\end{multline*}
In a similar manner, the bulk source/sink term takes the form
\begin{eqnarray*}
s_B\left(x,t\right)&=&\sum_{j=L,G}\left[REY_j\left(x,t\right)+DISS_j\left(x,t\right)\right],\\
REY_j\left(x,t\right)&=&-r_j\int_{a_j}^{b_j}\mathd{z}\,\delta u\,\delta w\frac{\mathd U_0}{\mathd z},\\
DISS_j\left(x,t\right)&=&-\frac{m_j}{Re}\int_{a_j}^{b_j}\mathd{z}\left[2\left(\delta u_x\right)^2+2\left(\delta w_z\right)^2+\left(\delta u_z+\delta w_x\right)^2\right],
\end{eqnarray*}
and finally, the interfacial source/sink $s_I\left(x,t\right)$ is given by
\begin{eqnarray*}
s_I\left(x,t\right)&=&TAN\left(x,t\right)+NOR\left(x,t\right),\\
TAN(x,t)&=&\left(\delta T^L_{zx}\,\delta u^L-\delta T^G_{zx}\,\delta u^G\right)_{z=0},\\
NOR(x,t)&=&\left(\delta T^L_{zz}\,\delta w^L-\delta T^G_{zz}\,\delta w^G\right)_{z=0}.
\end{eqnarray*}
There are no contributions to the energy balance from the perturbation turbulent stresses because these are neglected in the quasi-laminar approximation (see Section~\ref{sec:model:perturbations}).

\begin{figure}
\centering
\subfigure[]{\includegraphics[width=0.48\textwidth]{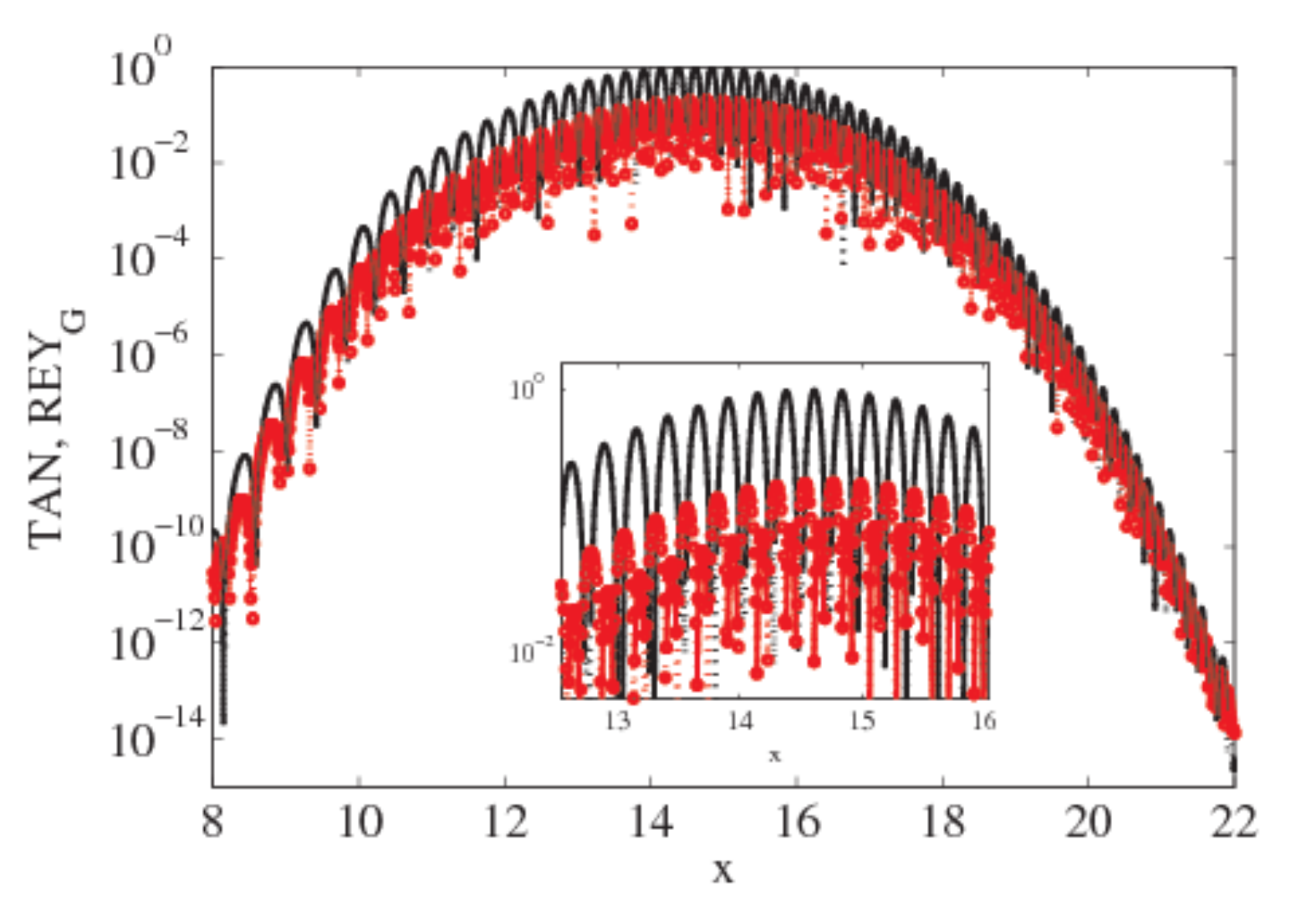}}
\subfigure[]{\includegraphics[width=0.48\textwidth]{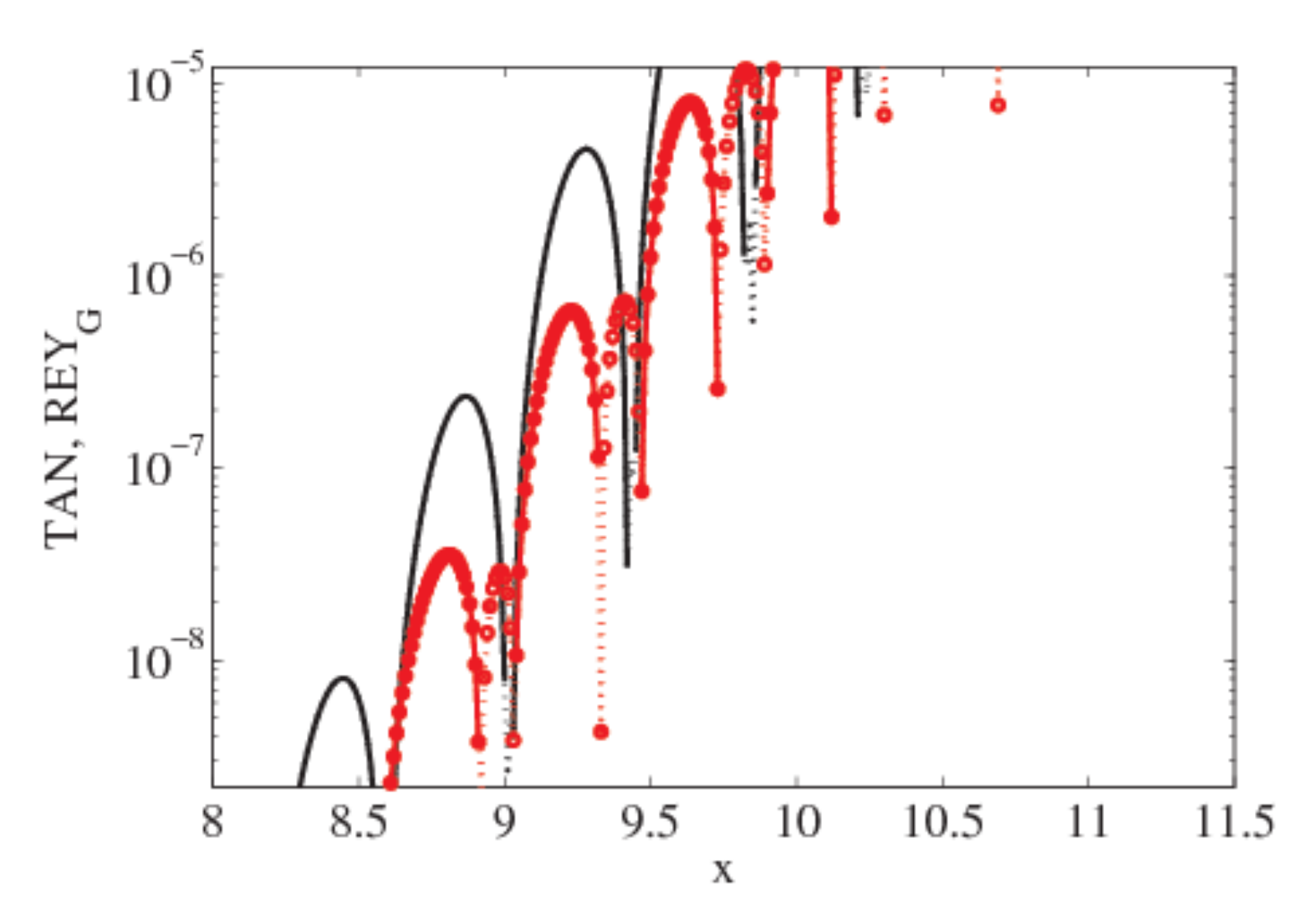}}
\subfigure[]{\includegraphics[width=0.48\textwidth]{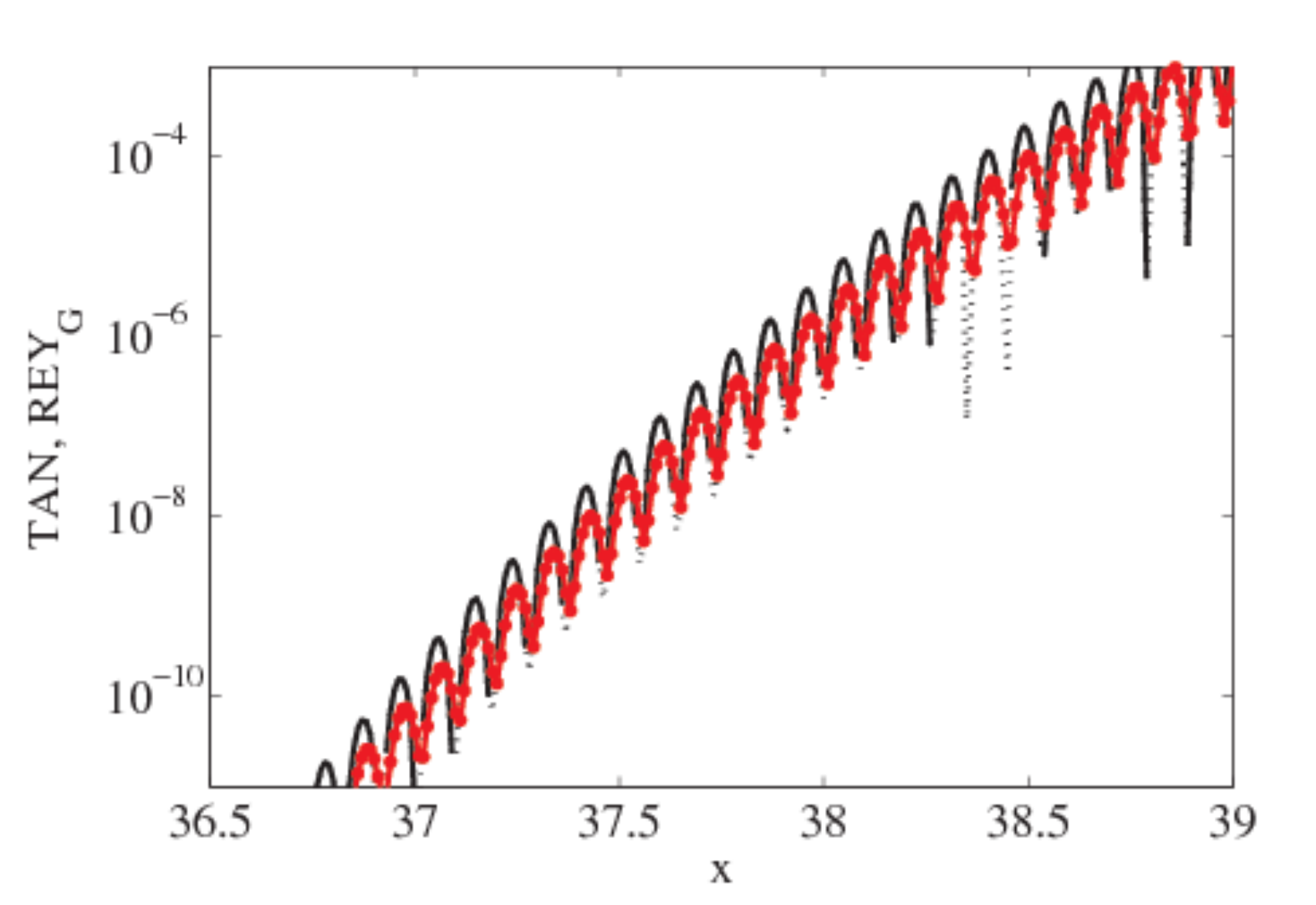}}
%\subfigure[]{\includegraphics[width=0.35\textwidth]{Re4000m2000x}}
\subfigure[]{\includegraphics[width=0.48\textwidth]{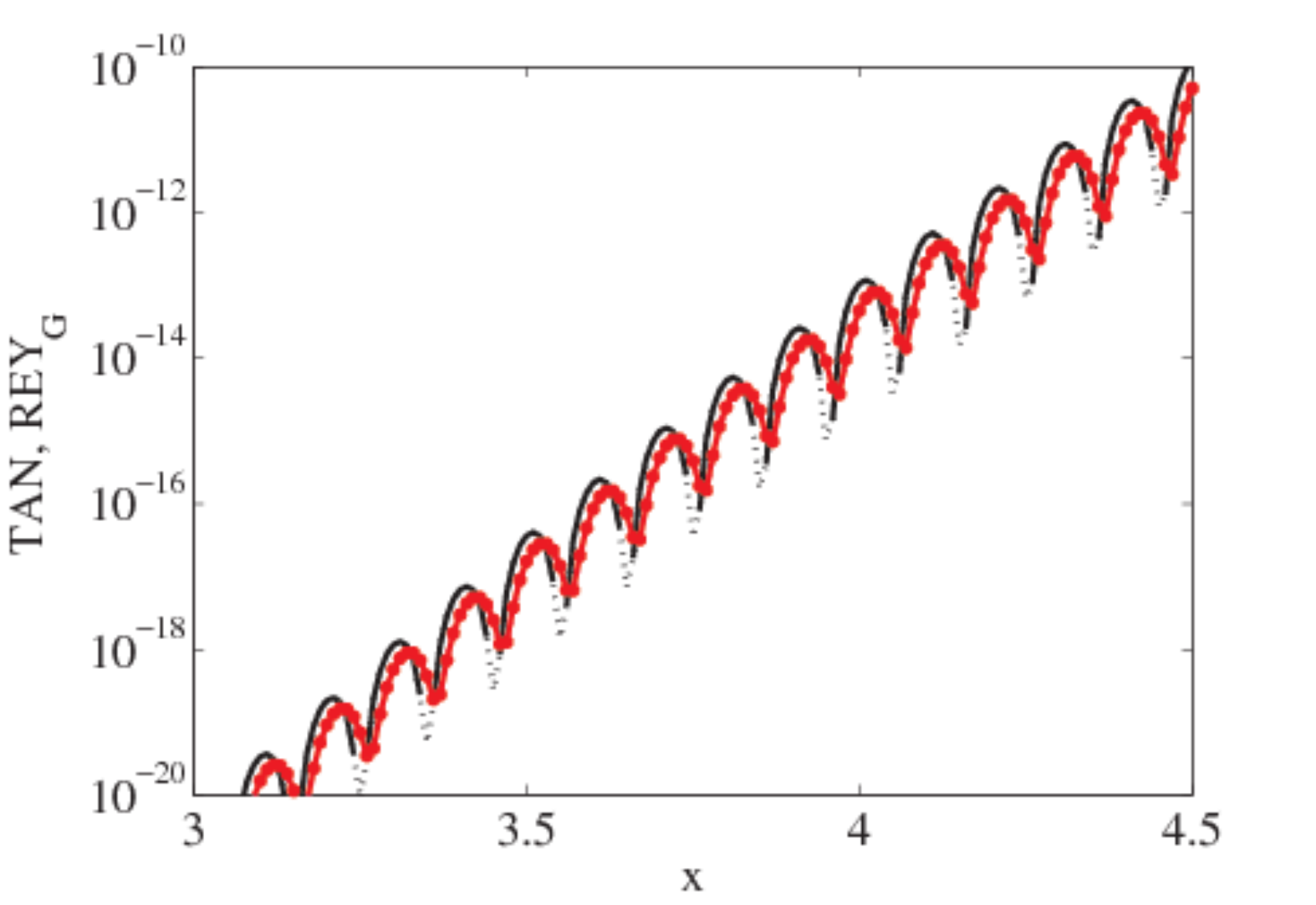}}\\
\subfigure[]{\includegraphics[width=0.48\textwidth]{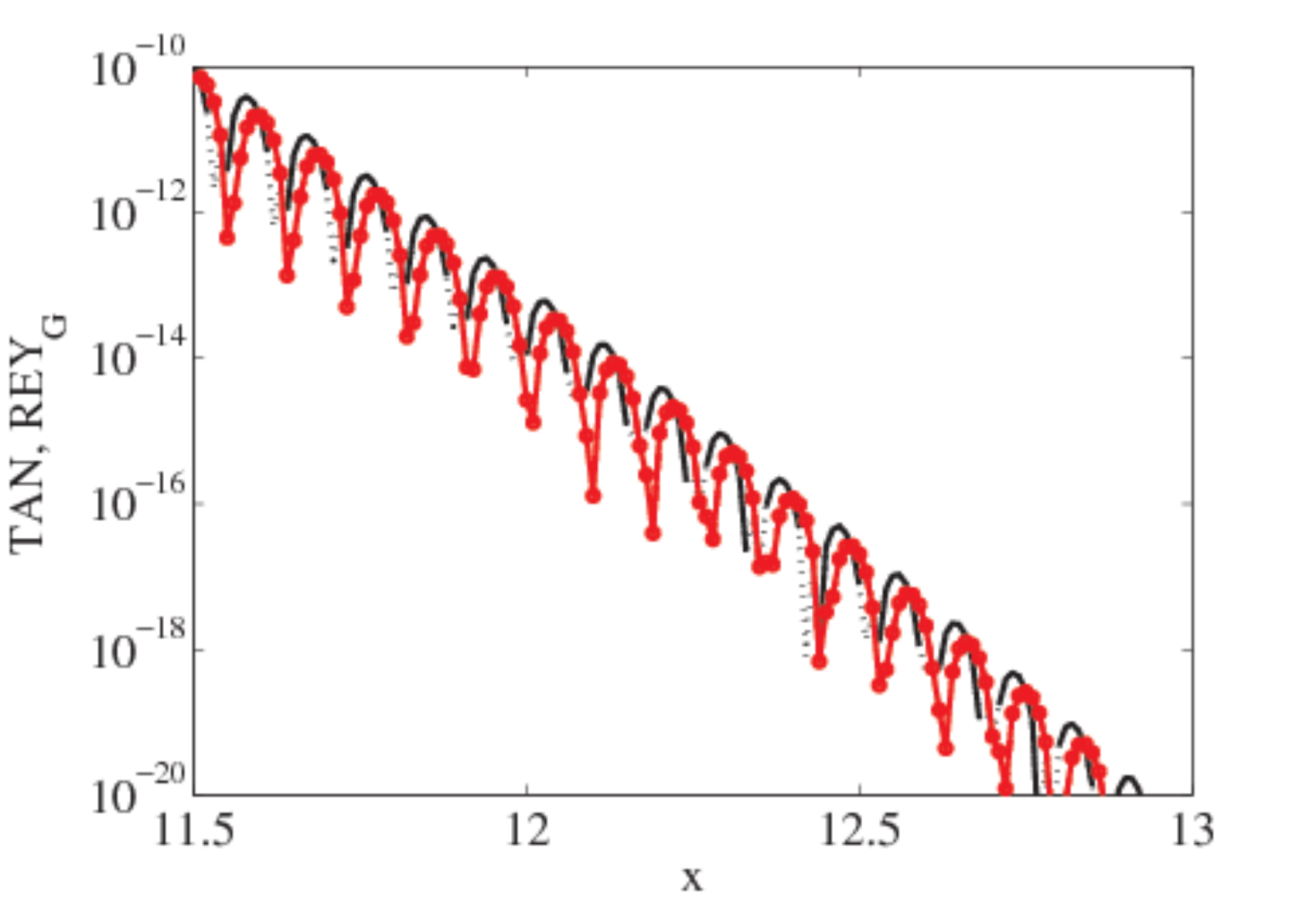}}
\caption{Energy budget for (a,b) $Re=2000$, $m=2000$ (convective); (c) $Re=4000$, $m=300$ (convective); (d-e), $Re=4000$, $m=2000$.  Solid line, $TAN$; Solid line with circles, $REY_G$; dotted line segments represent absolute value of the same variable as along the rest of the curve. All are at $t=50$ for $r=1000$, $\thickness =0.1$.}
\label{fig:energy_budgets_turb}
\end{figure}

As in previous studies of the purely temporal instability, the term $TAN$ is identified with the viscosity-contrast instability (`Yih mode')~\citep{Yih1967,Boomkamp1996}.  A positive value of this term indicates work done by the perturbations on the interface due to the viscosity jump across the interface.    Again, in analogy with the purely temporal case, the term $REY_L$ is due to an instability of Tollmien--Schlichting type in the bulk liquid flow, while positive values of $REY_G$ correspond to an instability of the Miles type~\citep{Miles1957} near the critical layer in the gas.  Equally, the terms $REY_L$ and $REY_G$ can be thought of as giving the rate at which energy is transferred from the mean flow to the disturbance via the wave-induced  Reynolds stresses.

We examine these terms in detail now (the sinks $DISS_{L,G}$ are not of interest, since they are necessarily stabilising).
To this end, we have selected three states $(Re,m)$ from  Figure~\ref{fig:CAturb}(b) that intersect the near-horizontal and near-vertical parts of the C/A transition. 
For these states, we plot the sources and sinks as a function of $x$ at a fixed point in time (the $t$-value is chosen such that all transience has been eliminated from the pulse).
Naturally, the curves exhibit oscillatory behaviour (the distribution of the phase of even a single temporal mode shows oscillations).  The energy budget for the purely temporal study is averaged over a single wave length, but spatial averaging is not conducted here, since the spatial distribution is the focus of the study.  Therefore,  we examine in Figure~\ref{fig:energy_budgets_turb}(a) a snapshot of the spatial distribution of the largest terms in the budget for $Re=2000$, $m=2000$, $\thickness =0.1$.  This parameter set is seen in Figure~\ref{fig:CAturb}(b) to be convectively unstable and to lie to the left of the C/A critical curve.   The term $TAN$ dominates throughout the pulse, followed by $REY_G$ (other, smaller terms are not shown). 
%
%
%An asymmetry in the distribution of $REY_G$ observed for the laminar case does not arise here. 
%
%
For  $Re=4000$, $m=300$, $\thickness =0.1$, this is also the case but, although $TAN$ reaches locally the largest values over a wave length, its sign changes, whereas $REY_G$ is positive virtually throughout the pulse (a zoomed view is shown in Figure~\ref{fig:energy_budgets_turb}(c)). Finally, crossing the C/A boundary (Figure~\ref{fig:energy_budgets_turb}(d-e)) causes the distribution of $REY_G$ to become asymmetric, while the features regarding the signs of $TAN$ and $REY_G$ previously observed for $m=300$ still hold.

Having characterized the turbulent base state in detail, we revisit the laminar problem
and investigate whether large-$r$ absolute instability is possible there.

\section{Revisiting the laminar problem}
\label{sec:laminar}

In this section we review the problem of interfacial instability where the upper layer is laminar.  Although the main conclusions for the turbulent case carry over here, some differences arise.
The base-state flow $U_0(z)$ is determined in a standard fashion by solving the momentum balance
\[
\mu_j\frac{d^2 U_0}{dz^2}-\frac{dP}{dL}=0,\qquad j=L,G,
\]
and is subject to continuity of velocity and shear stress across the interface at $z=0$.
To facilitate comparison with previous work on laminar flows, for this section only, we adopt the non-dimensionalization scheme of~\citet{Valluri2010}.  We set $\thickness=\dliq/(\dliq+\dgas)$ and $Re=\rho_G V (\dliq+\dgas)/\mu_G$, where the characteristic velocity $V$ is chosen to be the superficial velocity $(\dliq+\dgas)^{-1}\int_{-\dliq}^{\dgas} U_0(z)\mathd z$.  The gravity and surface tension are parameterized as $\mathcal{G}:= (\rho_L-\rho_G)g(\dliq+\dgas)^2/(\mu_G V)$ and  $\mathcal{S}:= \gamma /(\mu_G V)$ respectively; the value of these parameters is varied in the following parameter studies.

Because the model for the turbulent case relies on the quasi-laminar theory for the Orr--Sommerfeld perturbation equations, the streamfunction equations and the transient-DNS numerical method carry over directly to the laminar case, once the necessary rescaling and non-dimensionalization have been performed (e.g., using $\mathcal{G}$ and $\mathcal{S}$ in the normal stress condition).

\subsection{Parametric study~(1): $\mathcal{G}$ and $\mathcal{S}$ taken as constant}

We set $\mathcal{G}=0$ and take   $\mathcal{S}=0.01$, corresponding exactly to the paper of~\citet{Valluri2010}; this choice enables us to relate the current investigation to the results of~\citet{Valluri2010} for liquid/liquid systems. 
For these parameter values, the  ray analysis has revealed that the laminar base state is absolutely unstable for a significant portion of parameter space, even for a density ratio of $r=1000$. 
\begin{figure}
\centering
\subfigure[$\,\,\thickness=0.025$]{\includegraphics[width=0.49\textwidth]{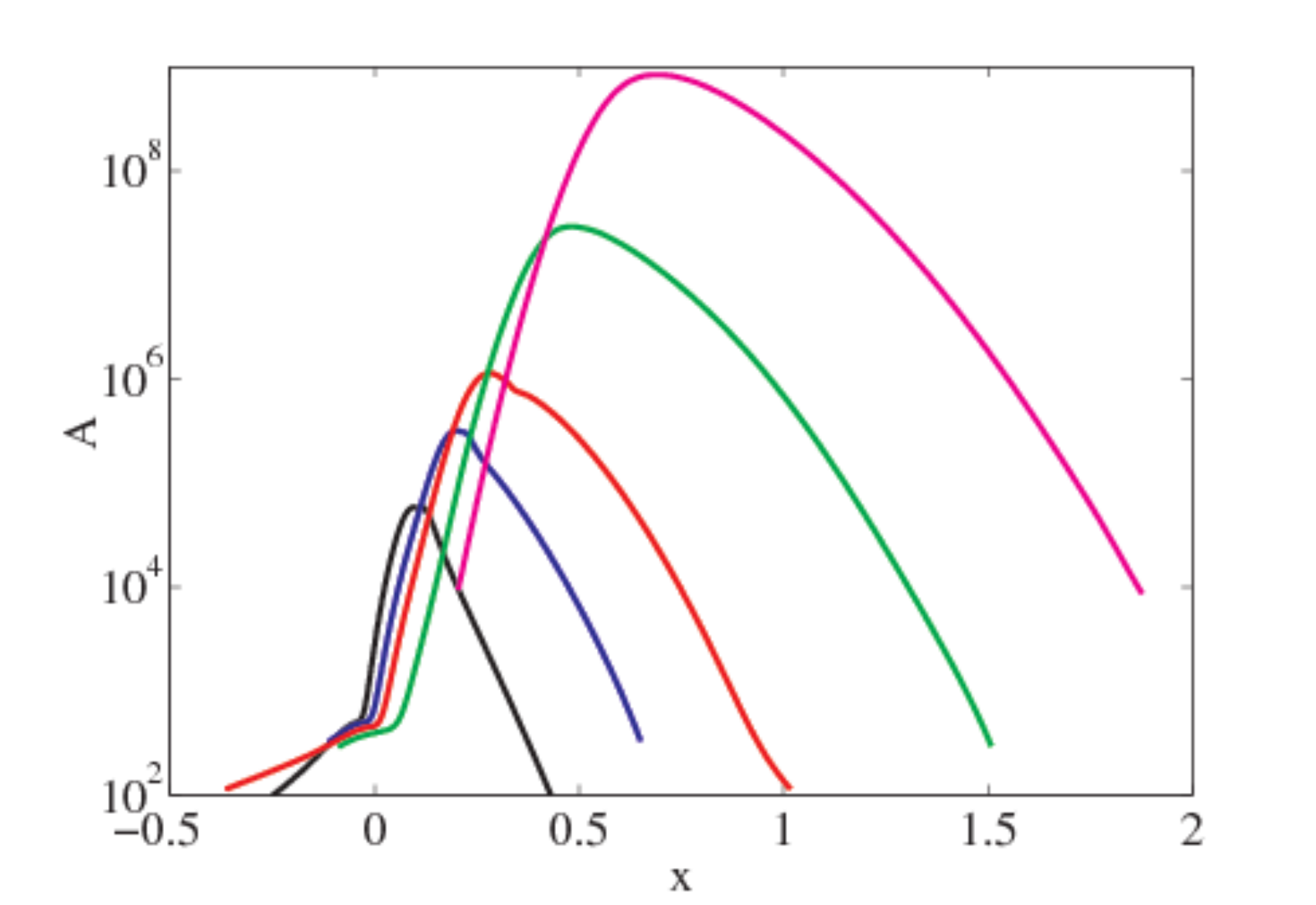}}
\subfigure[$\,\,\thickness=0.04$]{\includegraphics[width=0.49\textwidth]{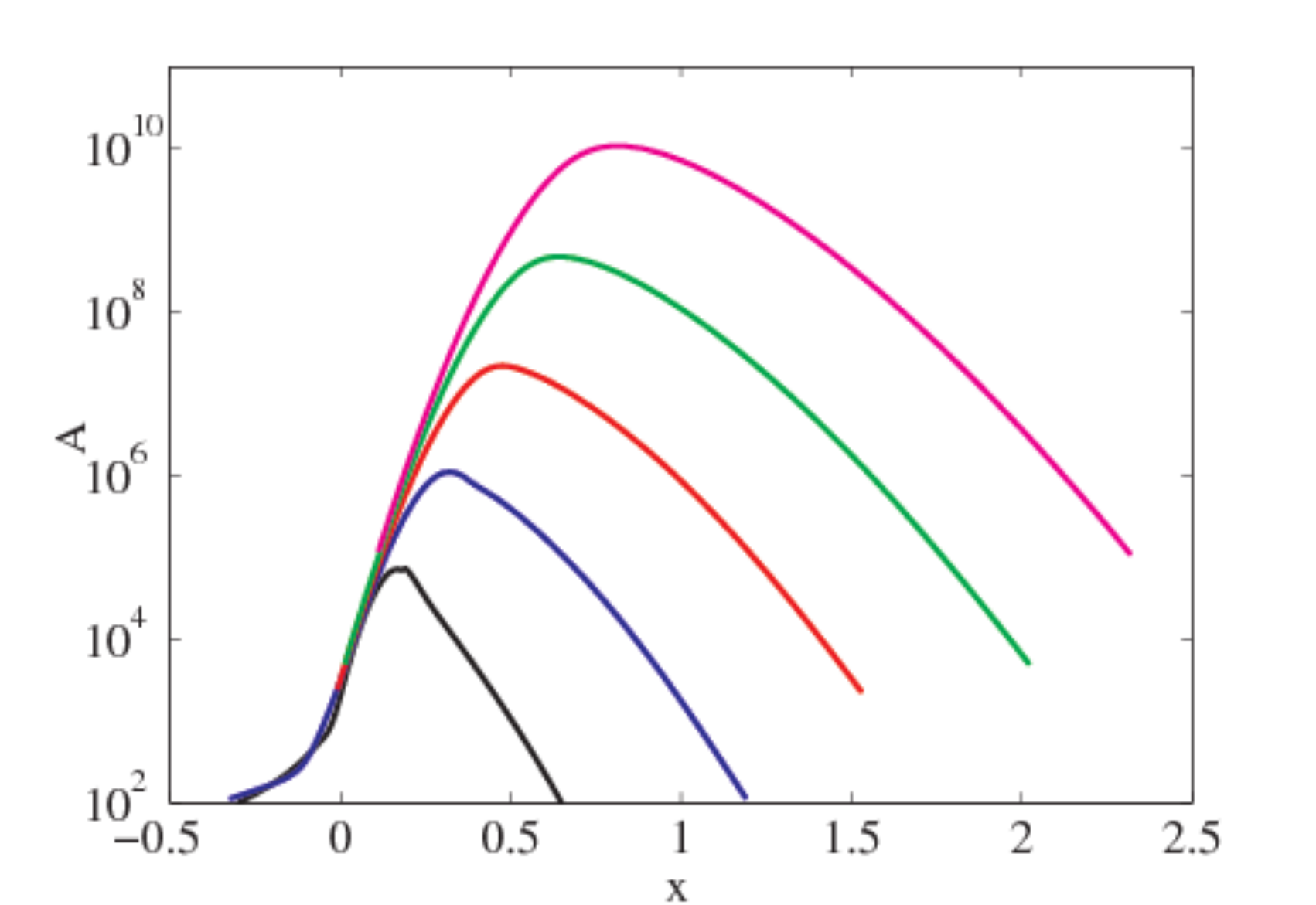}}
\subfigure[$\,\,\thickness=0.08$]{\includegraphics[width=0.49\textwidth]{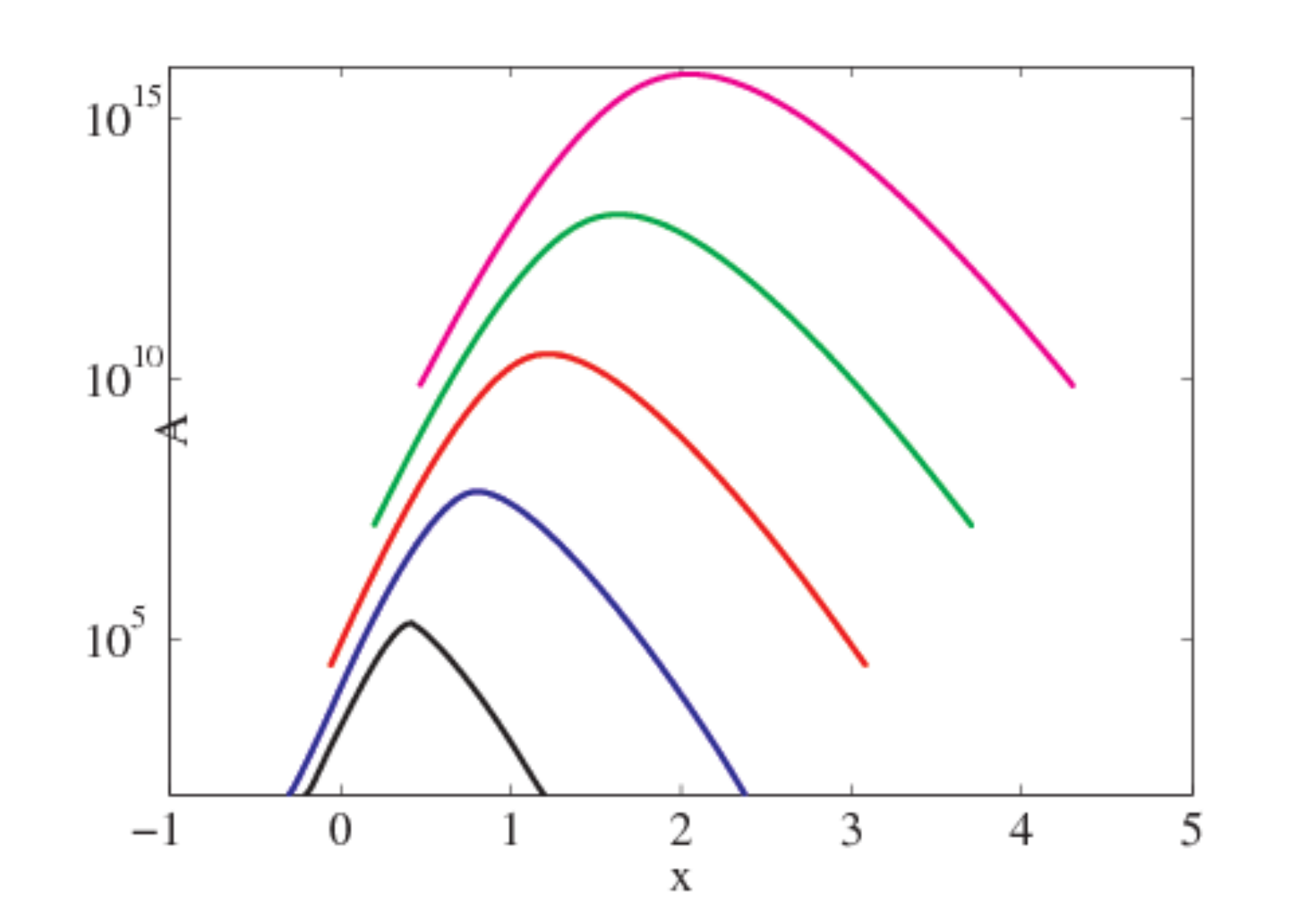}}
\subfigure[$\,\,\thickness=0.15$]{\includegraphics[width=0.49\textwidth]{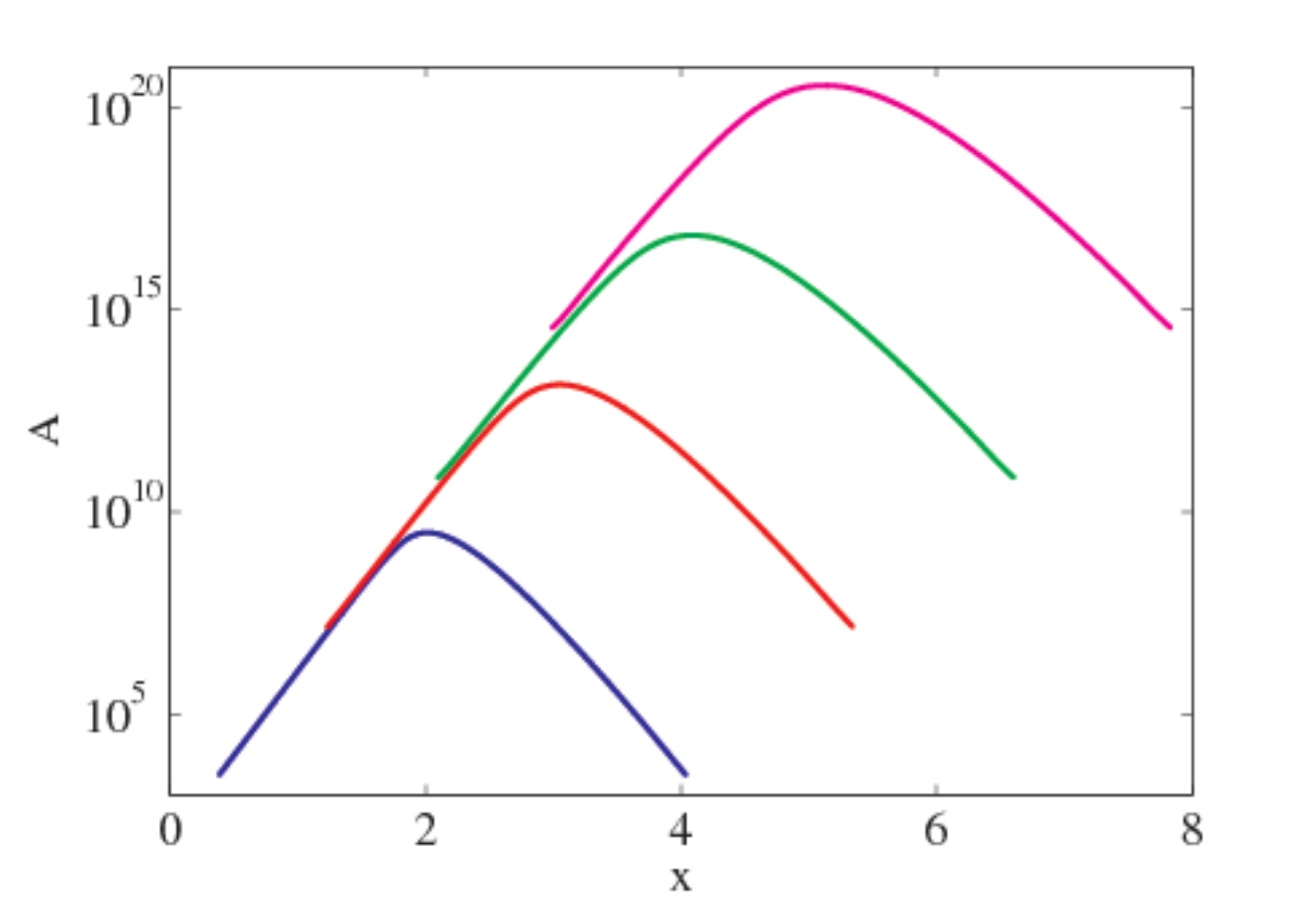}}
\subfigure[$\,\,\thickness=0.18$]{\includegraphics[width=0.49\textwidth]{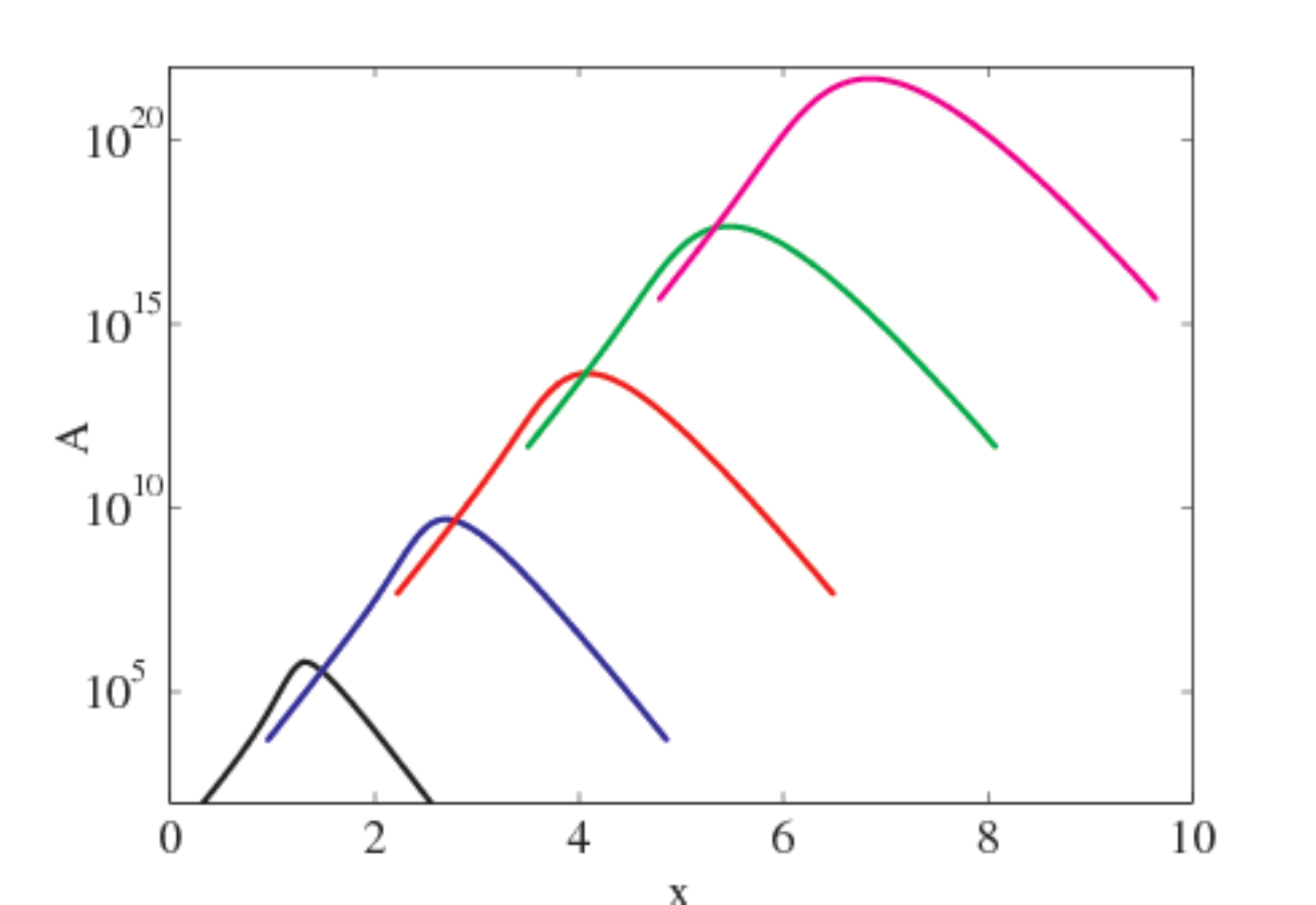}}
\caption{Snapshots of the norm $A(x,t)$ for different values the depth of $\thickness$ for the laminar base state. The parameters are $Re=1500$, $m=150$, $r=1000$, $\surft=0.01$, and $G=0$.  (a) $t=100, 200, 300, 500, 700$; (b)--(d) $t=100,200, 300, 400, 500$; (e) $t=50,100,150,200,250$.  Figures (a) and (e) are convectively unstable cases; (b)--(d) are absolutely unstable.
In all cases, $L_x=30$, $N_1=21$, $N_{2}=51$, the timestep is $\Delta t=0.1$, and the grid size $\Delta x$ in the $x$-direction is $0.003$.}
\label{fig:raytrans}
\end{figure}
This is seen in Figure~\ref{fig:raytrans}, where the $L^2$-norm  $A(x,t)$ associated with the broadband disturbance is shown at several times. These curves have been truncated at the points where they decrease below a specified fraction of their maximum value, as the finite working precision makes the results too uncertain far away from the pulse.
The existence of absolute instability at large density ratios  was not found in the modal analysis of~\citet{Valluri2010},  who only ascertained that density-matched fluids can become absolutely unstable. The cause for this oversight is that the magnitude of the wave number at the saddle point is very large, beyond the range searched by~\citet{Valluri2010}. 
\begin{figure}
\centering
\subfigure[]{\includegraphics[width=0.49\textwidth]{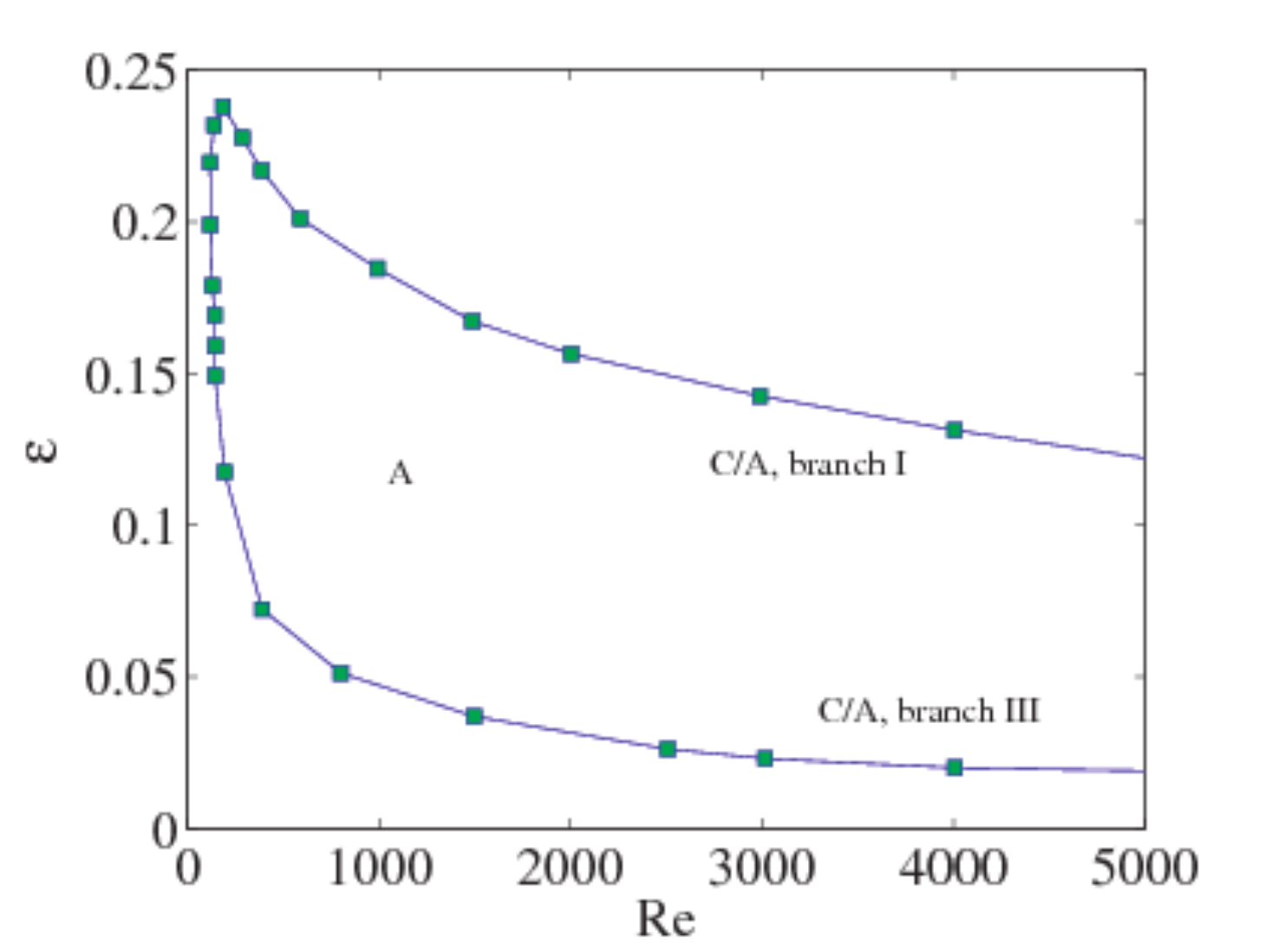}}
\subfigure[]{\includegraphics[width=0.49\textwidth]{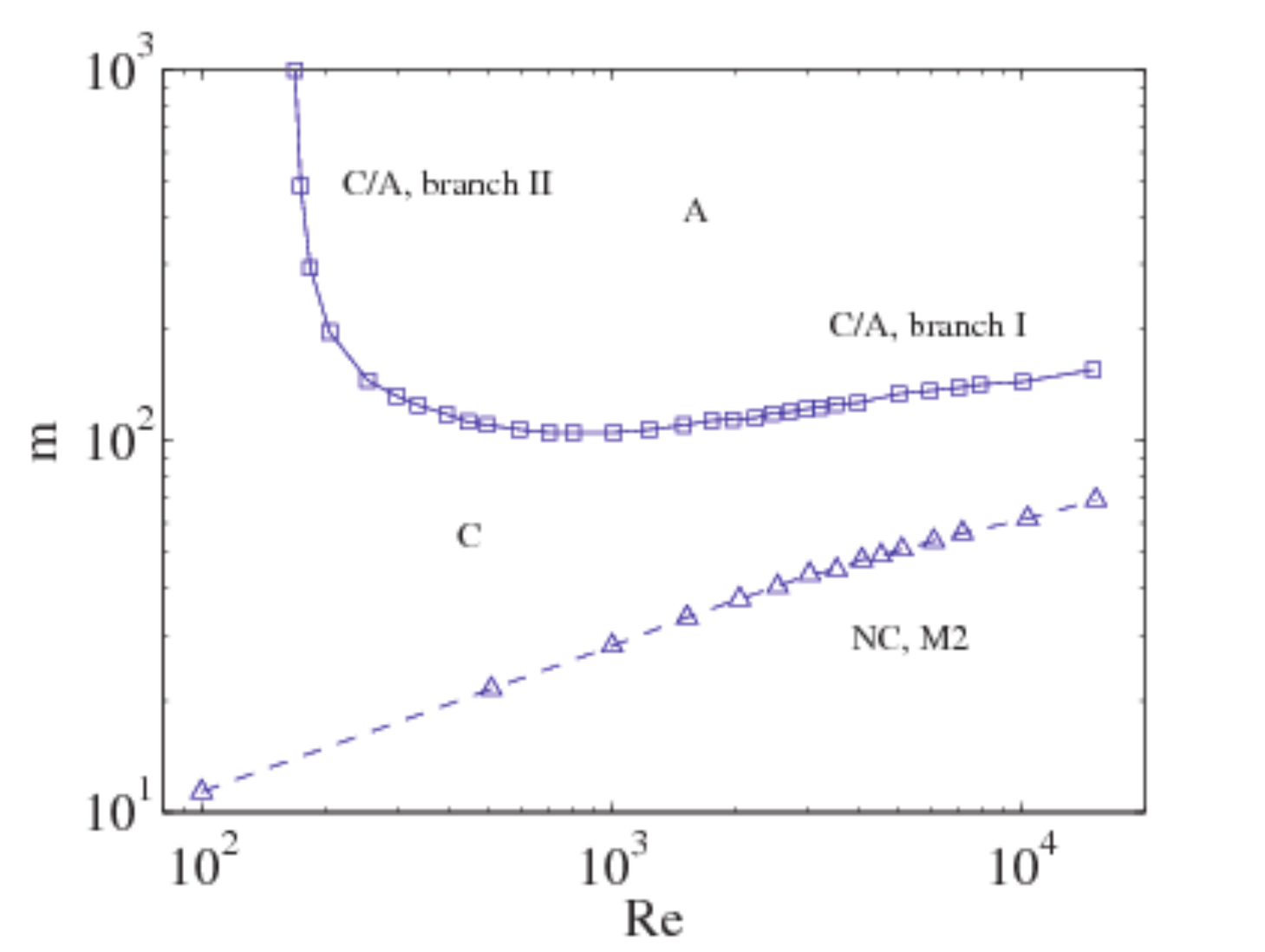}}
\caption{ Flow-regime map showing laminar base state in parametric study~(1).  Here, $r=1000$, $\mathcal{S}=0.01$, and $\mathcal{G}=0$.  The system is always either convectively (C) unstable or absolutely (A) unstable.  A second temporal mode is also unstable below the neutral curve (dashed curve labelled `M2', with triangles).  (a)  Variations in $\thickness$ at fixed $m=150$; (b) Variations in $m$ at fixed thickness $\epsilon=0.1$.}
\label{fig:CAplots}
\end{figure}
\begin{figure}
\centering
\includegraphics[width=0.49\textwidth]{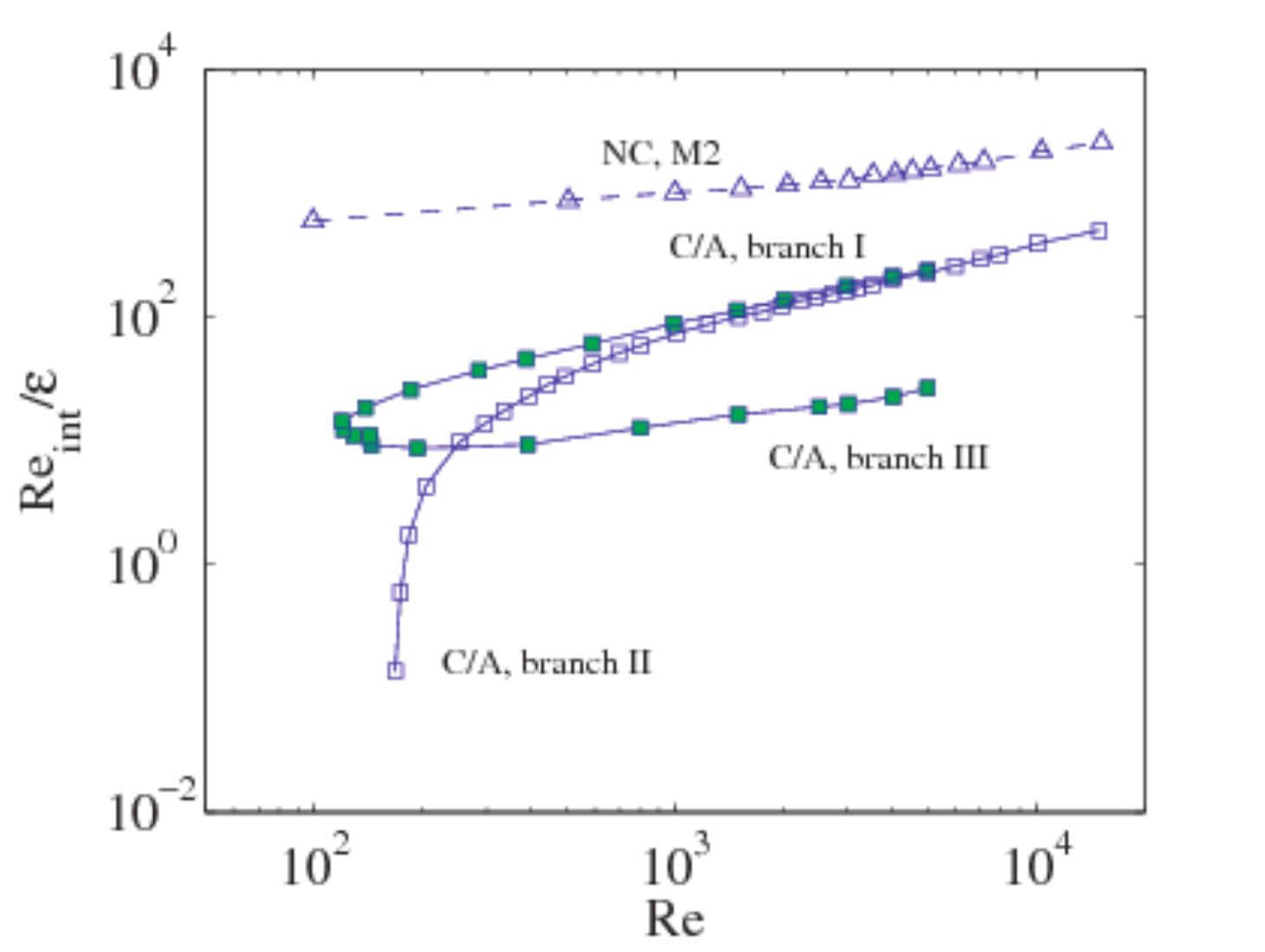}
\caption{Flow-regime map of parametric study~(1) in terms of the liquid Reynolds number $\Reint$.  Filled symbols: variations in $\thickness$ at constant $m=150$; open symbols: variations in $m$ at fixed $\thickness=0.1$.  As described in the text, a further mode (convectively unstable) exists for $Re\apprge 8000$, whose neutral curve is unaffected by changes in $\Reint$, $m$, and $\thickness$.}
\label{fig:CAplots_collapse1}
\end{figure}

Motivated by the plots in Figure~\ref{fig:raytrans} showing absolute instability, we revisit the otherwise more accurate modal analysis and perform a large scan through the complex $\alpha$-plane to obtain the C/A  boundaries. These boundaries are consistent with the results of the ray analysis shown in Figure~\ref{fig:raytrans}. Also, the results are similar to those in~\cite{Valluri2010} for $r=1$, suggesting that the transition found by~\citet{Valluri2010} extends to density ratios of at least $r=1000$. We have verified that, upon lowering the density ratio at a point in the absolute regime of Figure~\ref{fig:CAplots}, also identified as such in~\citet{Valluri2010}, the system remains absolutely unstable at intermediate values of $r$. In contrast to the $r=1$ case in~\citet{Valluri2010}, for $r=1000$, absolute instability occurs only at large  $m$-values.

As in the turbulent case, we have examined the flow-regime boundaries in the ($Re,\Reint$) plane, where $\Reint := r\thickness Re U_{\mathrm{int}}/m$ is a Reynolds number based on the liquid-film properties and the interfacial velocity $U_{\mathrm{int}}$ of the base state. In contrast to the turbulent case, the results of Figure~\ref{fig:CAplots} do not collapse, although the overall trends otherwise bear some resemblance to the turbulent case (e.g. Figure~\ref{fig:CAturb}): branch III corresponds to a critical value of $\Reint$ (consistent with~\citet{Valluri2010} for $r=1$), while branch II corresponds to a critical value of $Re$. These two branches shift when changing the value of $m$, respectively $\thickness$, thereby contracting or expanding the absolutely unstable regime.
Instead, we plot the regime boundaries in the ($Re,\Reint/\thickness$) plane in Figure~\ref{fig:CAplots_collapse1}. Although the two branches labelled `II' virtually coincide in this plane (this is discussed further below), the functional form $\Reint=\thickness f(Re)$ of the neutral curve differs substantially from that already encountered in the turbulent study.

\begin{figure}
\centering
\subfigure[]{\includegraphics[width=0.49\textwidth]{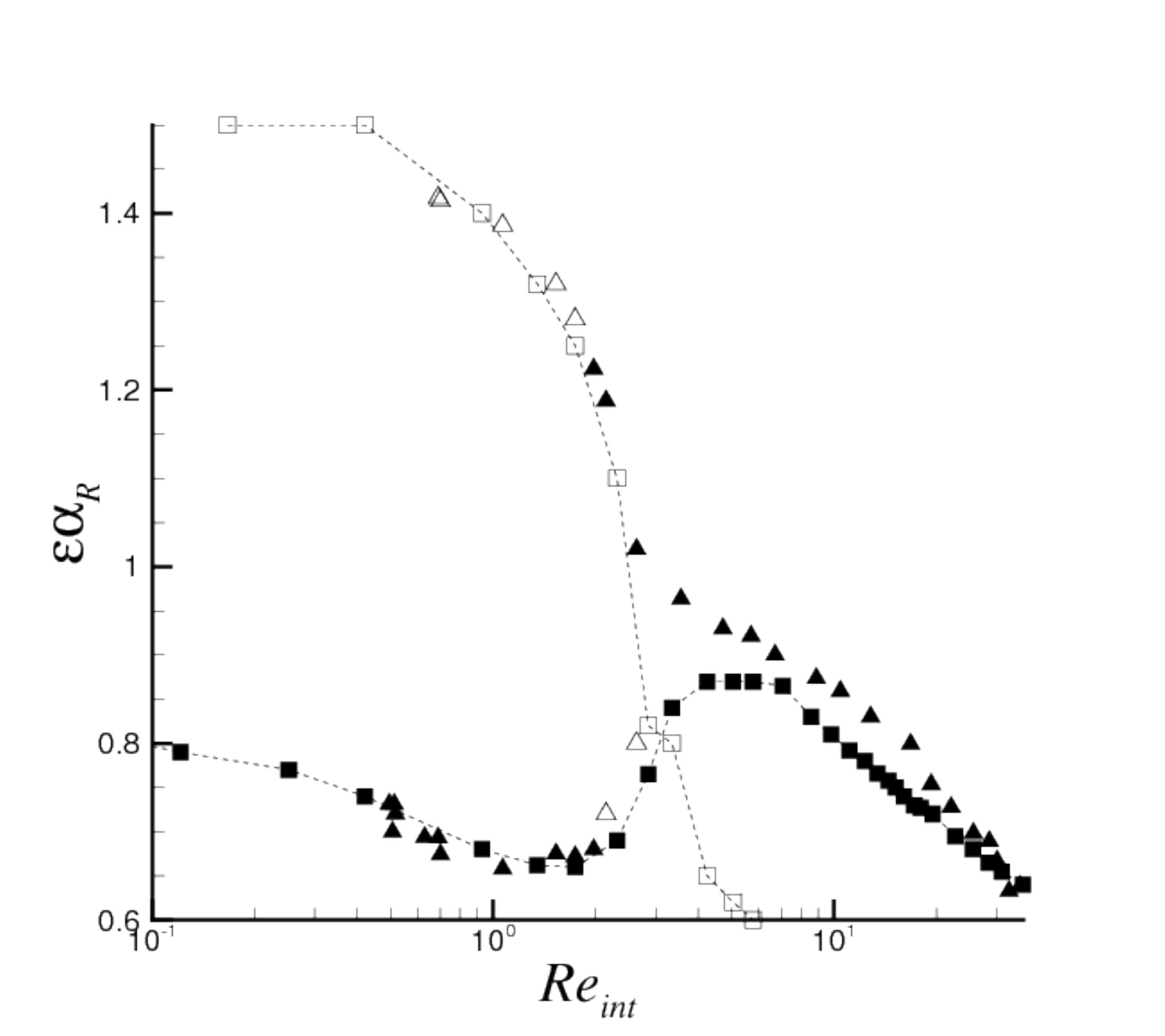}}
\subfigure[]{\includegraphics[width=0.49\textwidth]{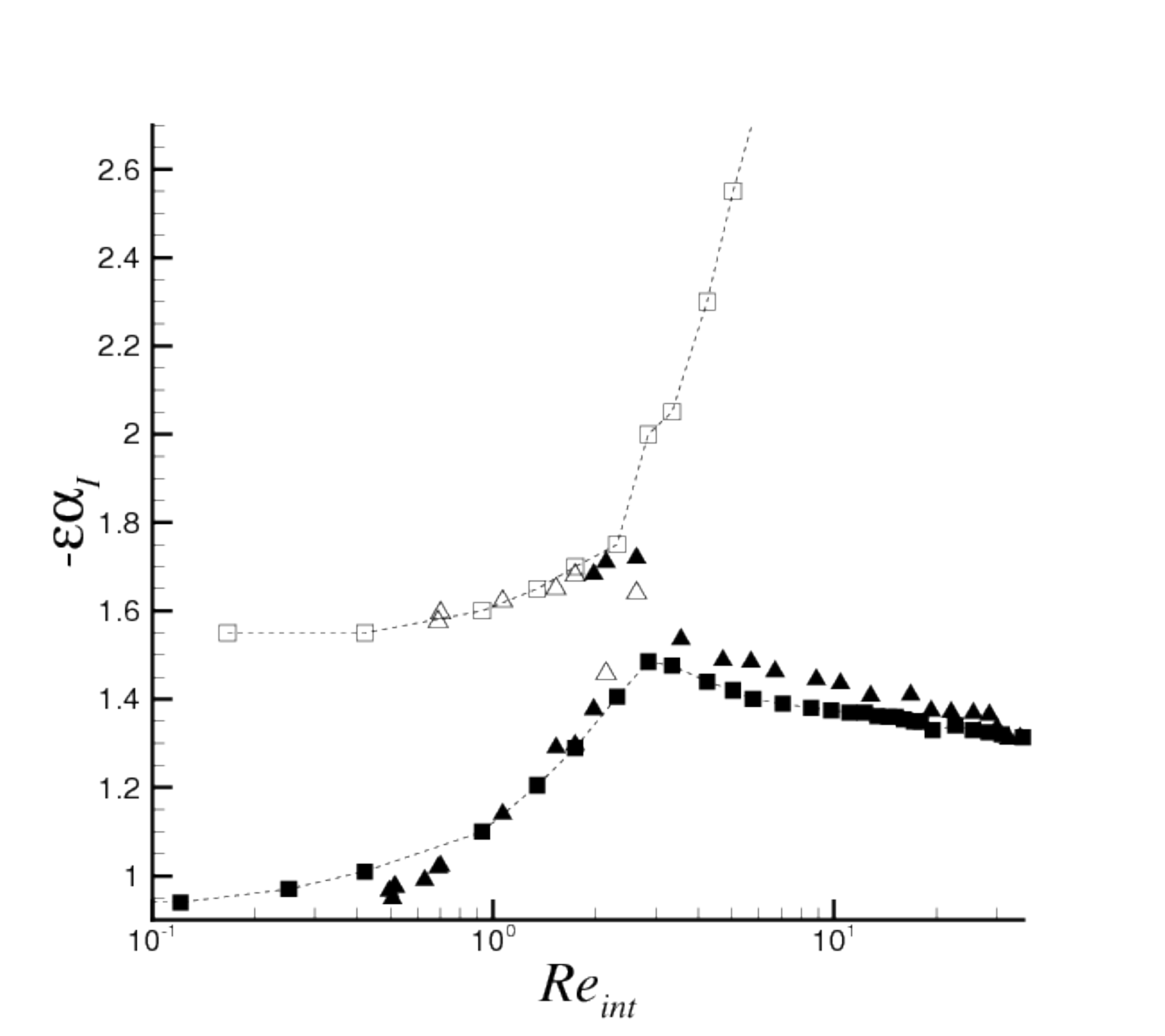}}
\caption{The real  (a) and imaginary (b) component of the wave number at the saddle point along all C/A transitions in Figure~\ref{fig:CAplots}(a) (squares) and  Figure~\ref{fig:CAplots}(b) (triangles), as functions of $\Reint$. The open symbols represent secondary saddle points that do not correspond to absolute instability.}
\label{fig:CAalpha1}
\end{figure}
We also investigate variations in the wave number at the saddle point (taken at the onset of absolute instability) in Figure~\ref{fig:CAalpha1}, where $\thickness(\ar,\ai)$ is plotted as a function of $\Reint$.   The contrast between these results and the earlier turbulent results is remarkable.
The wave number is $O(1/\thickness )$.  While the flow-regime map does not collapse in $(Re,\Reint)$-space, the wave numbers $\thickness (\ar ,\ai)$ depend mainly on $\Reint$, and not separately on $Re$. The wave number $\alpha_0$ at the saddle point follows from $(d\omega /d\alpha)_{\alpha_0} =0$, which therefore only involves a relatively simple scaling, whereas the flow regime is determined by $\omi(\alpha_0) >0$, which  involves more parameters. Nevertheless, the dependency of $\thickness\alpha$ on $\Reint$ is complex. Above $\Reint\approx 5$ (corresponding to branch I) the curves nearly collapse; a small difference remains, as is also seen in the corresponding conditions C/A transition curves in Figure~\ref{fig:CAplots_collapse1}. 
Upon decreasing the value of $\Reint$, two saddle points emerge.  However, only one of the saddle points produces absolute instability.  Most of the results in Figure~\ref{fig:CAalpha1} lie on the branch that corresponds to somewhat longer waves (note, the data represented by the filled triangles in Figure~\ref{fig:CAalpha1}(a) jump from the upper to the lower branch when $\Reint$ is decreased). 
These substantial differences between the turbulent and the laminar cases are now addressed in a further parametric study.

\subsection{Parametric study (2): $\mathcal{G}$ and $\mathcal{S}$ vary inversely with Reynolds number}

\begin{figure}
\centering
\includegraphics[width=0.49\textwidth]{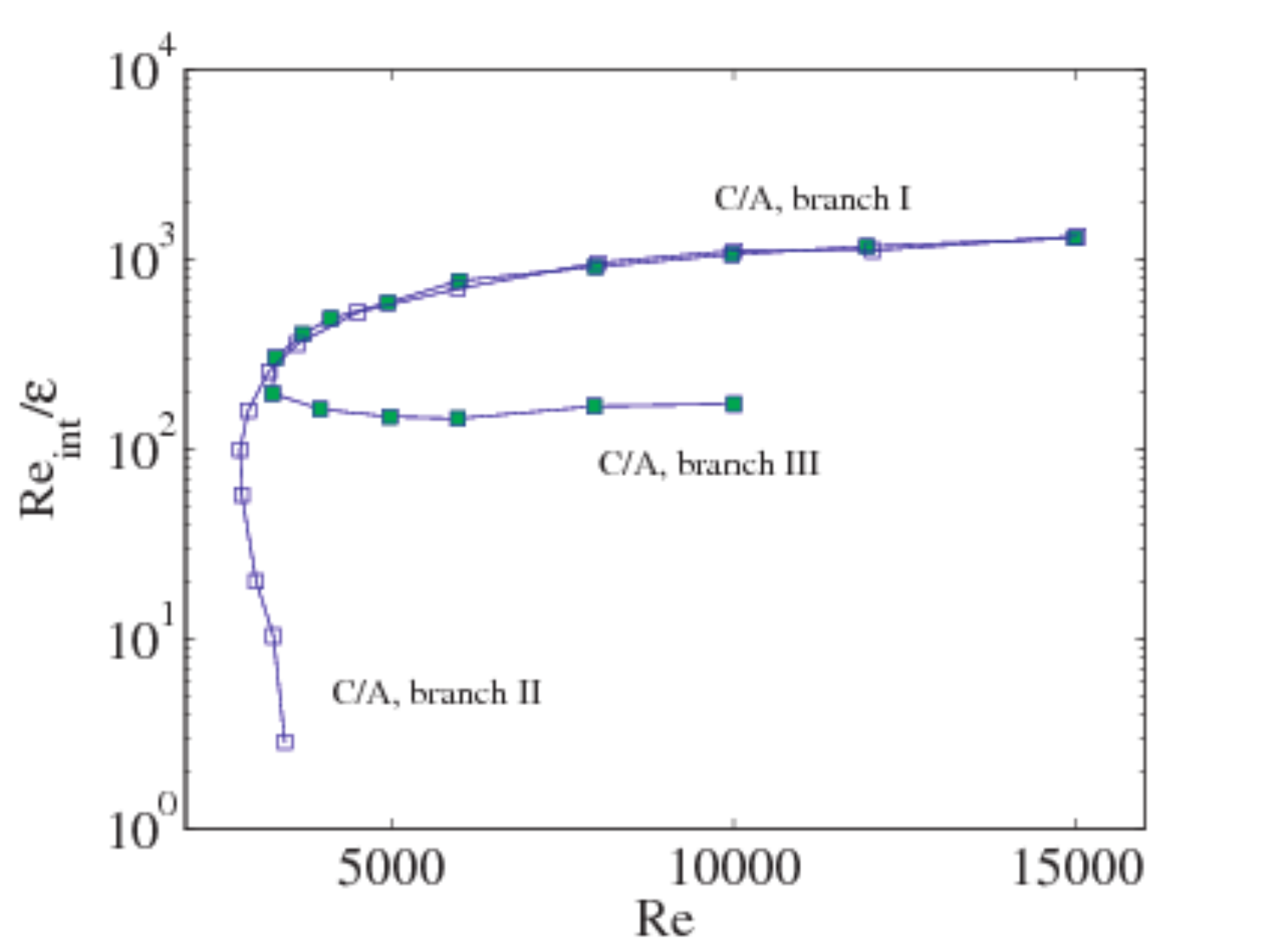}
\caption{Flow-regime map for parametric study~(2).   Filled symbols: variations in $\thickness$ at constant $m=100$; open symbols: variations in $m$ at fixed $\thickness=0.1$.}
\label{fig:CAplots_collapse2}
\end{figure}

 We extend parametric study~(1) to allow for effects of gravity and surface tension corresponding to systems with a large density ratio, and  we set $\mathcal{G}=\mathcal{G}_0 (r-1)/Re$ and $\mathcal{S}=\mathcal{S}_0/Re$.  This extension is more reflective of real systems because it corresponds to a fixed value of the dimensional surface tension $\gamma$, while allowing for variations in $\mathcal{S}$ through changes in the Reynolds number.  Here, the values of $\mathcal{G}_0$ and $\mathcal{S}_0$ are  the same as $\gravpre$ and $\surftpre$ (Section~\ref{sec:turb}) but such that these are for channels that are ten times smaller: $\mathcal{G}_0=\gravpre/10^3$ and $\mathcal{S}_0=\surftpre/10$ unless indicated otherwise (the effect of changing these values is discussed below).
The results of this parameter modification are shown in Figure~\ref{fig:CAplots_collapse2}.  A striking feature of this study is the blurring of the previously sharp distinction between the turbulent case (Figure~\ref{fig:CAturb_collapse}), and the laminar study (Figure~\ref{fig:CAplots_collapse1}).  Moreover, increasing $\mathcal{G}$ and $\mathcal{S}$ leads to a substantial modification of the absolute region in parameter space.
The role played by $\mathcal{G}$ and $\mathcal{S}$ in this modification is further highlighted in Tables~\ref{tab:lam1}--\ref{tab:lam2}.  Table~\ref{tab:lam1}  corresponds to branch I, and demonstrates that an increase in $\mathcal{G}$ or $\mathcal{S}$ at fixed $m$ and $\thickness$ calls for an increase in $Re$  to sustain absolute instability.   Similarly, Table~\ref{tab:lam2} corresponds to branch II, and demonstrates that an increase in $\mathcal{G}$  at fixed $Re$ and $\thickness$ requires a corresponding, destabilizing increase in $m$ to sustain absolute instability.  This produces a proportional decrease in $\Reint$.  The situation concerning increases in $\mathcal{S}$ is somewhat counter-intuitive (increases in $\mathcal{S}$ produce decreases in the critical value of $m$, which produce increases in the critical value of $\Reint$).  However, this agrees with the turbulent case, which has already been explained using the quadratic approximation. 
\begin{table}
\centering
\begin{tabular}{ccccc}
$\mathcal{G}_0/\mathcal{G}_{0,\mathrm{ref}}$ & $\mathcal{S}_0/\mathcal{S}_{0,\mathrm{ref}}$ & $Re$ & $\Reint$ & $\alpha$\\
\hline
0.1& 1 & 2050 & 0.17 & $1.8-0.65\imag$ \\
1& 1 & 3450 & 0.28 & $2.2-0.9\imag$\\
100& 1 & 9250 & 0.76 & $3.2-1.8\imag$\\
1& 0.1 & 2750 & 0.23 & $4.1-2\imag$\\
1& 1 & 3450 & 0.28 & $2.2-0.9\imag$\\
1& 100 & 5350 & 0.76 & $1.2-0.45\imag$\\
\end{tabular}
\caption{Dependence of the critical parameters on surface tension and gravity at  low $\Reint$ (branch I, $m=1000$, $\thickness =0.1$).
 The wavenumber at the saddle point is also stated. 
The uncertainty in $Re$ is $\pm 50$, which also introduces an uncertainty in the values of $\Reint$ and $\alpha$.
The symbols $\mathcal{G}_{0,\mathrm{ref}}$ and $\mathcal{S}_{0,\mathrm{ref}}$ refer to the reference values discussed at the beginning of the section.}
\label{tab:lam1}
\vspace{0.1in}
\begin{tabular}{ccccc}
$\mathcal{G}_0/\mathcal{G}_{0,\mathrm{ref}}$ & $\mathcal{S}_0/\mathcal{S}_{0,\mathrm{ref}}$ & $m$ & $\Reint$ & $\alpha$\\
\hline
0.1 & 1 & 82.5 & 120 & $4.1-3.1\imag$\\
1 & 1 & 87.5 & 107 & $4-3.2\imag$\\
100 & 1 & 167.5 & 29 & $4.1-3.1\imag$\\
1& 0.1 & 122.5 & 55 & $8.5-9.5\imag$\\
1& 1 & 87.5 & 107 & $4-3.2\imag$\\
1& 100 & 77.5 & 136 & $1.6-1.1\imag$\\
\end{tabular}
\caption{Dependence of the critical parameters on surface tension and gravity at low $Re$ (branch II, $Re=10^4$, $\thickness =0.1$).   The wavenumber at the saddle point is also stated. 
The uncertainty in $m$ is $\pm 2.5$, which also introduces an uncertainty in the values of $\Reint$ and $\alpha$.
}
\label{tab:lam2}
\end{table}

The scaling of the transition curves in Figure~\ref{fig:CAplots_collapse2} can again be elucidated with the quadratic approximation.  For example, for large $Re$ and $\Reint$, we have measured 
\begin{equation}
\max(\omitemp)\propto(\thickness/m^{1/2})Re^{1/4},\qquad 
\alpha_0-\alpha_{0\mathrm{c}}\propto(\thickness^{1/2}/m^{1/2})Re^{3/4},
\qquad \cg\approx U_{\mathrm{int}}\propto \thickness/m.
\end{equation}
Plugging these scaling rules into the quadratic approximation gives $\thickness Re/m^2=\mathrm{const.}$, or $\Reint/\thickness=\mathrm{const.}$ at large $Re$ and $\Reint$, which implies the existence of a critical value of $\Reint/\thickness$ along branch I.  The dramatic distinction between these scaling rules and those encountered in the turbulent case owes not to some deep distinction between the respective base states, but rather arises from the distinct non-dimensionalization schemes chosen in each case.  Indeed, as a final test, we have verified that a laminar temporal study, based on the non-dimensionalization scheme previously applied in the turbulent case, yields the scaling rules described in Section~\ref{sec:turb:scaling} (not shown).

\begin{figure}
\centering
\hspace*{0cm}\vspace{-2.5cm}

\subfigure[]{\includegraphics[width=0.49\textwidth]{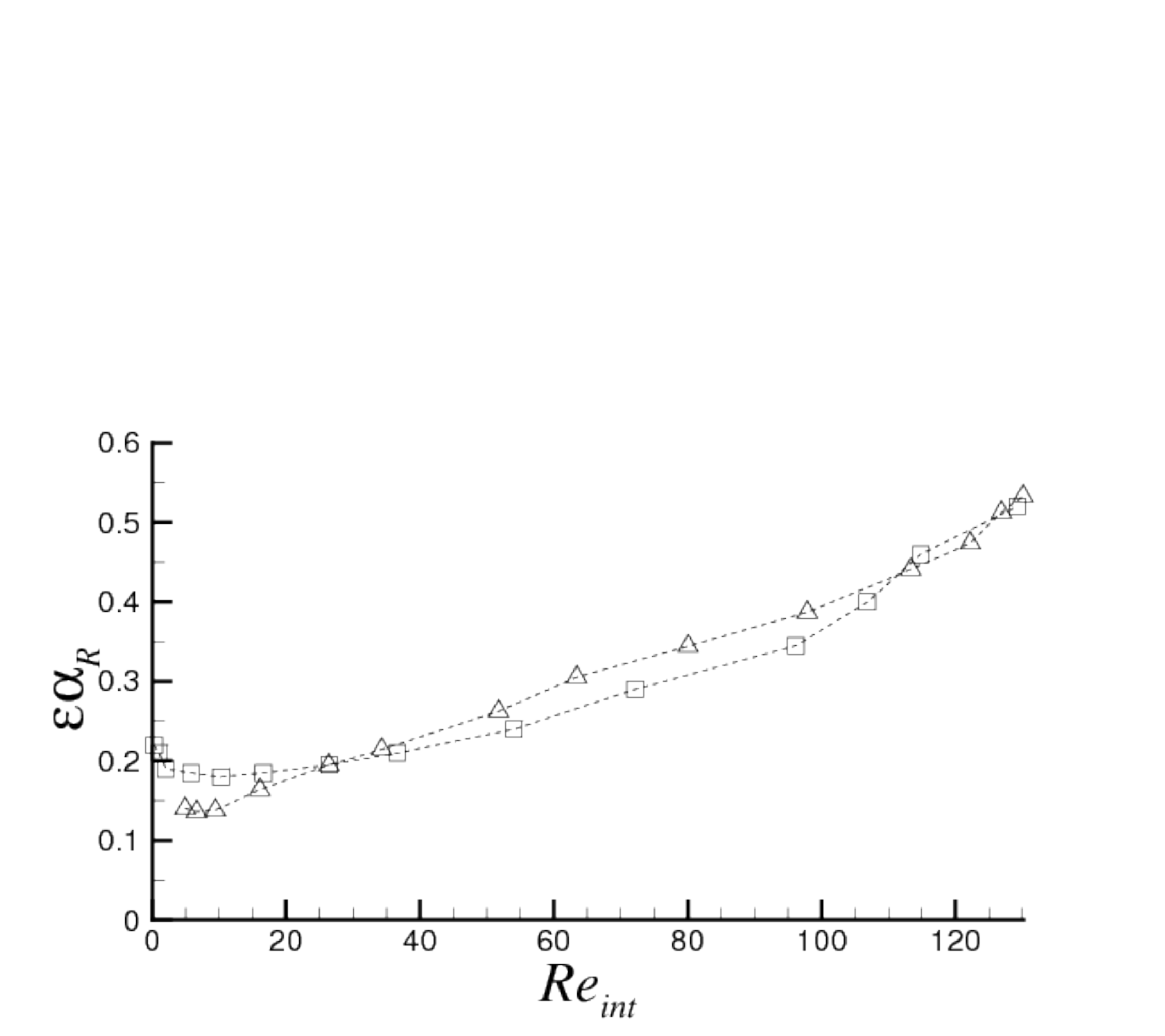}}
\subfigure[]{\includegraphics[width=0.49\textwidth]{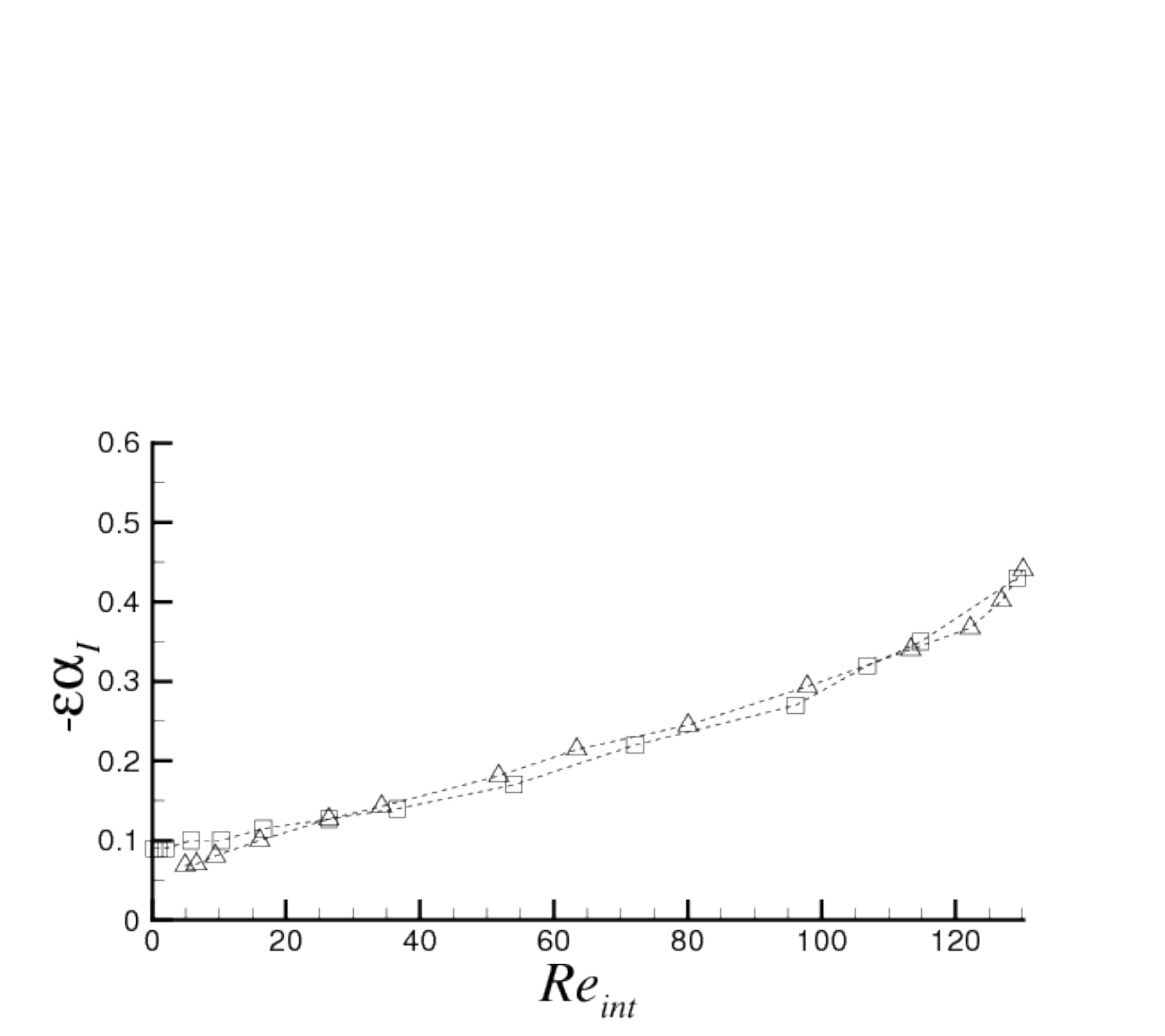}}
\caption{The real  (a) and imaginary (b) component of the wave number at the saddle point along all C/A transitions in Figure~\ref{fig:CAplots}(a) (squares) and  Figure~\ref{fig:CAplots}(b) (triangles), as functions of $\Reint$. The open symbols represent secondary saddle points that do not correspond to absolute instability.}
\label{fig:CAalpha2}
\end{figure}

In Figure~\ref{fig:CAalpha2} we examine again the wave number $\thickness(\ar,\ai)$ at the saddle point, at the inception of absolute instability.  The contrast between this figure and parametric study~(1) is remarkable.  In Figure~\ref{fig:CAalpha2}, the results resemble those for turbulence, with one significant difference: the saddle-point mode corresponds to a much longer wave.  
The reason for the contrast between parametric studies~(1) and~(2) is due to the fact that the wave length is controlled mostly by surface tension (as demonstrated by Table~\ref{tab:lam2}); the surface tension in study~(2) greatly exceeds that in study~(1), consequently, the most unstable wave length is longer.
%

% In Figure~\ref{fig:CAplots}(b), branch II (varying $m$) of the C/A transition `follows' closely the neutral stability curve for a % second temporal mode, M2 (the same neutral curve is also present in parametric study (2), but is further removed from the C/A transition curve).  As in the turbulent case, it is as if the existence of the M2 mode quenches the M1 absolute instability, and we have also found here  clear evidence of competition between spatio-temporal modes, in addition to temporal mode competition.  As before, this mode competition has been demonstrated in contour plots of $\omr$: as $m$ increases, the dominant saddle point connects to the M1 mode and absolute instability ensues.
%

\begin{figure}
\centering
\subfigure[$\,\,m=85$]{\includegraphics[width=0.32\textwidth]{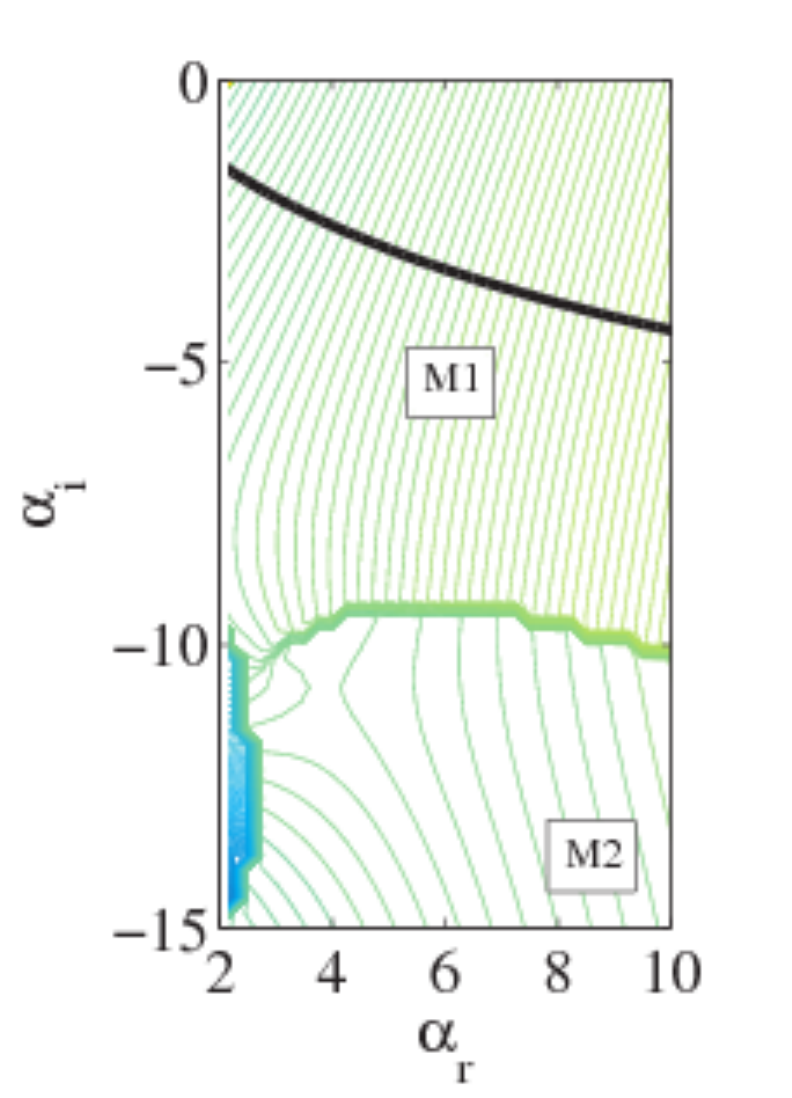}}
\subfigure[$\,\,m=110$]{\includegraphics[width=0.32\textwidth]{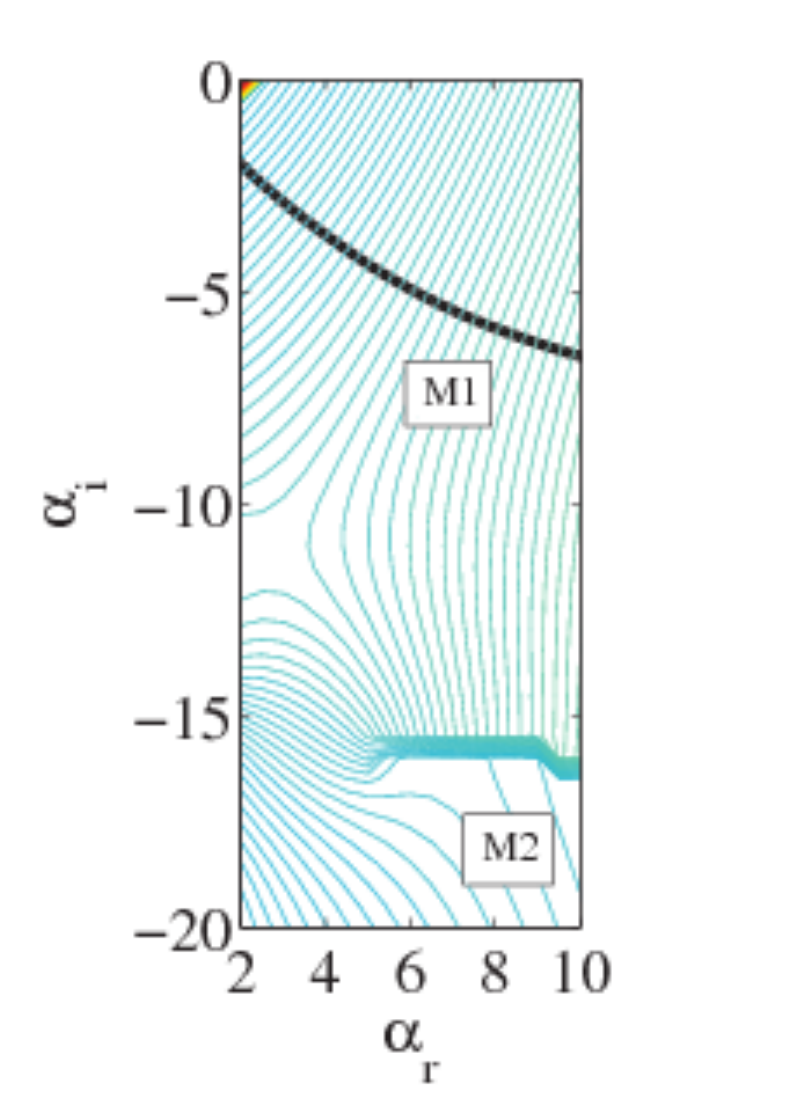}}
\subfigure[$\,\,m=130$]{\includegraphics[width=0.32\textwidth]{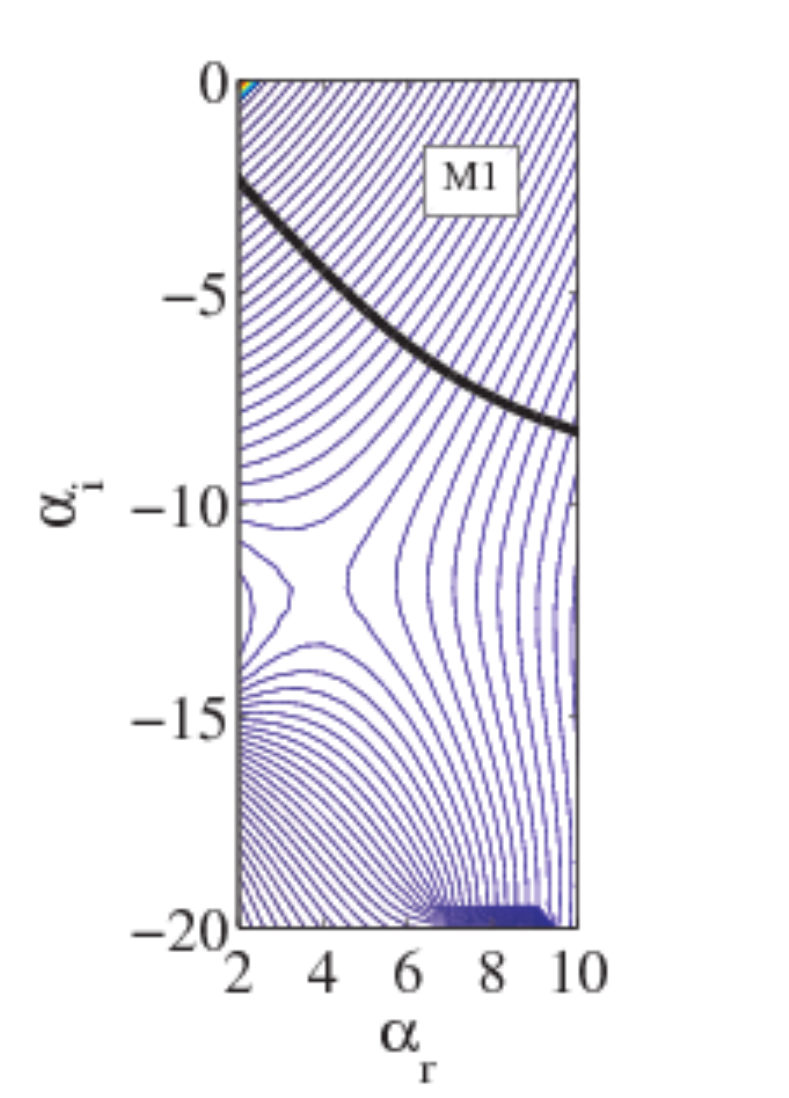}}
\caption{Competition between spatio-temporal modes for the laminar case.  Shown are contour plots of $\omr$ for the most dangerous mode in the complex $\alpha$ plane at $Re=20000$, $\thickness=0.1$ for different values of $m$.   The black lines represent the spatial curve $\omi=0$ of M1.
The sharp jumps in $\omr$ represent mode competition between M1 and M2.  The M2 temporal mode is unstable for $m < 75$, and the C/A transition occurs near $m=170$. }
\label{fig:competelam}
\end{figure}

Referring to parametric study~(1) and figure~\ref{fig:CAplots}(b), branch II (varying $m$) of the C/A transition `follows' closely the neutral stability curve for a second temporal mode, M2. The same neutral curve is also present in parametric study~(2), but is further removed from the C/A transition curve, and occurs at $\Reint/\thickness \approx 1.5\times  10^4$.   As in the turbulent case, it is as if the existence of the M2 mode quenches the M1 absolute instability.  Again, as in the turbulent case, there is clear evidence of competition between spatio-temporal modes, in addition to temporal mode competition. In Figure~\ref{fig:competelam}, $\omr$ is shown for the least stable mode at each complex $\alpha$.   Results are presented at different values of $m$ for $Re=20000$, $\thickness=0.1$, and $r=1000$.   The value $m=85$ describes a situation where M2 has just become stable while M1 is convectively unstable (the point of neutral stability for M2 is at $m=75$).  The C/A transition for M1 occurs at $m=170$. 
The spatial curve $\omi = 0$ of the most dangerous mode is identified with M1 by a Gaster-type analysis~\citep{Gaster1962}. The
%
%
%Spatio-temporal mode competition and the C/A transition: contour plots of $\omr$ for the most dangerous mode in the complex %$\alpha$-plane.    The black lines represent spatial curve $\omi=0$ of M1.   Here $Re=5000$, and $\thickness=0.1$.   The M2 temporal %mode is unstable  for $m < 800$, and the C/A transition occurs near $m=1100$.    The sharp jumps in $\omr$ represent mode competition %between M1 and M2, and the saddle point in (a),(b) does not correspond to M1.
%
contours $\omr = \mathrm{Const}.$ that connect orthogonally to the spatial curve are identified also
with M1.   In this
way, we have established that the saddle point in figure~\ref{fig:competelam}(a) does not correspond to M1.
Thus mode competition interferes most dramatically with the saddle point near the point of neutral temporal stability of M2 (near $m=75$), while an approach to the C/A transition from within the convective regime (i.e., an increase in $m$ from a low value) causes the mode competition to disappear gradually, thus producing a conventional single-mode  saddle point.  This strongly suggests that the proximity of the M2 neutral curve and the M1 C/A transition in the flow diagram is no coincidence, and is in agreement with the earlier results for the turbulent case.
The same phenomenon persists for lower Reynolds numbers (e.g. $Re=5000$).  However, we have presented the results for $Re=20000$ because this regime exhibits the mode competition most clearly.  At such high Reynolds numbers, a third unstable mode comes into existence (visible in Fig.~\ref{fig:competelam} in the neighbourhood $\ai=0$ and $\ar=2$).  This mode has a critical Reynolds number $Re\approx 8000$ and its neutral curve is almost independent of $\Reint$, $m$, and $\thickness$.
However, this third mode does not play any role in the spatio-temporal mode competition and is not discussed any further.

Finally, we have applied the spatio-temporal energy budget to the laminar flow for the cases corresponding to Figure~\ref{fig:raytrans}(a), (c) and (e). The results indicate that over most of the pulse width, $TAN$ dominates: this represents a source of energy for the instability that is associated with the viscosity-contrast mechanism, which dominates in temporal instability.  However,  $REY_G$ also plays a significant role. In all three cases, $REY_G$ is negative on the downstream side of the pulse, but positive on the upstream side. This is most visible in the absolutely unstable case.  
In contrast, for the turbulent base state, it is \textit{only} in the absolutely unstable case that $REY_G$ develops this asymmetry, where it even becomes the dominant term on the upstream side in the energy-budget analysis.

\subsection{Ray analysis revisited}

\begin{figure}
\centering
\subfigure[]{\includegraphics[width=0.48\textwidth]{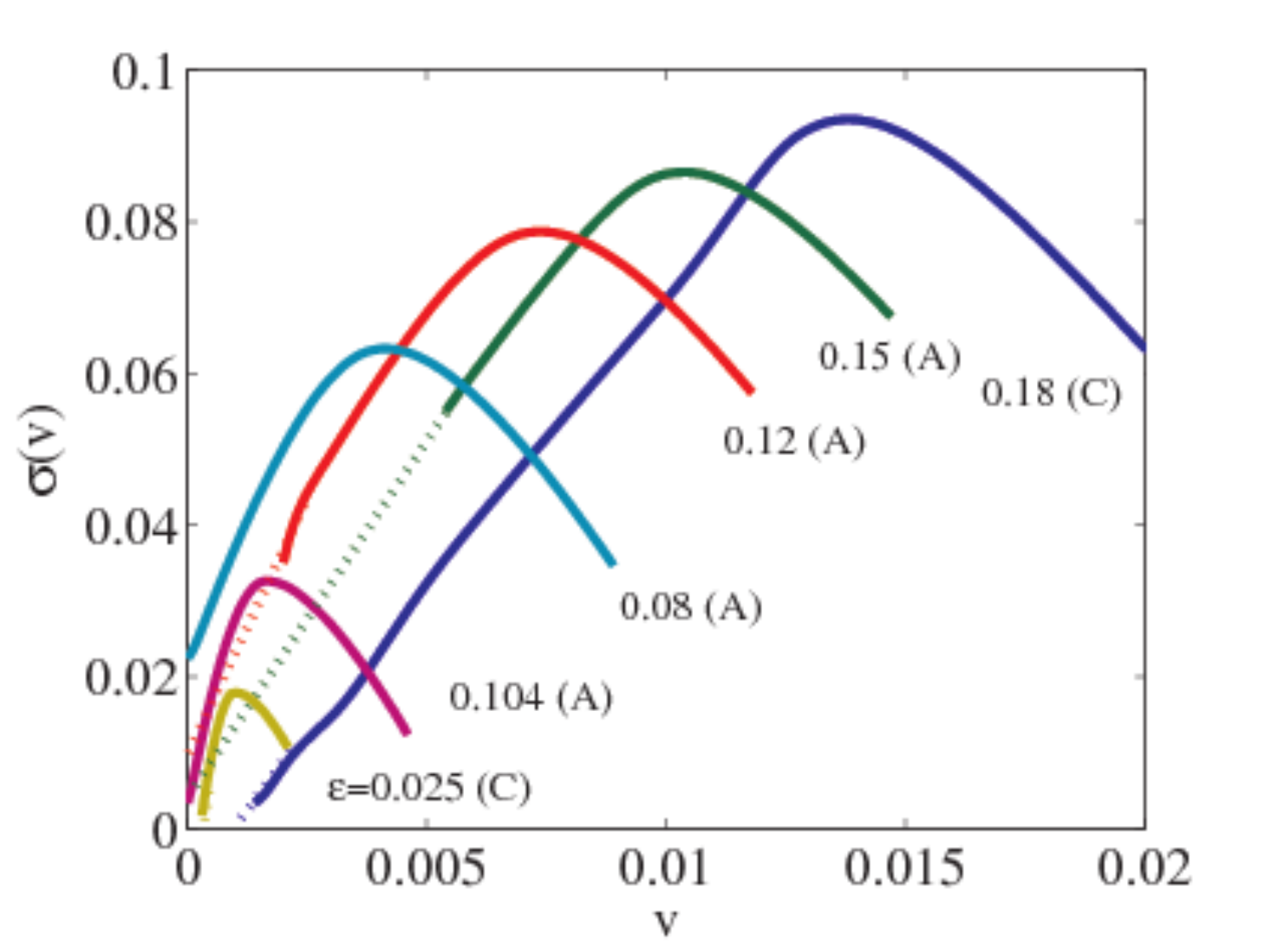}}
\subfigure[]{\includegraphics[width=0.48\textwidth]{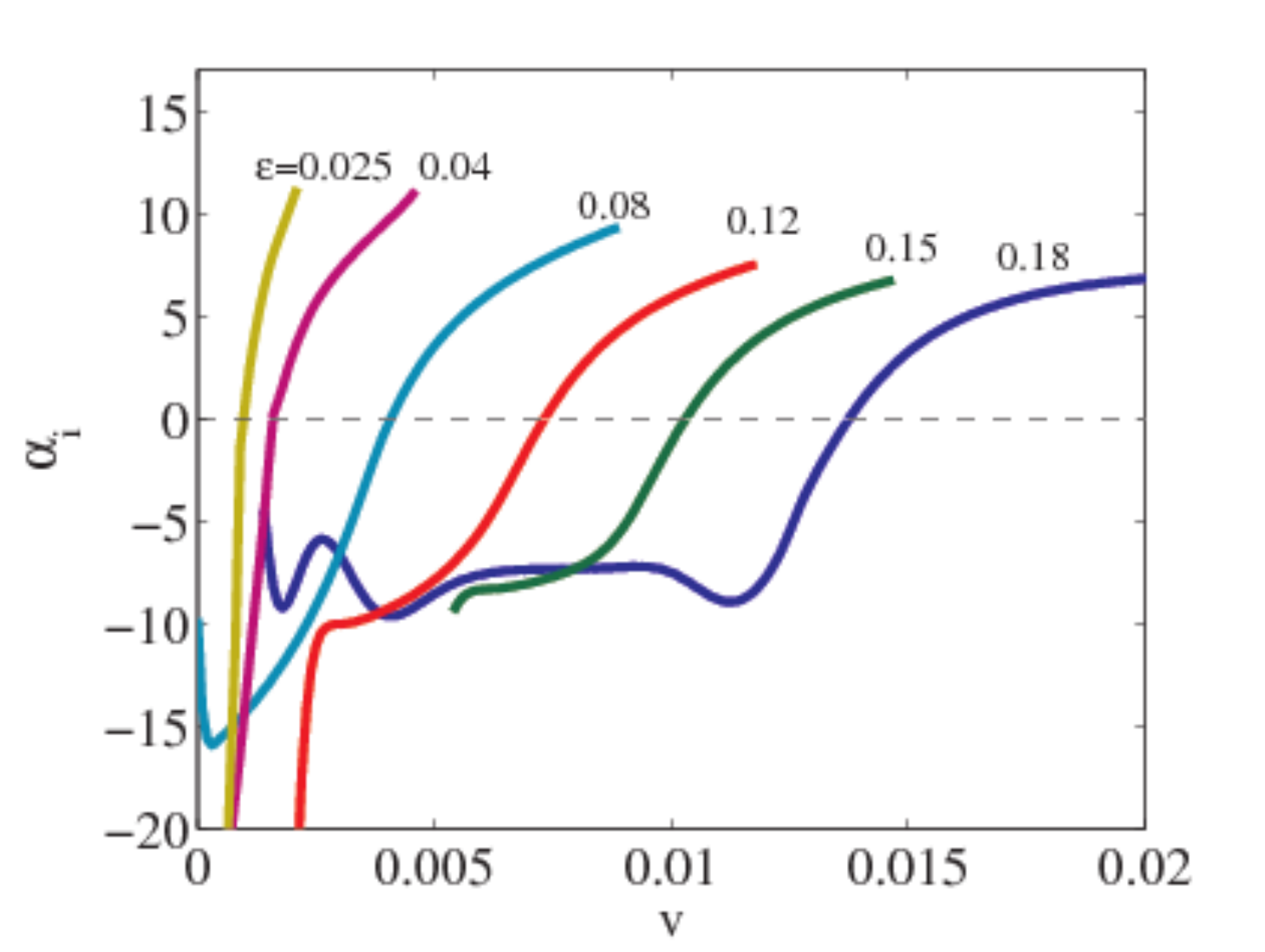}}
\subfigure[]{\includegraphics[width=0.48\textwidth]{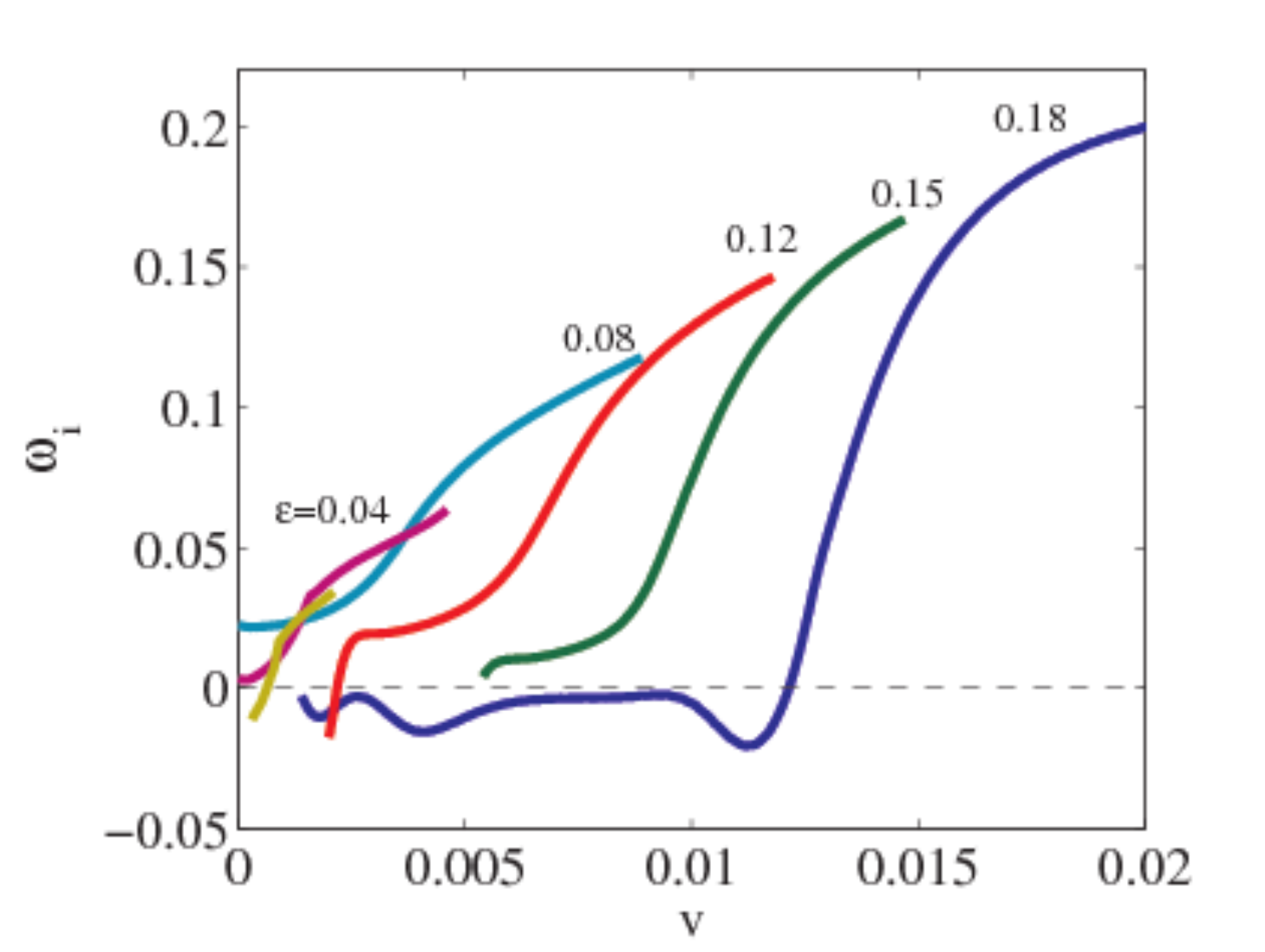}}
\caption{Ray analysis: (a) spatio-temporal growth rates, (b) spatial growth rates along rays; (c) temporal growth rates along rays, for laminar flow at various values of $\thickness$, at $r=1000$, $m=150$, $Re=1500$.   The rest of the parameters are the same as in Figure~\ref{fig:raytrans}.  Trend lines have been added in (a) merely to guide the eye.}
\label{fig:sigmalam}
\end{figure}

Some further information about the absolute instability of M1 is obtained from a ray analysis (Figure~\ref{fig:sigmalam}).
The parameters  are taken from the cases studied in Figure~\ref{fig:raytrans}.  The branches of the C/A boundary in Figure~\ref{fig:CAplots}(a) can be mapped on to families of $\sigma(v)$-curves in Figure~\ref{fig:sigmalam}: decreasing $\thickness$ from $\thickness\approx 0.18$ in the latter gives a family of $\sigma(v)$-curves that is associated with the upper branch I of the C/A transition in Figure~\ref{fig:CAplots}(a) ($m$ constant, $\thickness$ varying).   As  $\thickness$ is decreased, the $\sigma(v)$-curves shift
downwards in their entirety, until the growth rate $\sigma(v=0)$ eventually becomes positive,
indicating a switchover to absolute instability.
Conversely, increasing $\thickness$ from $\thickness\approx 0.025$ gives a family of $\sigma(v)$-curves that is associated with the lower branch III in Figure~\ref{fig:CAplots}(a).
These curves are steep for small $\thickness$-values, such that $\sigma(v)$ is negative.
As $\thickness$ increases, the slope diminishes, until $\sigma(v=0)$ is positive, and absolute instability
is attained from below. 
Furthermore, in a convectively unstable regime at large $\thickness$-values, $\sigma(v)$ is linear in $v$ for small $v$, and the corresponding $\ar$- and $\omi$-values approximate negative constants, consistent with Equations~\eqref{eq:tdns_growth}.
Again with reference to Equations~\eqref{eq:tdns_growth}, the upper branch in Figure~\ref{fig:CAplots}(a) corresponds to a shift in the value $v$ for which $\sigma=\sigma_{\rm max}$ (hence $\ai=0$), whereas the transition at the lower branch is associated with an increase in temporal relative to spatial growth.
Finally, for large values of $\thickness$, the $\sigma(v)$-curve is very wide, indicating a rapid spreading of an initial pulse, whereas for low values the pulse remains more localized. 

For the cases considered herein, we have verified that the wavenumber of the spatio-temporally most dangerous mode coincides with that of the temporally most dangerous mode (Table~\ref{tab:max_sigma}).
\begin{table}
\centering
\begin{tabular}{ccccc}
$\thickness$ & $\sigma_{\mathrm{max}}$(ST) & $\lambda_{\mathrm{max}}$(Modal) & $\alpha$(ST) & $\alpha$(Modal)\\
\hline
0.18& 0.0934 & 0.0936 & 15.2& 16.3\\
0.15& 0.0865 & 0.0866 & 16.9& 16.5\\
0.12& 0.0787 & 0.0788 & 17.6& 17.2\\
0.08& 0.0633 & 0.0633 & -- & 20.8\\
0.04& 0.0326 & 0.0326 & -- & 30.9\\
0.025& 0.0179 & 0.0179 & -- & 43.3\\
\end{tabular}
\caption{Comparison between modal and spatio-temporal analyses at various values of $\thickness$, at $r=1000$, $m=150$, and $Re=1500$.  The rest of the parameters are the same as in Figure~\ref{fig:raytrans}.}
\label{tab:max_sigma}
\end{table}
It was not possible to compute this wavenumber for all of the $\thickness$-values, and our finite-difference discretization of Equation~\eqref{eq:arv} is not always accurate.  It is for this reason that we did not include $\alpha_r$-$v$ plots in Figure~\ref{fig:sigmalam}.  Nevertheless, Figure~\ref{fig:sigmalam} still contains important information: a previous set of purely spatial and purely temporal analyses (see~\citet{Valluri2010} and Appendix~\ref{app:spatial}) found extremely large purely spatial growth rates (exceeding the least-negative purely spatial growth rate by an order of magnitude) that were not excited in a DNS.  From the spatial growth rates along rays in Figure~\ref{fig:sigmalam}(b), we see that such extremely large spatial growth rates are not selected, instead, the spatial growth rate comes from the same eigenmode as the temporally most dangerous mode.

\section{Discussion}
\label{sec:conc}

\subsection{Flow-regime maps}

\begin{figure}
\centering
\includegraphics[width=0.49\textwidth]{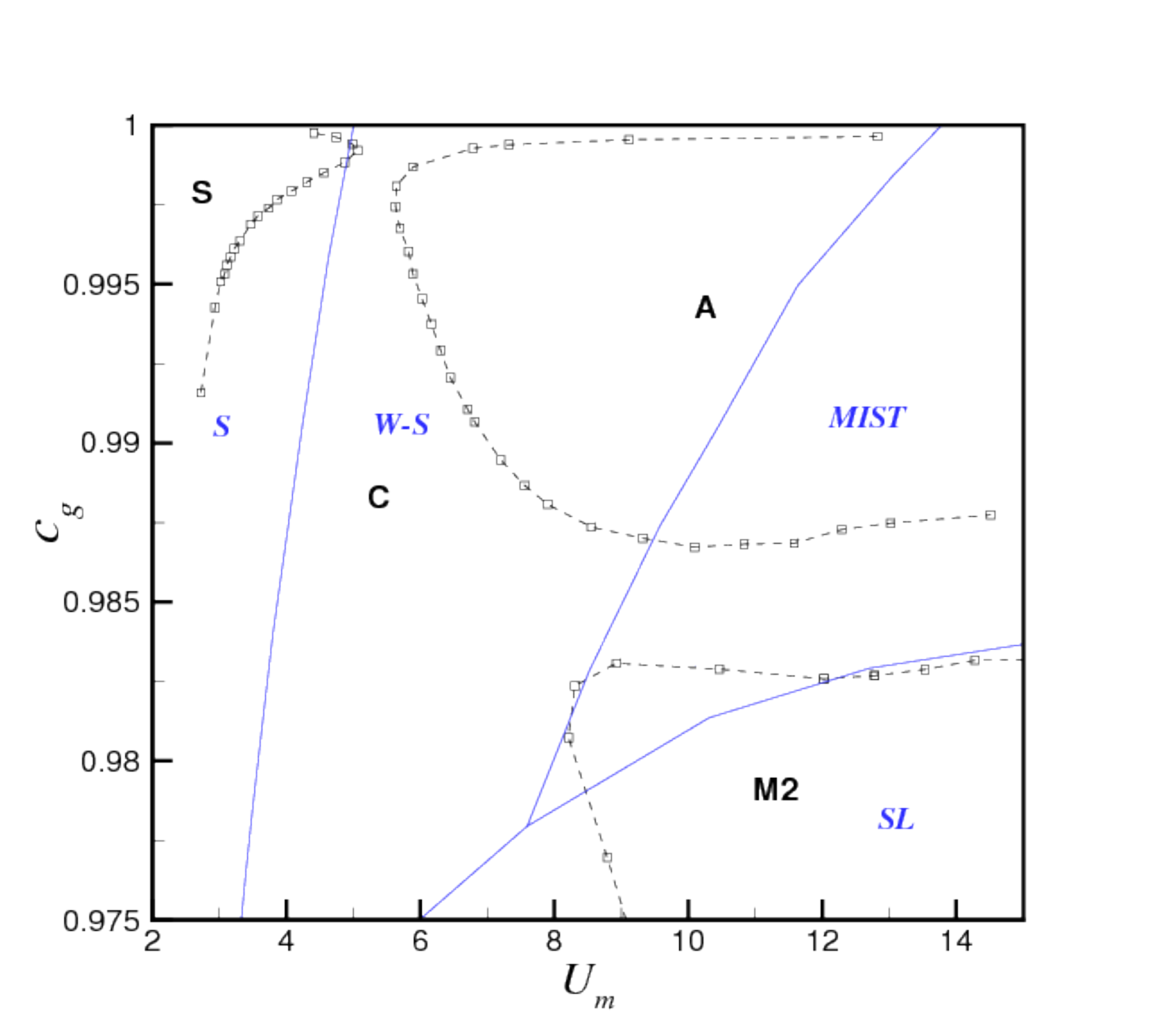}
\caption{Flow-regime map Figure~\ref{fig:CAturb} in terms of superficial gas and liquid velocities. The curves with symbols are the present work; other solid  lines and classifications in italics are experimental findings by \cite{Hoogendoorn59} for spindle oil/air flow in a 91mm diameter horizontal pipe wherein the following flow regimes are identified: stratified flow with a flat interface (S), wavy-stratified flow (W-S), wavy-stratified flow with suspended droplets (MIST), and slug flow (SL).}
\label{fig:Hoogendoorn}
\end{figure}

%\commentpdms{Need to check if there really are no oil/gas data for channels}
It is of interest to compare our regime boundaries (e.g., Figure~\ref{fig:CAturb_collapse}) with flow-regime maps determined from experiments.
Our theory is of course applicable to water/air systems, for which channel-flow experimental studies are available.  However, the results in Sec.~\ref{sec:turb} prove that water/air systems are usually convectively unstable, while the main focus of the present paper is on absolutely unstable systems.  Consequently, we have searched the literature for experiments concerning air and oil, for which the viscosity contrast is much higher, and for which absolute instability is expected.  We have found that such experiments were conducted on pipes (with the exception of the work by~\citet{Gondret99} on Hele-Shaw cells).  Nevertheless, a comparison between our channel-flow model and a pipe-flow experiment is justified, since a 2D analysis may be representative of the mid-plane in pipe flow, especially for thin films, linear transverse waves (in the third dimension) are usually overwhelmed by 2D waves, and in any case it is necessary to demonstrate that the boundaries determined in our study are not restricted to an obscure limit.

The results of Figure~\ref{fig:CAturb}(a) are therefore presented in Figure~\ref{fig:Hoogendoorn} in terms of superficial velocities $U_G$ and $U_L$, where $c_g=U_G/(U_G+U_L)$ and $U_m=U_G+U_L$ is the mixture velocity for parameter values approximately corresponding to the flow of air and spindle oil through a pipe of 91mm diameter, and flow-regime boundaries determined experimentally by \cite{Hoogendoorn59} are superposed (our data are for $m=1000$, $r=1000$ and the choice of $\surft$ and $\grav$ mentioned at the start of Section~\ref{sec:turb}, which are also representative of the experimental conditions stated in \cite{Hoogendoorn59}). The more widely used flow-regime map by \cite{Mandhane74} cannot be used here, as it is for water/air systems, but they observed good agreement with water/air data also published by \cite{Hoogendoorn59}.

We first note from Figure~\ref{fig:Hoogendoorn} that the neutral stability curve from the temporal analysis agrees reasonably well with the experimental data on the transition from stratified flow (i.e., no waves) to wavy stratified flow. This is expected based on the comparisons with experiments in our previous work on temporal instability, \citet{Onaraigh2011a}. We note here that the superficial velocities in a 2D system are expected to be larger than for a pipe flow at the same maximal velocities in a fluid, which may have reduced the difference between the results somewhat. On the other hand, the fact that the liquid layer at the centre of the pipe is expected to be smaller due to gas pushing the liquid to the side of the pipe is expected to be of less significance, as the liquid height is the parameter along the theoretical data in the figure.

The main C/A transition is predicted theoretically here at a total superficial velocity somewhat below the experimentally determined transition from wavy-stratified flow to mist flow (waves with droplets). A relation with the onset of atomization for very viscous liquids cannot be ruled out. In fact the experimental mist flow regime lies almost entirely in the absolutely unstable regime, since the critical value of the parameter $c_g$ for this regime is close to the high-$Re$ C/A transition from the theory (and even more so, the M2 neutral curve). In the experiments, slug flow is observed at $c_g$ below this critical value at large $U_m$. Hence slug flow is observed experimentally when according to the 2D theory the flow is convectively unstable.

\subsection{Potential use in global and non-linear analyses}
 Although we have identified and classified the nature of the linear instability for our stated two-phase stratified problem, we briefly outline extensions that follow naturally from the present work.

The stability analysis considered throughout this paper is a `local analysis' in the sense that the base flow varies on a length scale which is long relative to the wave length of the instability waves (this being true whether the analysis is linear or not). 
A more general approach is global stability analysis, wherein the whole of the physical domain is considered~\citep{huerre90a,Chomaz05}.  Under such a technique, spatial streamwise variation is accounted for in both the base flow and the perturbation terms, permitting the study of nonparallel flows, i.e., no restrictions are placed on spatial scalings. Of course, such studies require appropriate computational power and therefore have only been pursued recently.  In the context of the present work, such an approach permits changes in the thickness of the liquid layer which is important for many of the stated applications, but it means that ray analysis in the form used herein can no longer be employed, as periodic boundary conditions no longer apply. 

The theoretical analysis required to elucidate non-linear effects discussed herein is elaborate.  Thus, an important additional step might involve a numerical (e.g., DNS) study and experiments, with the aim of identifying which key non-linear effects dominate. Specifically, whether a transition to absolute instability in the nonlinear regime may precede the linear C/A transition (this has been observed by~\citet{Gondret99} in Hele-Shaw cells).
Another question that a full numerical simulation might address concerns the location of the dominant non-linear interactions, i.e., at the phase interface or in the bulk of the liquid layer.   A full numerical simulation  would also permit a comparison between  periodic boundary conditions (as used in the present work), and inlet conditions (as in a true channel flow).  In this paper, the ray analysis is conducted in Fourier space, and required periodic boundary conditions.  However,  the pulse-type initial conditions considered herein involve extremely long domains, and comparisons with even larger domains are self-consistent. Nevertheless, full numerical simulations would further validate these observations. Such a study and the subsequent non-linear analysis will form the basis of future work.

\subsection{Summary}

We have studied the linear stability of a liquid layer sheared by laminar or turbulent gas flow in a two-dimensional Cartesian geometry. A Poiseuille (laminar) or Reynolds-averaged (turbulent) model is used to describe the stratified two-phase base flow.  Results from a normal-mode analysis and a ray analysis show that the base flow is absolutely unstable to linear perturbations for large regions of parameter space. In particular, for density ratios of $O(1000)$, clear evidence is given of oil/gas flows being absolutely unstable for a sufficiently large dynamic viscosity ratio (at least $O(10^2)$ for the laminar and $O(10^3)$ for the turbulent base state), provided the Reynolds number of the flow is sufficiently high. Since the dynamic viscosity ratio for air/water systems is only $m\approx 55$ at $20^{o}$C, the present results for the turbulent case lead us to conclude that  laboratory experiments carried out for water/air may not be representative for oil/gas systems.

 In both turbulent and laminar base states, the flow-regime map collapses in the ($Re$, $\Reint$) plane, where $\Reint$ is the liquid-based Reynolds number (with a small subset of exceptional cases). A recently developed theory for C/A transitions~\citep{Onaraigh2012a} has enabled us to formulate a simple criterion for the onset of absolute instability, based on a competition between instability growth and convection by the group velocity.  A single number, based on the temporal growth properties, encodes the tendency for instability growth.  This number is governed by $\thickness$, $m$, and $Re$, but becomes parameter-free in certain limits, while the behaviour of the group velocity is governed simply by $\Reint$.  This criterion therefore explains  the importance of $\Reint$ in the C/A transition curves, and, moreover, when examined in detail, explains the collapse of the transition curves into universal forms.

Independent confirmation of our results from the modal analysis has been obtained using a ray analysis. The ray analysis benefits from our development of an efficient method of solution for linear differential-algebraic equations, which allows us  to extend drastically the simulation time over which the evolution of an initially pulsed disturbance can be traced, thereby eliminating the problems encountered with DNS by~\citet{Valluri2010}.  This method all but guarantees the detection of absolute instability, if it exists.
Additionally, the ray analysis leads naturally to an energy-budget method that determines the origin of the spatio-temporal instability.
Results from this method demonstrate the importance of the viscosity-contrast mechanism acting at the interface, as well as a transfer of energy from the bulk gas flow to the waves, as the sources of the instability.   Nevertheless, there exist several  internal modes (which also derive energy from  the interface) that are at least as significant in other cases.
A further application of the ray analysis is envisaged for flows wherein the complex dispersion relation contains singularities (e.g. if $\omega(\alpha)$ has a pole~\citep{Healey2007,Healey2009} or root-type behaviour~\citep{Aships1990} at some point $\alpha_0$).  In such problems, the application of the saddle-point criterion, with its associated construction of the steepest-descent path enclosing all the dynamically-relevant singularities, is a non-obvious task~\citep{Juniper2010}.  For these cases, a straightforward ray analysis could be used to verify the correctness of one's implementation of  steepest-descent method.
The ray analysis suffers from the amplification of numerical error when the spatial growth rates are large, meaning that only a  small region around the impulse centre can be used to extract meaningful information.  Thus, both methods are necessary to obtain a complete picture of the stability properties.
The ray analysis and the quadratic approximation are quite general, and we anticipate that a wide range of multiphase flow scenarios  can be tackled using these techniques.

\appendix

\section{Validity of the quadratic approximation as applied to the C/A transition of two-phase stratified flow}
\label{app:quadratic}

We review the so-called `quadratic approximation' developed by~\citet{Onaraigh2012a} and demonstrate that it approximates well the precise criteria for the onset of absolute instability.  The quadratic approximation is based on the following identity for the analytic continuation of the growth rate $\omi$ into the complex plane:
\begin{equation}
\omi(\ar,\ai)=\omitemp(\ar)+\sum_{n=0}^\infty \frac{(-1)^n}{(2n+1)!}\frac{d^{2n}\cg}{d\ar^{2n}}\ai^{2n+1}
+\sum_{n=0}^\infty \frac{(-1)^{n+1}}{(2n+2)!}\frac{d^{2n+2}\omitemp}{d\ar^{2n+2}}\ai^{2n+2},
\label{eq:app:omi_taylor}
\end{equation}
where $\cg$ is the group velocity $d\omr/d\ar$ in a purely temporal analysis.
This is a consequence of the Cauchy--Riemann conditions on $\omega(\alpha)$ viewed as a holomorphic function on an appropriate open subset in the complex plane, and has been derived elsewhere by~\citet{Onaraigh2012a}.  We truncate this series at quadratic order in $\ai$ and apply the conditions $\partial\omi/\partial\ar=\partial\omi/\partial\ai=0$ for a saddle point.  This yields the conditions
\begin{equation}
\frac{d\omitemp}{d\ar}+\frac{d\cg}{d\ar}\ai-\tfrac{1}{2}\frac{d^3\omitemp}{d\ar^3}\ai^2=0,\qquad
\cg(\ar)-\frac{d^2\omitemp}{d\ar^2}\ai=0.
\label{eq:app:ai_linear}
\end{equation}
Simultaneous solution of these equations yields the following $\ar$-value for the location of the saddle point:
\begin{equation}
%-\frac{d\omitemp}{d\ar}\frac{d^2\omitemp}{d\ar^2}=\cg(\ar)\frac{d\cg}{d\ar}.
\cg(\ar)\frac{d\cg}{d\ar}=-\frac{d\omitemp}{d\ar}\frac{d^2\omitemp}{d\ar^2}+\tfrac{1}{2}\cg^2(\ar)\left(\frac{d^3\omitemp}{d\ar^3}\bigg\slash\frac{d^2\omitemp}{d\ar^2}\right).
\label{eq:app:saddle}
\end{equation}
(Note that for a strictly quadratic approximation, as in the main part of the paper, the third derivative should be neglected here.)
At the onset of absolute instability, $\omi=0$ at the saddle point.  Substitution of this condition into the quadratic approximation to Equation~\eqref{eq:app:omi_taylor} yields the following criterion for the onset of absolute instability:
\begin{equation}
-\frac{d^2\omitemp}{d\ar^2}\bigg|_{\ar^*}\omitemp(\ar^*)=\tfrac{1}{2}\cg^2(\ar^*).
\label{eq:app:saddle_sign}
\end{equation}
where $\ar^*$ is the root of Equation~\eqref{eq:app:saddle}.  

\begin{figure}
\centering
\hspace*{0cm}\vspace{-2.5cm}

\subfigure[]{\includegraphics[width=0.49\textwidth]{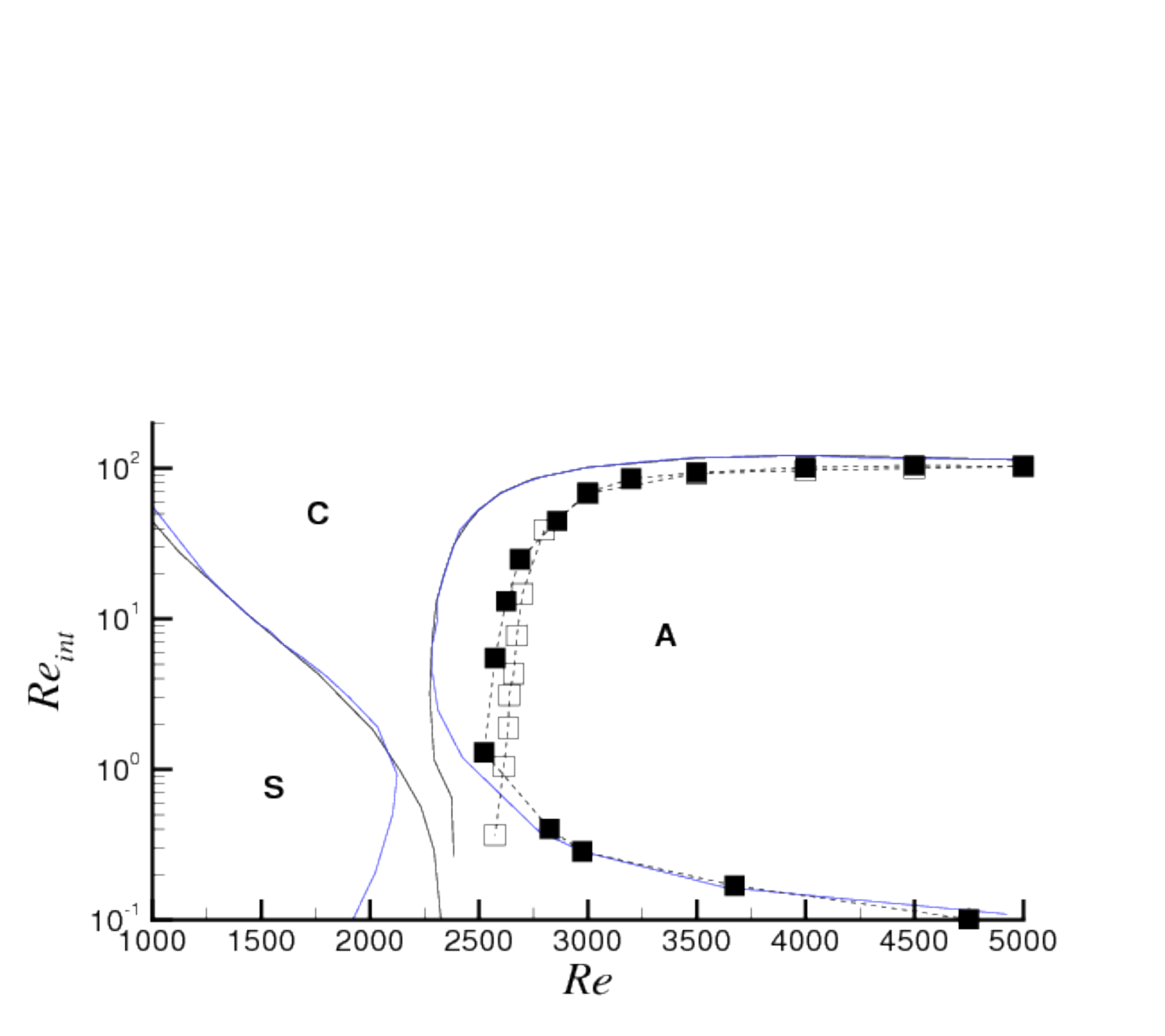}}
\subfigure[]{\includegraphics[width=0.49\textwidth]{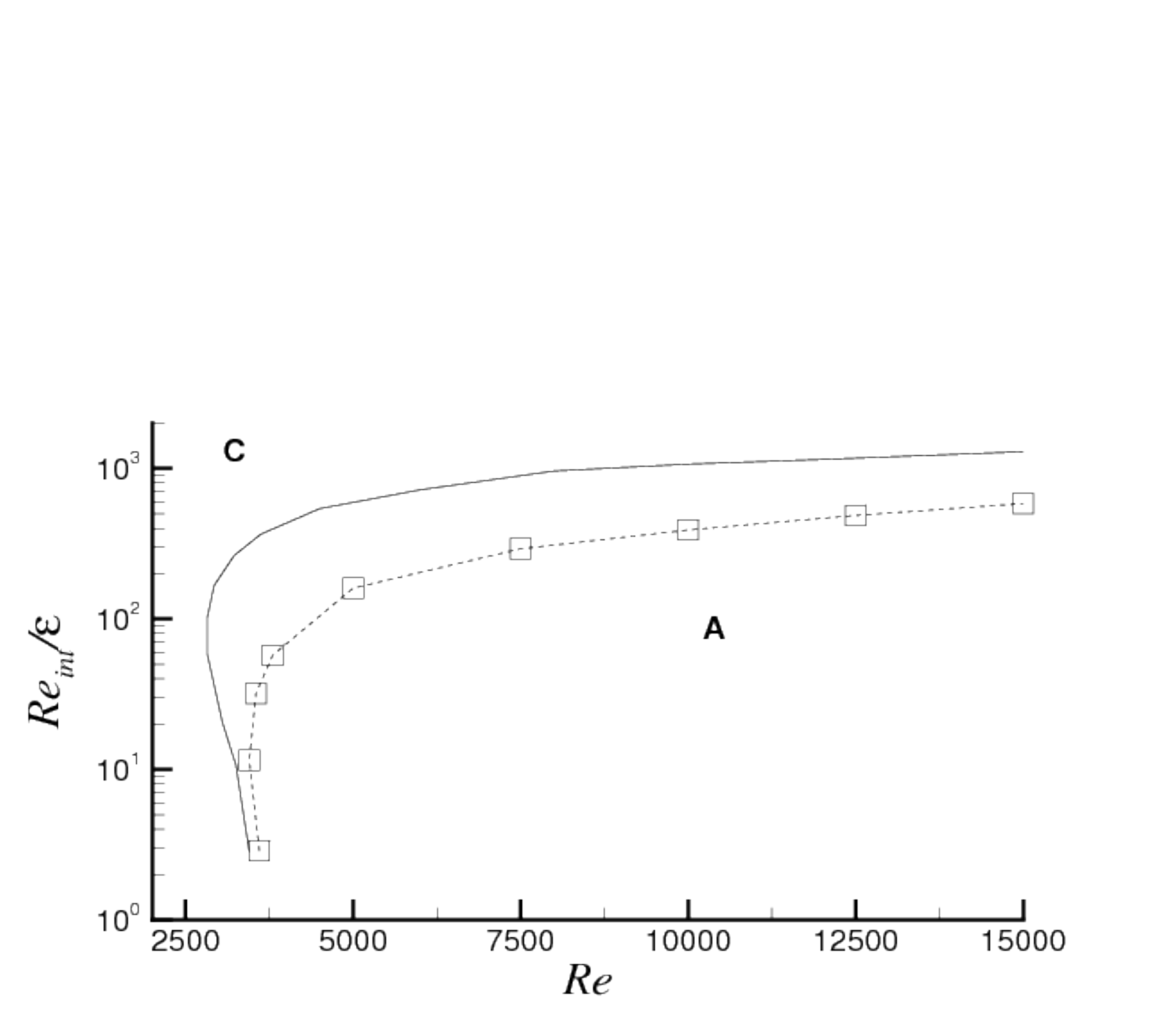}}
\caption{Comparison of flow-regime maps resulting from the quadratic approximation (open symbols, keeping $m$ constant; filled symbols, keeping $\epsilon$ constant) and the full modal analysis (solid lines) for the turbulent (a) and laminar (b) base state.}
\label{fig:quadcomp}
\end{figure}

In Figure~\ref{fig:quadcomp}, the results of the quadratic approximation are compared with the full modal results that have been taken from Figures~\ref{fig:CAturb_collapse} and~\ref{fig:CAplots_collapse2} for the turbulent and laminar base states, respectively. All 
branches are qualitatively correctly predicted, and quantitatively accurate results are even obtained at some, especially all results at large $Re$ in the turbulent case, and those for low $Re$ and a large viscosity ratio in both laminar and turbulent cases.

\section{Analysis of purely spatial modes}
\label{app:spatial}

The focus throughout the paper has been on spatio-temporal analysis.  In particular, our modal analysis has involved the extraction of saddle points from the solution of an eigenvalue problem in a complex wave number space.  In this section, we review the Gaster theory for the extraction of purely spatial growth rates.  The focus herein is on the turbulent base state.

The most basic description of linear instability is a temporal analysis, which involves the solution of the eigenvalue problem~\eqref{eq:os_abbrv} for $\alpha=\ar$ only.  This gives a dispersion relation $(\omr(\ar),\omi(\ar))$.  The pair $(\omr(\ar),\omi(\ar))$ that maximizes $\omi$ is called the {\textit{most dangerous temporal mode}}.  The maximum is taken over the complete spectrum of modes in the linear Orr--Sommerfeld (OS) equations.  The system is unstable if $\omi>0$ at the most dangerous temporal mode.
This case has been investigated exhaustively by the present authors~\citep{Onaraigh2010,Onaraigh2011a,Onaraigh2011b}.  

\begin{figure}
\centering
\subfigure[]{\includegraphics[width=0.45\textwidth]{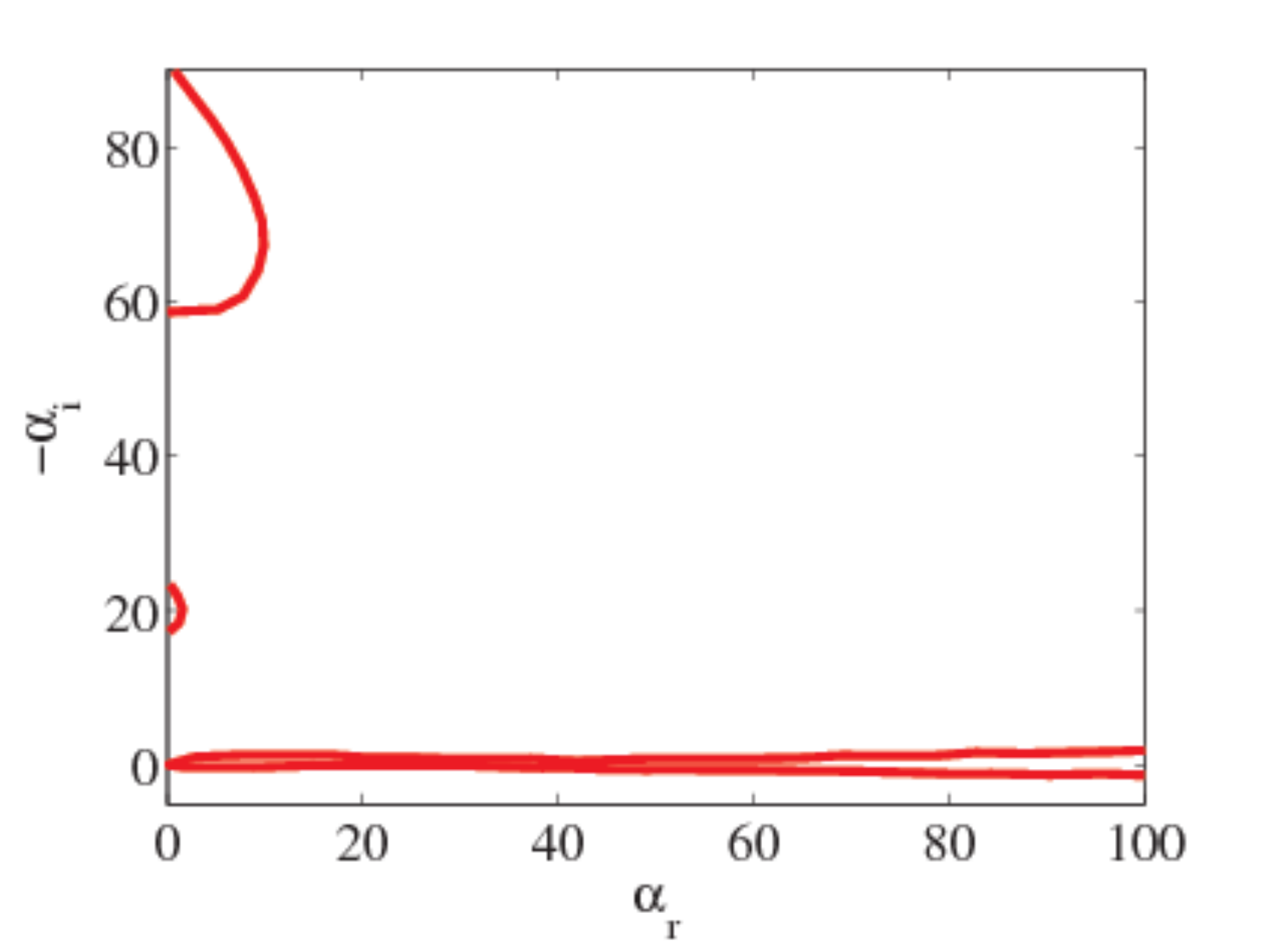}}
\subfigure[]{\includegraphics[width=0.45\textwidth]{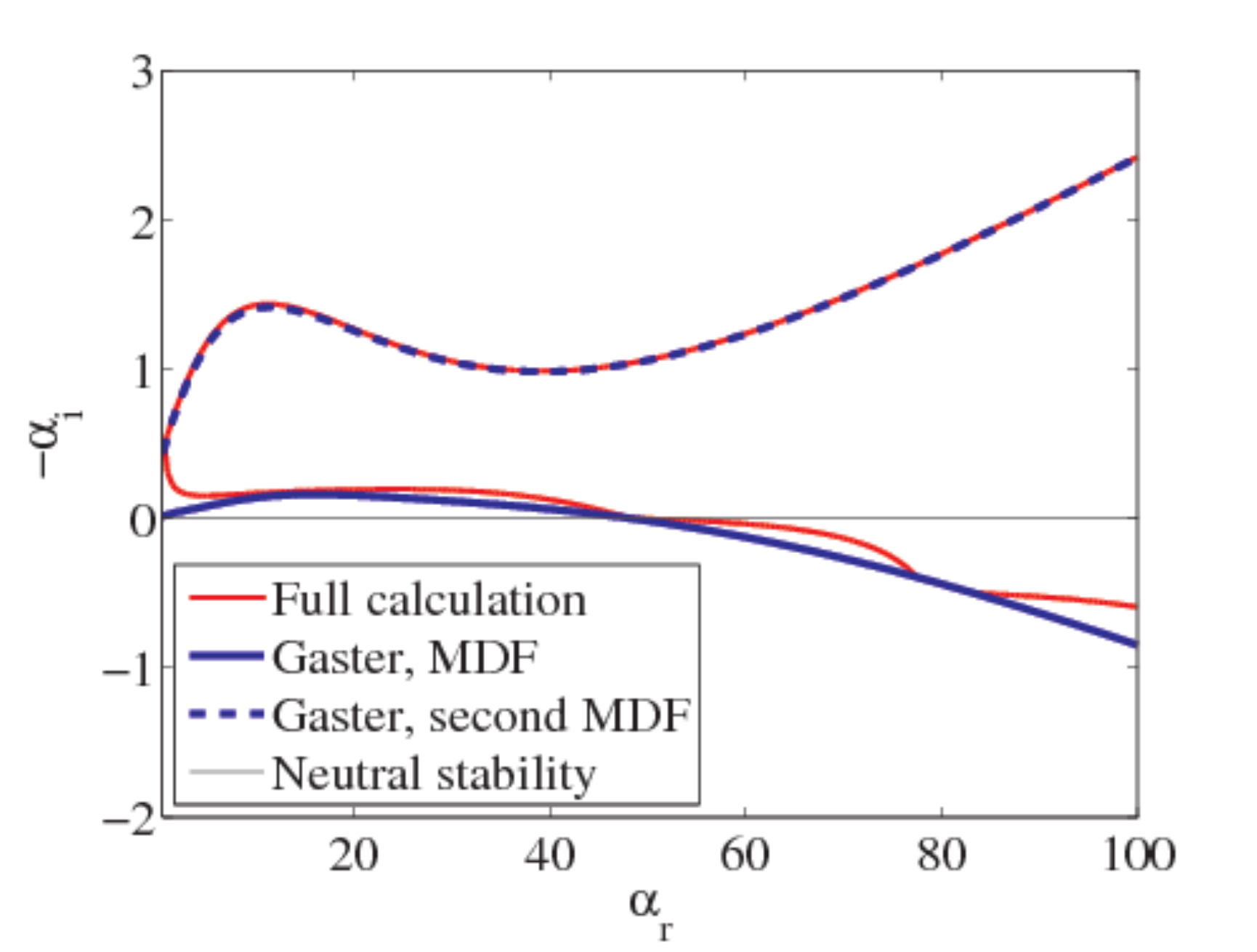}}
\caption{(a) Spatial stability study; (b) Comparison with Gaster analysis.  Here `MDF' refers to the `most-dangerous frequency', or, equivalently, the most-dangerous mode.}
\label{fig:dispersion_space}
\end{figure}

We therefore turn to the spatial case, wherein we are concerned with
the dispersion relation $\ai=\ai\left(\ar,\omi=0\right)$, which can be obtained by treating the OS problem~\eqref{eq:os_abbrv} as a non-linear eigenvalue problem in $\alpha$.  In practice, we do not solve this complicated problem.  Instead, we solve Equation~\eqref{eq:os_abbrv} in the complex $\alpha$-plane,  $\alpha:=\ar+\imag\ai$.   The result is a discrete spectrum of frequencies $\{\omega_n(\alpha)\}_n$, which we order according to the magnitude of $\Im\left[\omega_n(\alpha)\right]$: the \textit{most dangerous mode at complex wave number $\alpha$} is denoted by $\omega_C(\alpha)$, and is defined such that
\[
\Im\left[\omega_C(\alpha)\right]=\max_{n,\text{ all modes}} \Big\{\Im\left[\omega_{n}\left(\alpha\right)\right]\Big\}_{\alpha\in\mathbb{C}}.
\]
We plot the zero contour of $\Im\left[\omega_C(\alpha)\right]=0$, which gives the desired relation $\ai=\ai\left(\ar,\omi=0\right)$.  We repeat these steps  for the less dangerous modes.  This  results in multiple \textit{spatial curves}.  These  are displayed in   Figure~\ref{fig:dispersion_space}(a) for the parameter values $m=55$, $r=1000$, $\thickness=0.05$, with $\grav$ and $\surft$ given by Equation~\eqref{eq:fr_values}.  However, in this figure, we plot only the spatial curves produced by the most dangerous mode;  the less dangerous modes do not produce  the same large spatial amplification (spatial amplification corresponds to $\ai<0$), and are not shown.  The large spatial growth rates in the figure would appear to dominate in an evolving flow.  In practice, however, they are not observed, neither in turbulence  nor in laminar flows. 

Because the large spatial growth rates are not accessed, we focus our attention  on the spatial curves near $\alpha_\mathrm{i}=0$, for which an analytical theory is available.
\citet{Gaster1962} has shown that small spatial growth rates are related to the temporal ones through the formula
\begin{equation}
-\alpha_\mathrm{i}=\omega_\mathrm{i} \Big / \frac{\partial\omega_\mathrm{r}}{\partial\alpha_\mathrm{r}},
\label{eq:gaster}
\end{equation}
where the quantities on the right-hand side are derived from the purely temporal analysis.  We test the applicability of Equation~\eqref{eq:gaster} in Figure~\ref{fig:dispersion_space}(b).  We plot the Gaster curves associated with the temporally most dangerous mode and the second most dangerous mode in the figure, and compare the result with the contour in Figure~\ref{fig:dispersion_space}(a).  There is good agreement between the two theories.  However, the curve generated from the contour analysis exhibits `kinks' where the most dangerous mode switches from one eigenmode to another.
To obtain perfect overlap between the two curve sets, it is necessary to pick out all the eigenmodes from the contour analysis and to cut and re-connect the contours so as to obtain smooth curves.  This exercise is difficult but yields no new information, and we do not pursue it here.  Instead, we focus on the Gaster study, where two spatial curves overlap with the contour-generated ones.  The Gaster curves are obtained from the first and second temporally most dangerous modes.  Thus, the character of a spatial instability is identified with the character of the temporal instability through the Gaster formula~\eqref{eq:gaster}.  The Gaster curve suggests that the second most dangerous temporal mode  contributes most strongly to the spatial instability.  However, it is not obvious from this simple analysis which eigenmode is excited in an impulse-response scenario: this question is studied in detail in Section~\ref{sec:laminar}, where it is demonstrated that the spatial curve associated with the temporally most dangerous mode is the one that is selected.

\subsection*{Acknowledgements}
The authors would also like to thank J.-C. Loiseau for carrying out preliminary numerical investigations for this project.

\bibliographystyle{jfm}
%\bibliography{turbulence_bibliography}

\end{document}